\documentclass[traditabstract]{aa}
\usepackage{times}
\usepackage{longtable}
\usepackage{natbib}
\usepackage{rotate}
\usepackage{lscape}
\usepackage{amssymb}
\usepackage{graphicx}
\usepackage[draft]{hyperref} 

\bibpunct{(}{)}{;}{a}{}{,} 

\def\5{\footnotesize V\normalsize}
\def\4{\footnotesize IV\normalsize}
\def\3{\footnotesize III\normalsize}
\def\2{\footnotesize II\normalsize}
\def\1{\footnotesize I\normalsize}
\def\siv{\scriptsize IV\footnotesize}
\def\siii{\scriptsize III\footnotesize}
\def\sii{\scriptsize II\footnotesize}
\def\si{\scriptsize I\footnotesize}

\def\lam{$\lambda$}
\def\kms{$\mbox{km s}^{-1}$}
\def\p{$\phantom{:}$}
\def\q{$\phantom{?}$}

\def\v{$\phantom{^{l}}$}
\def\pp{$\phantom{-}$}
\def\o{$\phantom{1}$}

\newcommand{\vsini}{\ensuremath{v_{\rm e} \sin i}}

\begin{document}

\title{The VLT-FLAMES Tarantula Survey}
\subtitle{XVIII. Classifications and radial velocities of the B-type stars}

\author{C. J. Evans\inst{1}, M. B. Kennedy\inst{2}, P. L.
  Dufton\inst{2}, I. D. Howarth\inst{3}; N. R. Walborn\inst{4},
  N. Markova\inst{5},
  J. S. Clark\inst{6},\\ S. E. de~Mink\inst{7, 8, 9}\thanks{Einstein
    Fellow}, A. de Koter\inst{7,10},
  P. R. Dunstall\inst{2}, V. H\'{e}nault-Brunet\inst{11}, 
  J. Ma\'{i}z Apell\'{a}niz\inst{12}\thanks{Current address: Departamento de Astrof\'{i}sica, Centro de Astrobiolog\'{i}a (INTA-CSIC), Campus ESA, Apartado Postal 78, 28\,691 Villanueva de la Ca\~{n}ada, Madrid, Spain},\\
  C. M. McEvoy\inst{2}, H. Sana\inst{13},
  S.~Sim\'{o}n-D\'{i}az\inst{14,15}, W. D. Taylor\inst{1}, J. S,
  Vink\inst{16}} \offprints{C.~J.~Evans at chris.evans@stfc.ac.uk}
\authorrunning{C.~J.~Evans et al.}  \titlerunning{VFTS:
  Classifications \& radial velocities for the B-type stars}

\institute{UK Astronomy Technology Centre, 
           Royal Observatory, 
           Blackford Hill, Edinburgh, EH9 3HJ, UK
             \and 
           Department of Physics \& Astronomy,
           Queen's University Belfast,
           Belfast BT7 1NN, Northern Ireland, UK
            \and
           Dept. of Physics \& Astronomy, 
           University College London, 
           Gower Street, 
           London, WC1E 6BT, UK
             \and
           Space Telescope Science Institute,
           3700 San Martin Drive,
           Baltimore,
           MD 21218,
           USA
             \and
           Institute of Astronomy with NAO,
           Bulgarian Academy of Sciences, 
           PO Box 136, 4700 Smoljan, Bulgaria
            \and
           Department of Physics \& Astronomy, 
           The Open University, Walton Hall, 
           Milton Keynes, MK7 6AA, UK
            \and
           Astronomical Institute Anton Pannekoek, Amsterdam University, 
           Science Park 904, 1098 XH, 
           Amsterdam, The Netherlands
            \and
           Observatories of the Carnegie Institution for Science, 
           813 Santa Barbara St., 
           Pasadena, 
           CA 91101, 
           USA
            \and
           Cahill Center for Astrophysics, 
           California Institute of Technology, 
           Pasadena, 
           CA 91125, USA
            \and
           Institute voor Sterrenkunde, KU Leuven, 
           Celestijnenlaan 200D, 3001, 
           Leuven, Belgium
            \and
           Department of Physics,
           Faculty of Engineering \& Physical Sciences
           University of Surrey
           Guildford, GU2 7XH, UK
             \and
           Instituto de Astrof\'{i}sica de Andaluc\'{i}a-CSIC, 
           Glorieta de la Astronom\'{i}a s/n, 
           E-18008 Granada, Spain 
             \and
           ESA/STScI, 
           3700 San Martin Drive,
           Baltimore,
           MD 21218,
           USA
             \and
           Instituto de Astrof\'isica de Canarias, 
           E-38200 La Laguna, Tenerife, Spain 
             \and
          Departamento de Astrof\'isica, Universidad de La Laguna, 
          E-38205 La Laguna, Tenerife, Spain
             \and
          Armagh Observatory, College Hill, Armagh, BT61 9DG, UK
} 

\date{}

\abstract{We present spectral classifications for 438 B-type stars
  observed as part of the VLT-FLAMES Tarantula Survey (VFTS) in the
  30~Doradus region of the Large Magellanic Cloud.  Radial velocities
  are provided for 307 apparently single stars, and for 99 targets
  with radial-velocity variations which are consistent with them being
  spectroscopic binaries.  We investigate the spatial distribution of
  the radial velocities across the 30~Dor region, and use the results
  to identify candidate runaway stars.  Excluding potential runaways
  and members of two older clusters in the survey region (SL\,639 and
  Hodge\,301), we determine a systemic velocity for 30~Dor of
  271.6\,$\pm$\,12.2\,\kms\ from 273 presumed single stars.  Employing a
  3-$\sigma$ criterion we identify nine candidate runaway stars
  (2.9\,\% of the single stars with radial-velocity estimates).  The
  projected rotational velocities of the candidate runaways appear to
  be significantly different to those of the full B-type sample, with
  a strong preference for either large ($\ge$\,345\,\kms) or small
  ($\le$\,65\,\kms) rotational velocities.  Of the candidate runaways,
  VFTS\,358 (classified B0.5:~V) has the largest differential radial
  velocity ($-$106.9\,$\pm$\,16.2\,\kms), and a preliminary
  atmospheric analysis finds a significantly enriched nitrogen
  abundance of 12\,$+$\,log\,(N/H)\,$\gtrsim$\,8.5. Combined with a large rotational
  velocity (\vsini\,$=$\,345\,$\pm$\,22\,\kms), this is suggestive of
  past binary interaction for this star.}

\maketitle
%

\section{Introduction}
The VLT-FLAMES Tarantula Survey (VFTS) is a European Southern
Observatory (ESO) Large Programme that has obtained multi-epoch
optical spectroscopy of over 800 massive stars in the 30~Doradus
region of the Large Magellanic Cloud (LMC).  An overview of the
survey, including details of the stellar photometry, spectroscopic
observations and data reduction, was given by \citeauthor{vfts}
(\citeyear{vfts}; hereafter Paper~I).

The VFTS observations are the first optical spectroscopy for many of
the targets, and provide a wealth of new morphological information on
the high-mass population of 30~Dor, which will underpin quantitative
analyses of the spectra in future papers.  Detailed classifications
for the 352 O-type spectra (with discussion of notable morphological
groups) were presented by \citet{w14}, while classifications for the
smaller samples of Wolf--Rayet and late-type stars were given in
Paper~I. Here we present spectral classifications for 438 B-type
objects from the survey. Of these, only 74 (17\%) have previously published
classifications based on digital data, many of which were obtained at
a lower spectral resolving power.  

To complement the spectral classifications, we also present estimates
of radial velocities for the B-type sample.  These estimates provide
insights into the different populations within 30~Dor, and enable
identification of candidate runaway stars that have been ejected from
their birth sites. In the context of runaway stars, the size of the VFTS
sample is unprecedented for a single, young star-forming region.  The
survey has revealed $\sim$20 O-type runaway candidates \citep[][Sana
et al., in prep.]{e10}, with an apparent connection with rapid
rotation \citep[][Sana et al., in prep.]{w14}. There are also a number
of stars with moderate (projected) rotational velocities in the
peripheral regions of 30~Dor, which could perhaps be very rapid
rotators \citep[but with low inclinations; see][for a discussion of
their spectra]{w14}. For completeness, we note that two additional
O2-type stars near 30~Dor, Sk$-$68$^\circ$\,137 and BI\,253
(\,$=$\,VFTS\,072), were suggested as runaways by \citet{w02} on the
basis of their remote locations for their apparent youth.

In his pioneering study, \citet{b61} noted that the fractional
incidence of Galactic runaways is larger among O stars than among B
stars, a result substantiated by subsequent authors, and which also
emerged in the simulations from \citet{spz00}.  In this paper we
therefore complement the studies of the O-type runaways in the VFTS
with analysis of the B-type stars. In particular, the number and mass
spectrum of the ejected runaways from 30~Dor will provide important
constraints on models of cluster formation and evolution
\citep[e.g.][]{fpz11}.  The estimated radial velocities presented
here, combined with those from the O-type stars \citep{s13}, will also
provide an important input for comparisons with theoretical
predictions, e.g. in the context of simulations to investigate the
initial conditions leading to the distinct clump identified to the
north-east of R136 \citep{sabbi12}.

In Section~\ref{data} we summarise the relevant features of the VFTS
data and their post-processing. Section~\ref{class} sets out the
framework within which the spectra were classified and discusses
notable objects.  Section~\ref{RVs} details the methods we used to
estimate radial velocities for each target, followed by a discussion
of members of the two oldest clusters in the survey region (Hodge\,301 and
SL\,639) in Section~\ref{older_clusters}.  Finally, in
Section~\ref{sec_runaways}, we use our velocity estimates to identify
candidate single-star and binary runaways and discuss their nature and
potential origins.  In parallel to this study, Dunstall et al. (in
prep.)
have exploited the multi-epoch VFTS observations to investigate the
multiplicity properties of the B-type sample.

\section{Observations and data processing}\label{data}

All of the VFTS data discussed here were obtained using the
Medusa--Giraffe mode of the Fibre Large Array Multi-Element
Spectrograph (FLAMES) instrument on the Very Large Telescope (VLT).
The Medusa fibres (which subtend 1\farcs2 on the sky) were used to
relay light from up to 130 targets simultaneously to the Giraffe
spectrograph \citep[see][]{flames}.

Full details of the observational strategy and reduction of the data
are given in Paper~I. In brief, the ESO Common Pipeline Library FLAMES
recipes were used to undertake bias subtraction, fibre location,
summed extractions of each object, division by a normalised
flat-field, and wavelength calibration.  Subsequent processing
included correction of the spectra to the heliocentric frame, sky
subtraction, and rejection of significant cosmic rays.

The Medusa observations used three of the standard Giraffe settings
(LR02, LR03, and HR15N), providing spectral coverage of
\lam3960-5071\,\AA\ (at resolving powers, $R$, of 7\,000 to 8\,500),
and \lam6442-6817\,\AA\ (at $R$\,$=$\,16\,000).  These resolutions are
greater than for most published standards so, for the purposes of
classification, the data were degraded to an effective resolving power
of $R$\,$=$\,4\,000, ensuring consistency with the approach to the
classification of the O-type stars by \citet{w14}.

For stars that appear to be single (Dunstall et al., in prep.)  the
LR02 and LR03 spectra were normalised then merged, with the
$\sigma$-clip levels for cosmic rejection and relative weightings
based on the signal-to-noise (S/N) ratios of the individual spectra.
Finally, for classification, the merged LR02 and LR03 spectra were
combined, then smoothed and rebinned to $R$\,$=$\,4\,000. For stars
identified as having significant but relatively modest velocity
variations ($\Delta\,v_{\rm r}$\,$\le$\,40\,\kms), the individual
spectra were similarly co-added and degraded (as the variations were
within the effective velocity resolution of the rebinned data).  For
single-lined binaries with $\Delta\,v_{\rm r}$\,$>$\,40\,\kms, the
single-epoch LR02 and LR03 spectra with the best S/N ratio were
combined and then smoothed/rebinned.  We were careful in these cases
to ensure that any velocity offsets between features in the small
overlap region between the LR02 and LR03 data were not misinterpreted.
The small number of double-lined binaries were classified from
inspection of the individual spectra.

\section{Spectral classification}\label{class}

The spectra were classified by visual comparison to the B-type
Galactic standards from \citet{sot11} and Sana et al. (in
prep.), 
while also taking into account the effects of the reduced metallicity
of the LMC compared to the Galaxy \citep[which can affect the
appearance of both the metal and helium absorption lines;
e.g.][]{mar09}. This was achieved for the luminous, low-gravity
objects by interpolating between the available Galactic standards and
those for supergiants in the Small Magellanic Cloud \citep{l97}, and
with reference to the framework developed for the LMC by
\citet{f88, f91}.  Classifications of the dwarfs and giants were
also informed by comparisons with the standards from \citet{wf90} and
the LMC stars classified by \citet{f2}.

The primary diagnostic lines are the ionisation ratios of silicon,
while also taking into account the appearance of the helium and
magnesium (\lam4481) lines.  For reference, the primary
temperature-sequence criteria used to classify the VFTS sample are
summarised in Tables~\ref{criteria_sg} and \ref{criteria_dwarfs} (for
supergiants and dwarfs, respectively). Example temperature sequences
for supergiants, giants and dwarfs are shown in
Figures~\ref{bsgs_fig}, \ref{giants_fig} and \ref{dwarfs_fig},
respectively.  At a given spectral type, luminosity classes were
  assigned from the width of the Balmer lines and (at the earliest
  types) from the intensity of the silicon absorption lines, while
  also taking into account the possible effects of rotational
  broadening on the spectra.  Example luminosity sequences for B0.2,
B1, and B2.5 types are shown in Figures~\ref{seq_b02}, \ref{seq_b1},
and \ref{seq_b25}, respectively.

The supergiants and bright giants (i.e. class I and II objects) have
rich absorption-line spectra.  Given the importance of spectral
peculiarities in the context of CNO abundances, and the influence of
metallicity effects, these luminous objects were classified
independently by three of the authors (CJE, NM, and NRW) with a
typical consistency of 0.5 subtypes/luminosity class from the first
pass; final types were adopted following discussion of individual
objects as required. The larger sample of dwarfs and giants, which
exhibited a considerable range in data quality, was classified by CJE.
In many instances these classifications are less precise than those
possible for the luminous objects, but they should provide sufficient
morphological information to enable robust subdivisions of the sample
in future analyses.

\begin{table}
\begin{center}
\caption{\footnotesize Primary temperature-sequence criteria for B-type supergiants.}\label{criteria_sg}
\begin{tabular}{p{0.5cm}p{7.5cm}}
\hline\hline
Type & Criteria \\
\hline
B0   & Si\,\siv\,\lam\lam4089,\,4116\,$>>$\,Si\,\siii\,\lam4553\,$>$\,He\,\sii\,\lam4542 \\
B0.2 & Si\,\siv\,\lam4089\,$>$\,Si\,\siii\,\lam4553\,$>>$\,He\,\sii\,\lam4542 \\
B0.5 & Si\,\siv\,\lam4089\,$>$\,Si\,\siii\,\lam4553; He\,\sii\,\lam4542 absent/marginal \\
B0.7 & Si\,\siv\,\lam4089\,$\sim$\,Si\,\siii\,\lam4553; He\,\sii\,\lam4686 absent/marginal \\
B1   & Marginal Si~\siv\,\lam4116; Si\,\siii\,\lam4553\,$>$\,Si\,\siv\,\lam4089 \\ 
B1.5 & Weak/marginal Si~\siv\,\lam4089, weak Mg\,\sii\,\lam4481 \\ 
B2   & Si\,\siii\,\lam4553\,$>$\,Mg\,\sii\,\lam4481; Si~\siv\,\lam4089 absent \\
B2.5 & Si\,\siii\,\lam4553\,$\sim$\,Mg\,\sii\,\lam4481; weak Si\,\sii\,\lam\lam4128-32 \\
B3   & Si\,\siii\,\lam4553\,$<$\,Mg\,\sii\,\lam4481; Si\,\sii\,\lam\lam4128-32 present \\
B5   & He\,\si\,\lam4471\,$>$\,Mg\,\sii\,\lam4481; Si\,\sii\,\lam\lam4128-32\,$\sim$\,He\,\si\,\lam4121 \\
B8   & He\,\si\,\lam4471\,$\sim$\,Mg\,\sii\,\lam4481; Si\,\sii\,\lam\lam4128-32\,$\sim$\,He\,\si\,\lam4144 \\
B9   & He\,\si\,\lam4471\,$<$\,Mg\,\sii\,\lam4481; Si\,\sii\,\lam\lam4128-32\,$>$\,He\,\si\,\lam4144 \\
\hline
\end{tabular}
\tablefoot{\footnotesize Example spectra are shown in Figure~\ref{bsgs_fig}.}
\end{center}
\end{table}

\begin{table}
\begin{center}
  \caption{\footnotesize Primary temperature-sequence criteria for B-type dwarfs.}\label{criteria_dwarfs}
\begin{tabular}{p{0.5cm}p{7.5cm}}
\hline\hline
Type & Criteria \\
\hline
B0   & Si\,\siv\,\lam4089\,$>$\,Si\,\siii\,\lam4553$>$\,He\,\sii\,\lam4542 \\
B0.2 & Si\,\siv\,\lam4089\,$>$\,Si\,\siii\,\lam4553; He\,\sii\,\lam4542 marginal \\
B0.5 & He\,\sii\,\lam4542 absent; Si\,\siii\,\lam4116 marginal/absent \\
B0.7 & He\,\sii\,\lam4686 marginal/absent; Si\,\siii\,\lam4116 marginal/absent \\
B1   & Si\,\siii\,\lam4553\,$>$\,Si\,\siv\,\lam4089; Si\,\siv\,\lam4116, He\,\sii\,\lam4686 absent\\
B1.5 & Si\,\siv\,\lam4089 absent; Si\,\siii\,\lam4553\,$\sim$\,Mg\,\sii\,\lam4481\\
B2   & Si~{\scriptsize III}\,\lam4553\,$<$\,Mg~{\scriptsize II}\,\lam4481\\
B2.5 & Si\,\siii\,\lam4553 absent; Si\,\sii\,\lam\lam4128-32 marginal \\
\hline
\end{tabular}
\tablefoot{\footnotesize Example spectra are shown in Figure~\ref{dwarfs_fig}.}
\end{center}
\end{table}

Each merged spectrum was classified in the framework described above.
After initial classification of all the spectra, careful checks were
undertaken to ensure self-consistency within the sample. An
overlapping sample of ten B0-type stars was classified independently
(by NRW) for the study by \citet{w14}, with excellent agreement (to
within one luminosity class) in all instances.

The catalogue of classifications and radial-velocity estimates for the
B-type objects from the VFTS is presented in
Table~\ref{classifications} (available online). For convenience,
supplementary information is also provided for each star on its binary
status (from Dunstall et al., in prep.), H$\alpha$ morphology (see Section~3.1),
optical photometry (from Paper~I), alternative identifications
used in the literature, and previous spectral classifications.

\begin{figure*}
\begin{center}
\includegraphics{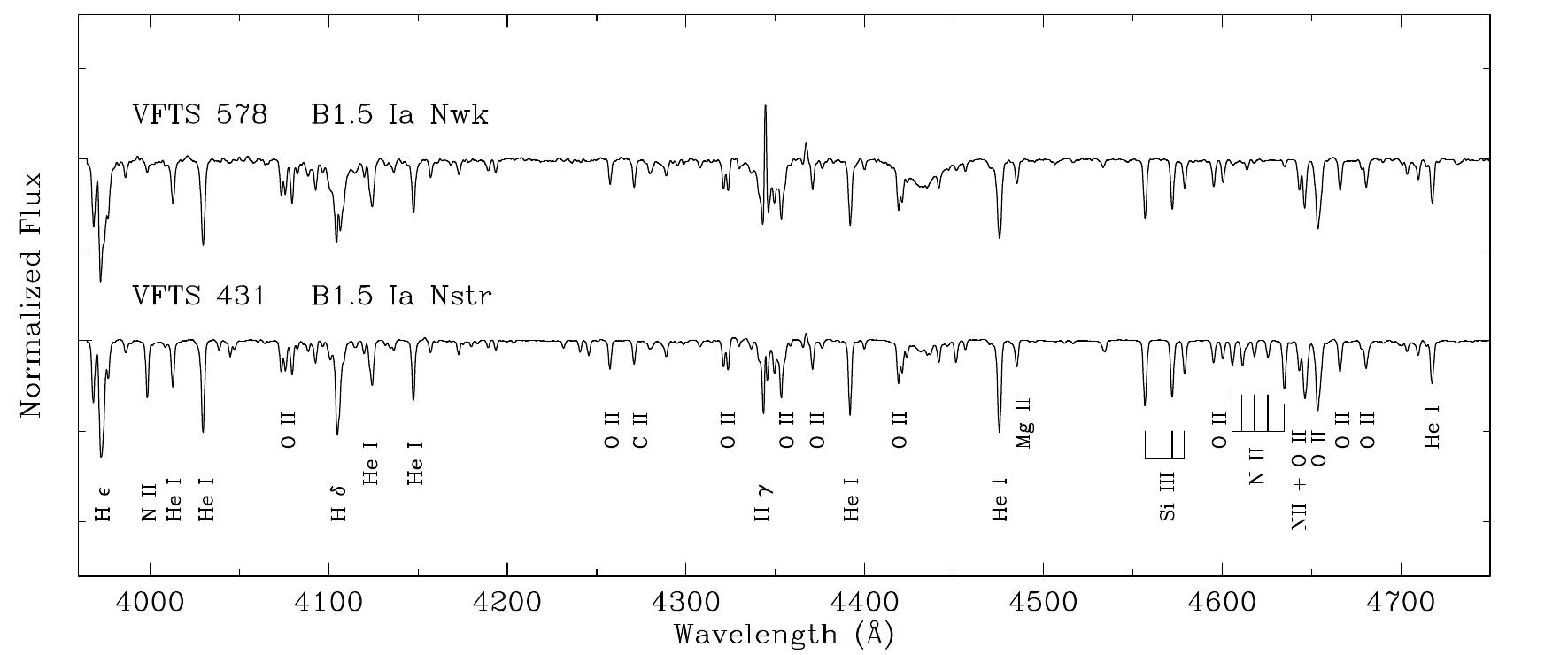}
\caption{\footnotesize Differing CNO morphologies in early B-type supergiants
  illustrated by VFTS\,431 (`Nstr') and VFTS\,578 (`Nwk'). In
  addition to the Balmer lines, the absorption features identified are:
  C~{\scriptsize II} \lam4267; He~{\scriptsize I} \lam\lam4009, 4026,
  4121, 4144, 4388, 4471, 4713; Mg~{\scriptsize II} \lam4481;
  N~{\scriptsize II} \lam\lam3995, 4601-07-14-21-31, 4640-43 (blended
  with O~{\scriptsize II}); O~{\scriptsize II} \lam\lam4070, 4254,
  4317-19, 4350, 4367, 4414-17, 4591-96, 4650, 4661, 4674-76;
  Si~{\scriptsize III} \lam\lam4553-68-74.}\label{bsgs_n}
\end{center}
\end{figure*}

The S/N ratios of the supergiant spectra were particularly good
(typically $>$\,100 for both the LR02 and LR03 regions; see
Figure~\ref{bsgs_fig}).  This allowed us to provide consistent
indications of the nitrogen enhancement/deficiency
compared to morphologically-normal stars (via Nstr and Nwk
qualifiers), as diagnosed from the CNO absorption features in the
\lam\lam4640-4650 region \citep{w76}. The significant differences in
CNO morphology are illustrated by the two B1.5~Ia spectra in
Figure~\ref{bsgs_n}.  Atmospheric analysis of the supergiants,
including estimates of nitrogen abundances and a thorough study of the
line-broadening in their spectra, will be presented by \citet{cmm} 
and Sim\'{o}n-D\'{i}az et al. (in prep.), respectively.

Given the wide range in quality of the dwarf and giant spectra, we do
not comment on CNO morphologies in their classifications.  Equally,
while we employ the `n' and `(n)' indications of broadening \citep[see
e.g. Table~3 from][]{sot11} at all luminosity classes, there are
likely to be some lacunae \citep[cf. the quantitative \vsini\
estimates from][]{duf13}.

\subsection{Emission-line objects}\label{emstars}

Where appropriate, spectra were initially classified as `Be$+$' on the
basis of Fe~\2 emission lines in the LR02 and LR03 spectra \citep[see
discussion by][]{f2}; the Fe~\2 lines were preferred over the
blue-violet Balmer lines owing to the significant (and variable)
contamination of the latter by nebular emission.

Independently of the blue-region classifications, we inspected the
H$\alpha$ profiles of every target to investigate their morphologies.
The greater resolution of the H$\alpha$ observations and the extended
wings from disc emission made it easier to distinguish contributions
from Be-type Balmer emission and nebular contamination than in the
blue, but these classifications should still be treated with some
caution.  The H$\alpha$ morphologies are encoded in the fourth column
of Table~\ref{classifications} (with the notation explained in the
corresponding footnote).

Objects classified as `Be$+$' from the LR02/LR03 spectra were seen to
display H$\alpha$ emission in all cases.  Approximately 30 additional stars
display Be-like H$\alpha$ emission without Fe~\2 emission.  The
H$\alpha$ emission in luminous supergiants most likely arises from
their stellar winds, but for less luminous objects such emission is
the primary Be diagnostic.  To preserve the distinction of the lines
seen in emission, the standard `e' suffix is used in the
classifications for the (non-supergiant) stars where only
H$\alpha$ is seen in emission.

The complex star-formation history of 30~Dor means that our sample
will span a range of stellar ages, so we are unable to comment on the
incidence of the Be-phenomenon for a single, co-eval population.
Nonetheless, we note that the incidence of the Be-type classifications
for the dwarf and giant stars is 18\% (70/388 objects). This is in
good agreement with the Galactic fraction \citep[$\ge$\,17\%,][]{zb},
and that reported for the field population near NGC\,2004 in
the LMC \citep[of 16 and 17.5\%, from Evans et al. 2006 and][respectively]{mhf}.

\subsubsection{B[e]-type spectra}

Two targets in the Medusa--Giraffe sample show forbidden-line emission
characteristic of the B[e] category. VFTS\,698 is a double-lined
binary system with a B[e]-like spectrum, comprising what appears to be
an early B-type secondary in orbit around a veiled, more massive
companion \citep[see][]{d12}.  The second B[e] object is VFTS\,822,
shown in Figure~\ref{be_fig} together with VFTS\,1003 (the B[e]-like
object from Paper~I).  Given the intensity of its Si~\2
\lam\lam4128-32 and Mg~\2 \lam4481 absorption, the spectrum of VFTS\,822 was
classified as mid-late B[e] ($\sim$B5-8); further discussion of its
possible pre-main-sequence nature was given by \citet{k14}.

Although the nature of VFTS\,1003 remains uncertain at present, the number
of evolved B[e] stars (at most two) in the VFTS is clearly small
compared to the number of normal supergiants.  This is in keeping
with the small number of sgB[e] stars in Galactic clusters \citep[see
the appendix of][]{crn13}, arguing that the B[e] phenomenon in evolved
supergiants must be a relatively brief phase or originate from a rare
process.

\subsubsection{VFTS\,766: A peculiar emission-line star}\label{vfts766}

\begin{figure*}
\begin{center}
\includegraphics{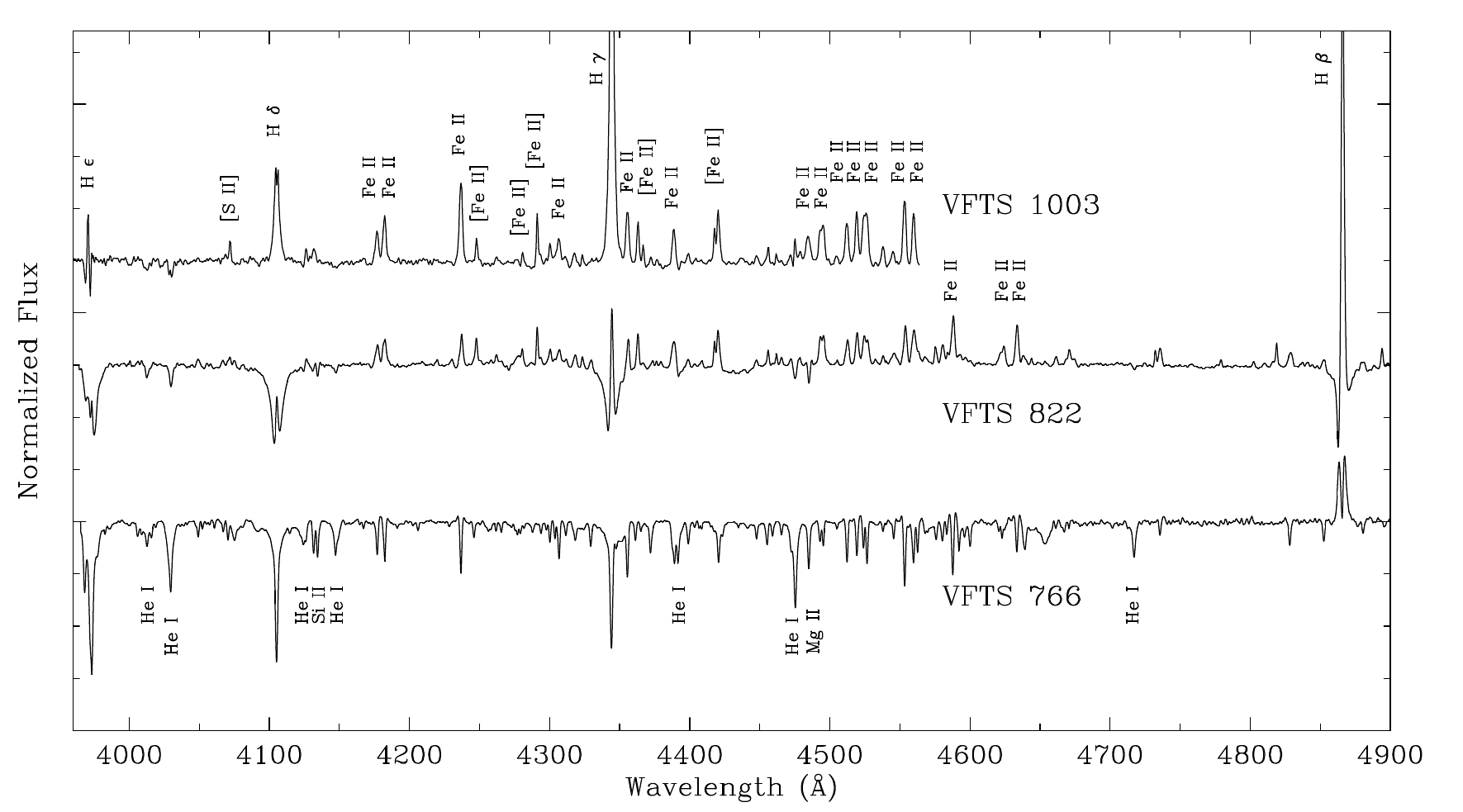}
\caption{\footnotesize Spectra of three peculiar Be-type stars from the VFTS, in
  which each spectrum has been smoothed and rebinned to
  $R$\,$=$\,4\,000 for clarity. In addition to the Balmer lines, the
  emission lines identified in VFTS\,1003 are [S~{\scriptsize{II}}]
  \lam4069; Fe~{\scriptsize{II}} \lam\lam4173, 4179, 4233, 4303, 4352,
  4385, 4481, 4491, 4508, 4515, 4520-23, 4549, 4556;
  [Fe~{\scriptsize{II}}] \lam\lam4244, 4277, 4287, 4358-59, 4414-16.
  Emission lines identified in VFTS\,822 are Fe~{\scriptsize{II}}
  \lam\lam4584, 4620.5, 4629.  Absorption lines identified in
  VFTS\,766 are He~{\scriptsize{I}} \lam\lam4009, 4026, 4121, 4144,
  4388, 4471, 4713; Si~{\scriptsize{II}} \lam\lam4128-32;
  Mg~{\scriptsize{II}} \lam4481.}\label{be_fig}
\end{center}
\end{figure*}

VFTS\,766 displays strong H$\alpha$ emission and inspection of its
blue-region spectrum (see Figure~\ref{be_fig}) revealed it as a
peculiar object.  At first glance its spectrum appears to be that of a
mid-late B-type supergiant (B5-B8, as traced by e.g. Si~\2, Mg~\2),
but with superimposed metallic absorption lines, which are consistent
with a cooler component. This seemingly composite appearance can
indicate a circumstellar shell or disc observed edge-on, and this
hypothesis is supported by the twin-peaked H$\beta$ shell-like
emission profile.  Its near- and mid-IR colours\footnote {VFTS\,766:
  $J$\,$=$\,15.14, $H$\,$=$\,15.01, $K_{\rm s}$\,$=$\,14.76 (from
  Paper~I); [3.6]\,$=$\,14.28, [4.5]\,$=$\,14.06, [5.8]\,$=$\,13.64
  \citep[from][]{sage}.} place it in the region in the $J$,\,$J-[3.6]$
colour-magnitude diagram which is occupied by early-type stars
associated with free-free/bound-free emission, and we find no evidence
of hot dust \citep[cf.][]{ab_lmc}.

Similar emission profiles are observed in some low-luminosity
supergiant B[e] stars \citep[e.g. S59,][]{gzw95}, with examples that
also appear to lack dusty discs \citep[][although none have yet been
discovered in the edge-on orientation required to yield shell
profiles]{glo12}.  However, we speculate that VFTS\,766 is a classical
Be star, in which the twin-peaked emission in the wings of the Balmer
profiles gives rise to the supergiant-like appearance. An example of
this effect is the development of a circumstellar disc in the
classical edge-on Be star $o$~And, which conspires to yield a narrower
absorption feature than the underlying photospheric profile
\citep{ctp03}.  Another example is provided by the Be star 48\,Lib
\cite[see][and references therein]{rcm13}, which has previously been classified
as a supergiant \citep{hsm88}.

This possible explanation for the nature of VFTS\,766 is supported by
its magnitude relative to other stars in our sample. Spectroscopic
monitoring to identify an episode of disc loss, during which the
(uncontaminated) photospheric Balmer line profiles might be observed,
would clarify its luminosity type and true evolutionary status.

\subsubsection{Blue hypergiants}

We identifed five blue hypergiants (BHGs) in our sample, namely:
VFTS\,003 (B1 Ia$^+$), 424 (B9 Ia$^+$p), 430 (B0.5 Ia$^+$((n))\,Nwk),
458 (B5 Ia$^+$p) and 533 (B1.5 Ia$^+$p Nwk). Recent theoretical and
observational studies suggest that early-type BHGs
might be massive stars which have passed through a blue supergiant
phase and are about to enter a luminous blue variable (LBV) phase
\citep{cnn12,g14}. This suggests that VFTS\,003, 430, and 533 might be
on the verge of LBV-like behaviour, although we note that VFTS\,430
appears somewhat subluminous compared to the blue supergiants in
30~Dor, and is significantly fainter than VFTS\,533 (even accounting
for its apparently greater line-of-sight extinction, see photometry in
Table~\ref{classifications}).  Later-type BHGs have been observed to
undergo LBV-like photometric and spectroscopic behaviour
\citep[e.g.][with potential implications for VFTS\,424 and 458
here]{svdk04,cnn12}, suggesting that these two evolutionary phases are
essentially synonymous for such stars.

On further inspection of their spectra, broad emission wings were
identified in the H$\beta$ profiles of each of the BHGs, except for
VFTS\,430. This H$\beta$ morphology is seen in some LBVs and
  other luminous supergiants \citep[e.g.][]{wf00}, and is also
present in VFTS\,739 in our sample (classified as A0~Ip but, given its
omission from Paper I, included here in the B-type sample). The
spectra and evolutionary status of these potential LBV
  precursors/candidates will be discussed in more depth by Walborn et al. (in
prep.).

\subsubsection{Comparisons with X-Ray observations}
We compared our sample with the point-source detections from
observations in 30~Dor with the {\em Chandra X-ray Observatory}
\citep{lt06,lt14}; none of the VFTS B-type stars is detected in the
extant X-ray data. A new X-ray Visionary Project is now underway,
`The Tarantula Revealed by X-rays (T-ReX)', which will observe
30~Dor with {\em Chandra} for 2\,Ms.  This will push far deeper than
the existing data, and we will revisit the VFTS sample once the T-ReX
observations and analysis are complete.

\section{Stellar radial velocities}\label{RVs}

\subsection{Single stars}
Radial-velocity estimates for the stars identified as apparently
single by Dunstall et al. (in prep.)
were obtained from Gaussian fits to selected
absorption lines in the merged LR02 and LR03 spectra of each object (at
the native resolution of the data, without the degrading applied for
classification purposes).  This approach is consistent with that used
by \citet{s13} for analysis of the O-type stars, albeit (necessarily)
using a different subset of absorption lines.  We note that the choice
of Gaussian profiles may not be appropriate for fitting the wings of
the lines in the presence of different broadening mechanisms, but they
should suffice for robust estimates of the centres of our chosen lines.

We also considered a cross-correlation method to estimate the radial 
velocities.  However, given the range in temperature and
luminosity of the sample, as well as the varying data quality and
problems of nebular contamination, the selection of suitable template
spectra for cross-correlation was not trivial and would have introduced
additional uncertainties (relating to e.g. the projected rotational
velocities, \vsini). Thus, we did not pursue this approach further.

\subsubsection{Line selection}

The large range in \vsini\ and effective temperature for the sample
meant that it was not possible to use a single set of diagnostic lines
for all stars.  Moreover, the large intrinsic width and nebular
contamination of the Balmer lines precluded their use; similarly,
nebular emission affects some of the stronger He~\1 lines. Following
the approach taken by \citet{duf13}, we used different sets of
diagnostic lines (as summarised in Table~\ref{vr_lines}) depending on
the \vsini\ of the star.

For narrow-lined spectra (defined as \vsini\,$\lesssim$\,150\,\kms), we
used three non-diffuse He~\1 lines and four isolated metal
lines listed as `Set~1' in Table~\ref{vr_lines}. Where possible,
radial velocities were estimated from Si~\3 \lam4553 in both the merged LR02
and LR03 spectra, giving up to eight individual estimates for the
narrow-lined objects.

We note that He~\1 \lam4121 is potentially affected by blending with
O~\2 \lam\lam4120.28/54. In many instances these features could
be resolved and a double Gaussian profile was fitted to estimate the
He~\1 and O~\2 line centres; in cases where the O~\2 lines were weaker
or unresolved, they do not appear to have unduly influenced the
results (see Table~\ref{vr_lines}, in which the residual for He~\1
\lam4121 is consistent with the other lines).

Radial-velocity estimates for stars with broader lines were more
difficult, particularly for spectra with significant nebular
contamination.  In these cases, velocities were estimated by fitting
the five He~\1 lines listed as `Set~2' in Table~\ref{vr_lines}, for
which the profiles are expected to be
effectively symmetric\footnote{He~{\scriptsize I} $^3$P\,$\rightarrow$\,$^3$D
  transitions were excluded because of the presence of significant
  $^3$P\,$\rightarrow$\,$^3$F components.}.

\begin{figure}
\begin{center}
\includegraphics[scale=1.0]{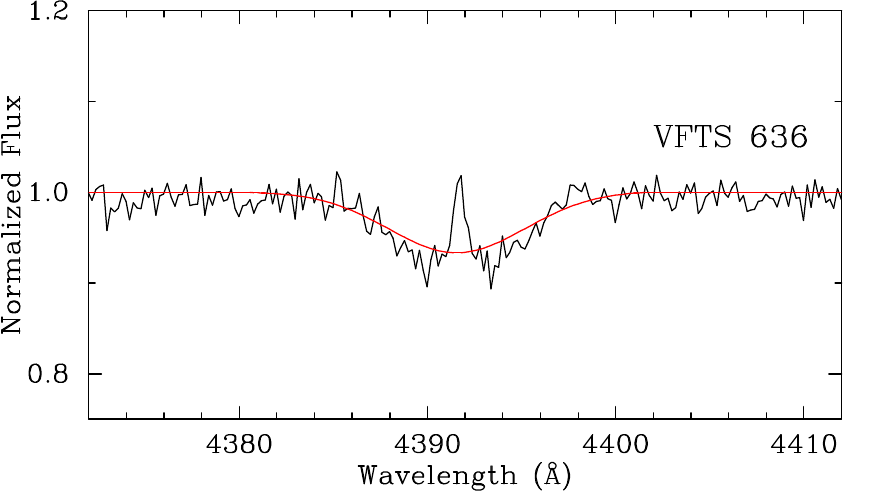}
\includegraphics[scale=1.0]{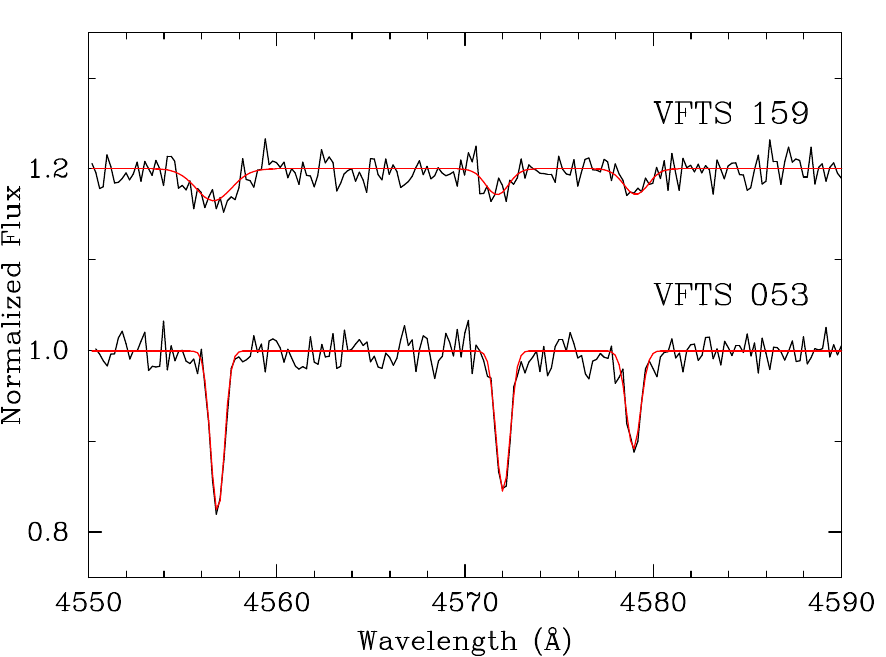}
\caption{\footnotesize Examples of Gaussian fits used to estimate radial velocities
  ($v_{\rm r}$) for the B-type sample.  {\it Upper panel:} Fit to
  He~{\scriptsize I} \lam4388 in a rapidly-rotating object (VFTS\,636)
  with nebular contamination.  {\it Lower panel:} Fits to the
  Si~{\scriptsize III} triplets (\lam\lam4553-68-74) of VFTS\,053 and
  159.}\label{rv_fits}
\end{center}
\end{figure}

Representative examples are shown in Figure~\ref{rv_fits}.  The lower
panel shows fits to the Si~\3 triplet for two narrow-lined stars,
VFTS\,053 and 159, which have \vsini\,$\le$\,40 (i.e. less than the
velocity resolution of the data) and 96\,$\pm$\,9\,\kms, respectively
\citep[from][]{duf13}.  As an example of a spectrum of a rapidly-rotating
star which also suffers nebular contamination, the upper panel of
Figure~\ref{rv_fits} shows the Gaussian fit to He~\1 \lam4388 for
VFTS\,636 \citep[\vsini\,$=$\,371\,$\pm$\,35\,\kms; ][]{duf13}.

For three stars with later/peculiar types -- VFTS\,272 (possible shell
star), 739 (A0~Ip), and 766 (B5-8e, see Section~\ref{vfts766}) -- we employed
weak Fe~\2 and Ti~\2 absorption lines to estimate velocities, although
we note that in VFTS\,272 and 766 they might not be representative of
the true (stellar) radial velocity, as these lines could trace cooler
features associated with a disc of material \citep[see e.g. the
discussion of VFTS\,698 by][]{d12}.

\begin{table}
\begin{center}
  \caption{\footnotesize Rest wavelengths used to estimate radial velocities
    ($v_{\rm r}$) for the narrow- and broad-lined stars (Sets~1 and 2,
    respectively). Values in the final column are average
    differences ($\Delta$) between the velocity estimates for each
    line and the final velocity for each star.}\label{vr_lines}
\begin{tabular}{lcl}
\hline\hline
Ion & \lam [\AA]  & \pp$\Delta$ [\kms] \\
\hline
\multicolumn{3}{c}{Set~1}\\
 He~\1 & 4120.82  & $-$0.8\,$\pm$\,6.7 \\ 
  C~\2 & 4267.13  & \pp1.1\,$\pm$\,5.1 \\ 
 He~\1 & 4437.55  & $-$0.4\,$\pm$\,5.4 \\
 Si~\3 & 4552.62  & \pp0.3\,$\pm$\,6.1 (LR02) \\
       &          & $-$1.6\,$\pm$\,5.8 (LR03) \\
 Si~\3 & 4567.84  & $-$0.8\,$\pm$\,6.4 \\
 Si~\3 & 4574.76  & $-$1.7\,$\pm$\,7.0 \\
 He~\1 & 4713.15  & \pp2.1\,$\pm$\,4.5 \\
\hline
\multicolumn{3}{c}{Set~2}\\
He~\1  & 4009.26  & $-$0.8\,$\pm$\,11.3 \\
He~\1  & 4143.76  & \pp0.1\,$\pm$\,9.0\o \\
He~\1  & 4387.93  & \pp2.7\,$\pm$\,9.1\o \\
He~\1  & 4713.15  & \pp2.8\,$\pm$\,7.1\o \\
He~\1  & 4921.93  & $-$3.5\,$\pm$\,8.7\o \\
\hline
\end{tabular}
\tablefoot{\footnotesize Wavelengths are taken from the NIST atomic spectra database
  \citep{nist}; the adopted value for C~{\sc ii} is the average of the two
  stronger transitions.}
\end{center}
\end{table}

\subsubsection{Mean velocities}\label{methods_single}

Radial-velocity estimates for the single stars were obtained by
calculating an average of the $n$ individual measurements ($v_i$),
weighted by the estimated central depth ($d_i$) from the
Gaussian fit of each line, i.e.

\begin{equation}
v_{\rm r}~=~\frac{\sum_{i\,=\,1}^n v_{i}\,d_{i}}{ \sum_{i\,=\,1}^n d_{i}}~.\label{mean_vr}
\end{equation}

The sample standard deviation ($\sigma$) for each star was similarly calculated
as a weighted average:

\begin{equation}
\sigma^2~=~\frac{\sum_{i\,=\,1}^n (v_{i} - v_{\rm r})^2\,d_{i}} {\sum_{i\,=\,1}^n d_{i}}~.\label{sigma_vr}
\end{equation}

In cases where the central depths of the absorption lines were less
than twice the inverse of the continuum S/N ratio, the individual
estimates were excluded from the above calculations (to avoid their
larger uncertainties degrading the results from stronger lines). For
stars with estimates from three (or more) absorption lines, we present
their estimated mean velocities and standard deviations in columns~7
and 8 of Table~\ref{classifications}. For completeness, the
line-by-line estimates from Sets~1 and 2 for the single stars are
detailed in Tables~\ref{rv_set1} and \ref{rv_set2}, respectively; the
mean uncertainty of the estimates for the 307 (presumed) single stars is
7.8\,\kms\ (and with a median uncertainty of 6.2\,\kms).  

As a check on our results, we used the methods described above to
analyse a sample of the late O-type stars from the VFTS, finding
excellent agreement between our velocity estimates and those from
\citet[][see discussion in Section~\ref{xcheck_sana}]{s13}.  We also
investigated the internal consistency of our results from the selected
lines by calculating the average difference ($\Delta$) of the
individual measurements for each line compared to the adopted mean
velocity for each target. The means and standard deviations of
these differences are given in column~3 of Table~\ref{vr_lines}, with
no significant shifts seen for any of the diagnostic lines compared to
the adopted radial velocities.

\subsection{Radial velocities of the binaries}\label{methods_binaries}

To complement the analysis of the single stars, we also investigated
the radial velocities of the single-lined binaries in the sample.  We
obtained single-epoch velocity estimates ($v_{\rm single}$) for each
system using measurements from the best LR02 spectra (in terms of S/N
ratio of the co-added exposures from a given night, which span three
hours at most).

Given the reduced spectral coverage of the LR02 data compared to the
full LR02\,$+$\,LR03 range, we used all of the available lines from
Table~\ref{vr_lines} to estimate $v_{\rm single}$ (except for He~\1
\lam4121, given the nearby O~\2 feature and the lower S/N ratio of the
single-epoch spectra compared to the quality of the co-added spectra
of the single stars). Line-by-line estimates and the date for which
they were measured are listed in Table~\ref{rv_binaries}. Mean
velocities and standard deviations for the selected epoch were then
estimated using the same methods as for the single stars.

The mean single-epoch velocities serve as an absolute reference point
for the inter-epoch differential velocities calculated by
Dunstall et al. (in prep.) yielding absolute estimates for each epoch.  We then
calculated the mean and standard deviation of the multi-epoch
estimates to obtain a velocity estimate for each single-lined
binary. These values are presented in Table~\ref{classifications}, and
are flagged as `B' (binary) in the ninth column.  We note that the
typical uncertainty on the cross-correlation results from
Dunstall et al. was less than 5\,\kms, while the typical error on
the absolute estimates determined here from the single-epoch data was
9\,\kms; i.e. the uncertainties in our estimates will generally
be dominated by the analysis presented here.

Full orbital solutions are presently unavailable for the B-type
binaries, so we strongly caution the reader that the velocity
estimates in Table~\ref{classifications} are not their centre-of-mass,
systemic velocities (although, as discussed in the Appendix, in many
instances they may provide reasonable estimates).

\subsection{Spatial distribution}\label{rvs_spatial}

\begin{figure*}
\begin{center}
\includegraphics[scale=0.85]{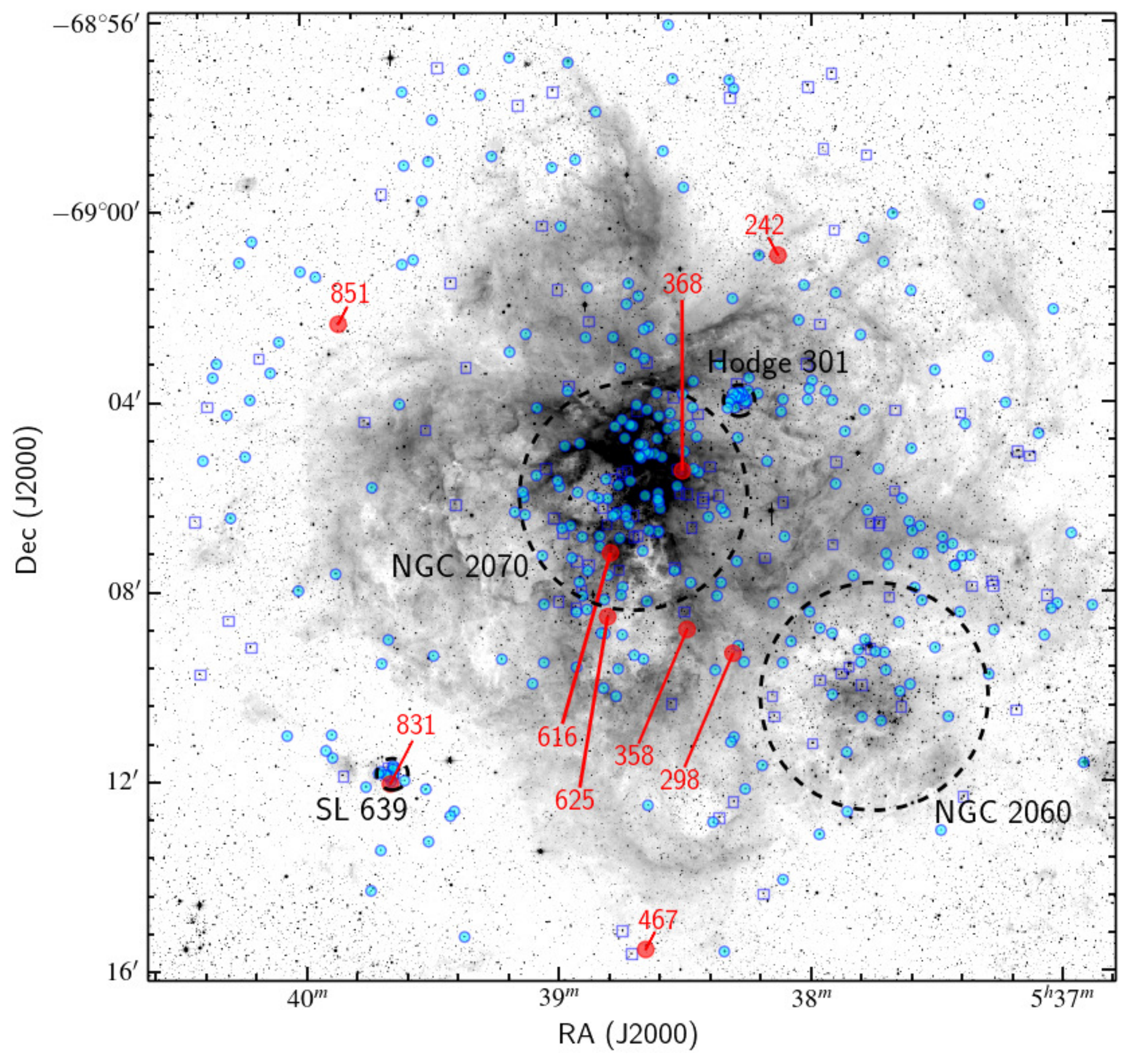}
\caption{\footnotesize Spatial distribution of the B-type single stars (cyan
  circles) and spectroscopic binaries (open blue squares) with
  radial-velocity estimates; the nine candidate runaways,
    labelled by their VFTS identifiers, are marked in red.  The
  spatial extents of NGC\,2070, NGC\,2060, SL\,639, and Hodge\,301 (as
  defined in Table~\ref{results_spatial}), are indicated by the
  overlaid dashed circles. The underlying image is from a $V$-band
  mosaic taken with ESO's Wide Field Imager on the 2.2\,m telescope at
  La Silla.}\label{runaways_spatial}
\end{center}
\end{figure*}

In addition to NGC\,2070 (the main 30~Dor cluster, which includes R136
at its core) and NGC\,2060 (the association to the south-west), there
are two smaller clusters in the VFTS survey: Hodge\,301 \citep{h88}
and SL\,693 \citep{sl63}; their relative locations are shown in
Figure~\ref{runaways_spatial}. Both clusters appear older than
NGC\,2060 and 2070 as they contain evolved supergiants, from B-
through to M-types.

We calculated weighted mean velocities ($\overline{v_{\rm r}}$, and
their standard deviations) for the full sample of single stars, for
the subsamples associated with the four distinct stellar assocations
in the survey region (NGC\,2060, NGC\,2070, Hodge\,301, and SL\,639,
as defined in Table~\ref{results_spatial}), and for the field
population of stars outside the four clusters. These results,
which exclude candidate runaway stars (see Section~\ref{sec_runaways}), 
are summarised in Table~\ref{results_spatial}.

The results for the two younger star-forming regions (i.e. NGC\,2060
and 2070) are in excellent agreement with the field population of
B-type stars in the region, but the two older clusters (Hodge\,301 and
SL\,639) appear at slightly lower recession velocities, suggesting
that they are kinematically distinct from the rest of the sample.  The
weighted mean velocity for the 273 single stars excluding these two clusters
(and potential runaways) is $\overline{v_{\rm
    r}}$\,$=$\,271.6\,$\pm$\,12.2\,\kms.

\begin{table*}
\begin{center}
  \caption{\footnotesize Mean radial velocities (\,$\overline{v_{\rm r}}$\,) and standard deviations
    ($\sigma$) of the stellar associations identified within 30
    Doradus.}\label{results_spatial}
\begin{tabular}{lccccc}
\hline\hline
Region & No. of stars & Radius & $\alpha$ & $\delta$ & $\overline{v_{\rm r}}$\,$\pm$\,$\sigma$ \\
& & [$'$] & \multicolumn{2}{c}{[J2000.0]} & \\
\hline
All                        & 298  & \ldots &       \ldots &          \ldots & 270.4\,$\pm$\,12.4 \\
\hline
NGC\,2070                  &  78  & 2.40   & 05\,38\,42.0 & $-$69\,06\,02.0 & 271.2\,$\pm$\,12.4 \\
NGC\,2060                  &  22  & 2.40   & 05\,37\,45.0 & $-$69\,10\,15.0 & 273.7\,$\pm$\,13.4 \\
Hodge\,301                 &  14  & 0.33   & 05\,38\,17.0 & $-$69\,04\,01.0 & 261.8\,$\pm$\,\o5.5 \\
SL\,639                    &  11  & 0.33   & 05\,39\,39.4 & $-$69\,11\,52.1 & 253.5\,$\pm$\,\o3.9 \\
Field                      & 173  & \ldots &       \ldots &          \ldots & 271.3\,$\pm$\,11.8 \\
\hline
All (excl. H\,301/SL\,639) & 273  & \ldots &       \ldots &          \ldots & 271.6\,$\pm$\,12.2 \\
\hline
\end{tabular}
\tablefoot{\footnotesize These statistics exclude the nine candidate runaways
  identified in Table~\ref{runaways}.  For consistency, the central
  positions adopted for NGC\,2060 and NGC\,2070 are those used by
  \citet{s13}.  Approximate centres for Hodge\,301 and SL\,639 were
  adopted using the coordinates of WB97-5 \citep{wb97} and VFTS\,828,
  respectively.}
\end{center}
\end{table*}

\section{Spectral content of the older clusters}\label{older_clusters}

The VFTS data comprise the first comprehensive spectroscopy in Hodge\,301
and SL\,639, so we briefly discuss these clusters in turn.

\subsection{Hodge\,301}

There are 20 VFTS targets within a 20$''$ radius (chosen to
conservatively identify likely members, and corresponding to
$\sim$4.9\,pc, assuming a distance modulus of 18.5\,mag). Fifteen of
these were classified as B-type and their likely membership of the
cluster is indicated in the final column of
Table~\ref{classifications}\footnote{The five others are: VFTS\,310
  \citep[O9.7 V:, ][]{w14} and VFTS\,271, 281, 289, and 294 (A7~II,
  Mid-late K, Late G/Early K, and A0~Ib, respectively, from Paper~I).
  Three additional stars (VFTS\,263, 291, and 317) are also nearby, at
  radii from the adopted centre of $\sim$25$''$.}.

From analysis of {\em Hubble Space Telescope (HST)} imaging,
\citet{gc00} estimated Hodge\,301 to be 20-25\,Myr old. From
narrow-band imaging they also identified 19 Be-type stars, eight of
which (VFTS\,272, 276, 279, 282, 283, 287, 293, and 301) have
spectroscopy from the VFTS. The H$\alpha$ spectra of each of these
contain significant nebular emission (with multiple components in all
sightlines except towards VFTS\,293).  Nonetheless, broad Be-like
H$\alpha$ emission is present in each object, except for VFTS\,276
(GC00-Be2). 

\citet{gc00} noted two potential blue stragglers in the cluster, which
they argued were too luminous for the main-sequence population (but
less luminous than expected for supergiants).  The first, VFTS\,270
(WB97-2), is a seemingly unremarkable B3~Ib star.  The second is
VFTS\,293 (WB97-9, GC00-Be1), classified here as B2 III-II(n)e.  From
two previous observations, \citet{wb97} noted small shifts of its
absorption features (relative to superimposed nebular emission) and
possible weak He~\2 \lam4686 emission at one epoch.  If the He~\2
emission were real, \citeauthor{wb97} speculated that it might be an
X-ray binary.  There is no evidence of $\lambda$4686 emission in the
LR03 FLAMES spectra (albeit all taken on the same night, see Paper~I),
but small line-profile variations are seen in the absorption features
in the multi-epoch LR02 spectra.

More intriguing is VFTS\,310 on the periphery of the cluster and
classified as O9.7~V: \citep{w14}.  With such a classification,
VFTS\,310 would be expected to be significantly younger than the other
cluster members, as shown by its location in the
Hertzsprung--Russell (H--R) diagram of the cluster
(Figure~\ref{clusters_fig}, see Section~\ref{hrd} for details).  The
star appears otherwise unremarkable, with a low projected rotational
velocity \citep[\vsini\,$\le$\,40\,\kms, ][]{duf13} and a radial
velocity \citep[$v_{\rm r}$\,$=$\,272.3\,$\pm$\,2.8\,\kms,][]{s13}
which is consistent with those in NGC\,2070 and the local field
population. Moreover, it does not appear to be a runaway from
Hodge\,301 (see Section \ref{sec_runaways}).  Whether this is a true
blue straggler or simply a line-of-sight coincidence remains
uncertain at this time.

\subsection{SL\,639}

Adopting a 20$''$ radius ($\sim$4.9\,pc), sixteen of the VFTS targets
are within SL\,693.  Fourteen are in the B-type sample, with their
membership of SL\,639 indicated in the final column of
Table~\ref{classifications}\footnote{The two others are VFTS\,820
  and 828 (A0~Ia and Early~M, respectively, from Paper~I).  The only
  other VFTS target within a 35$''$ (8.5\,pc) radius is VFTS\,839 at
  $\sim$25$''$ (6\,pc), classified as G-type in Paper~I. All three of these
  cooler objects have radial velocities consistent with 
  membership of the LMC.}. Ten of these stars are early-B dwarfs/giants,
with six of them displaying Be-like H$\alpha$ emission.  The remaining
four are comprised of two supergiants (VFTS\,827, B1.5~Ib; VFTS\,831,
B5~Ia) and two bright giants (VFTS\,826, B1~IIn; VFTS\,829,
B1.5-2~II).

There are previous age estimates for SL\,693.  From
consideration of its main-sequence turn-off, \citet{hodge83} estimated
its age to be 18\,Myr (although he noted a large uncertainty compared
to the theoretical models of the time). A larger estimate of 30\,Myr
was given by \citet{sbc95} from its reddening-corrected, integrated
$(U\,-\,B)_{0}$ colour and adopting an empirical calibration of
colour vs. age from \citet{bsa90}.

\subsection{H--R diagram}\label{hrd}

Motivated by the apparent similarities of the two clusters in terms of
their spectral content, we constructed an H--R diagram in order to
compare the properties of their likely members.  Ahead of detailed
quantitative analysis of the VFTS spectra and pending extinction
analysis of the whole sample, we adopted effective temperatures
($T_{\rm eff}$) on the basis of the spectral types in Table~3 using
the calibrations of \citet{tr07}, and interpolating between the
Galactic and SMC temperatures for A-type supergiants from
\citet{eh03}; cooler objects were omitted given the larger
uncertainties in their spectral classifications. Stellar luminosities
were estimated using the optical photometry from Paper~I, intrinsic
colours from \citet{f70}, bolometric corrections from \citet{bal94},
and assuming $R_V$\,$=$\,3.5 \citep[see Appendix~C of][]{d13}.

\begin{figure}
\begin{center}
\includegraphics[scale=0.52]{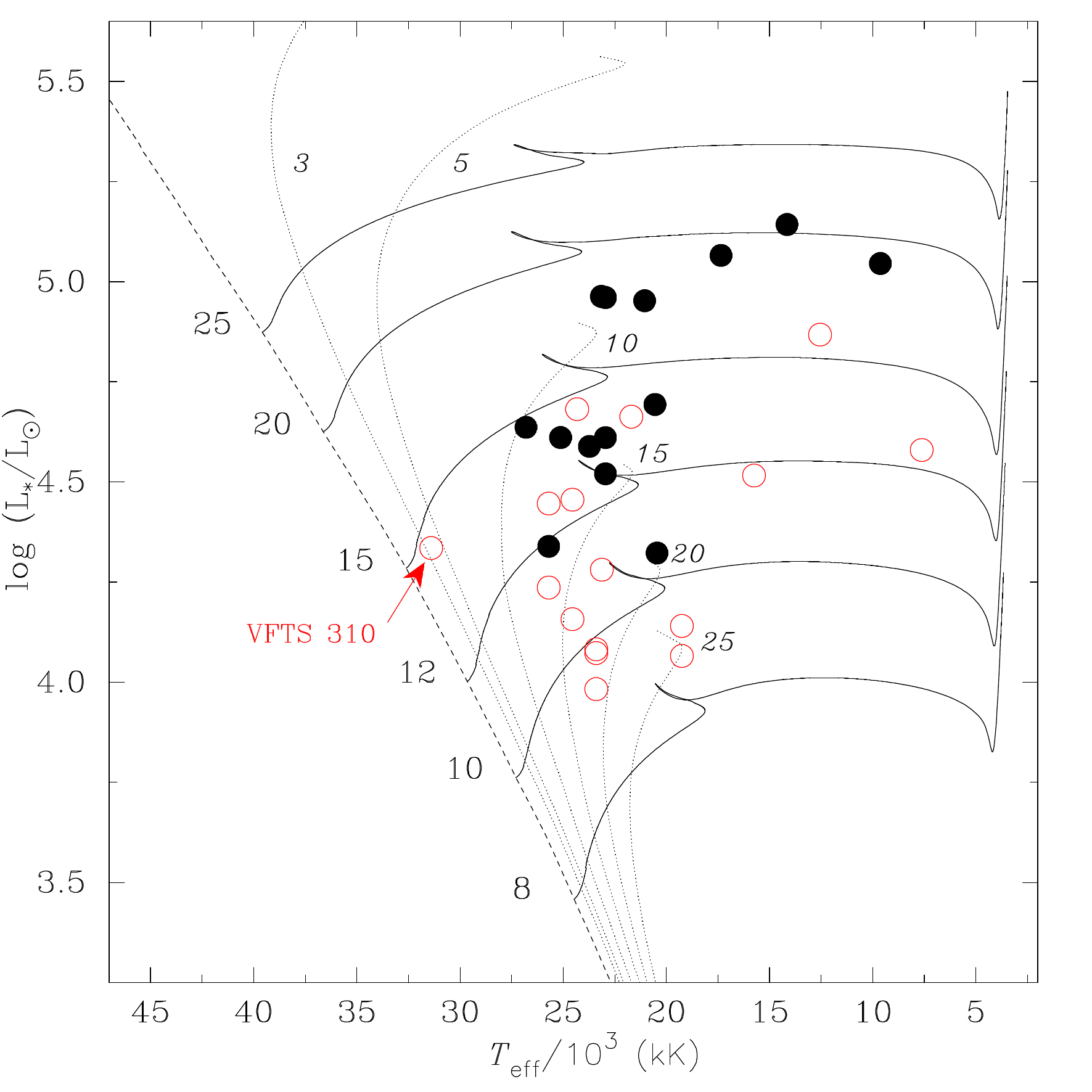}
\caption{\footnotesize Hertzsprung--Russell diagram for the VFTS objects observed in
  SL\,639 (black circles) and Hodge\,301 (red, open circles).  Also
  plotted are evolutionary tracks for initial masses of 8, 10, 12, 15,
  20, 25\,M$_{\odot}$ (solid lines), the zero-age main sequence
  (dashed line), and isochrones (3, 5, 10, 15, 20, and 25\,Myr, dotted
  lines) for the non-rotating models from \citet{b11a}. The apparent
  blue straggler in Hodge\,301 (VFTS\,310) is a late O-type
  star.}\label{clusters_fig}
\end{center}
\end{figure}

The H--R diagram is shown in Figure~\ref{clusters_fig}, with SL\,639
and Hodge\,301 members indicated by the filled black and open red
circles, respectively.  Evolutionary tracks and isochrones (for
non-rotating models) from \citet{b11a} are also plotted for
comparison.  As noted earlier, the late O-type star in Hodge\,301,
VFTS\,310, stands out as a blue straggler with $T_{\rm
  eff}$\,$>$\,30\,kK.

Tailored atmospheric and extinction analyses of these objects should
help to provide further insight in due course.  Nonetheless, the stars
in SL\,639 appear marginally younger than those in Hodge\,301. From
qualitative comparison with the isochrones, we estimate an age in the
range of 10-15\,Myr for SL\,639, significantly younger than the
estimate of 30\,Myr from \citet{sbc95}. There is a larger spread in
the inferred ages for the stars in Hodge\,301 of 15\,$\pm$\,5\,Myr,
overlapping with the lower estimate of 20\,Myr from the photometric
study of \citet{gc00}.  Adopting isochrones calculated for the
rotating models from \citeauthor{b11a} had little impact on these
results; e.g. assuming an intial rotational velocity of 301\,\kms, the
inferred ages were only marginally younger (by $\sim$1\,Myr).

\section{Candidate runaways}\label{sec_runaways}

\subsection{Single stars}

\begin{table*}
\begin{center}
  \caption{\footnotesize Candidate runaway stars identified from the
    radial-velocity analysis of the VFTS B-type spectra.}\label{runaways}
\begin{tabular}{llcccccc}
\hline\hline
VFTS & Classification & $v_{\rm r}$ & ($v_{\rm r}$\,$-$\,$\overline v_{\rm r}$) & $|v_{\rm r}$\,$-$\,$\overline v_{\rm r}|$/$\sigma$ & \vsini & \multicolumn{2}{c}{$r_{\rm d}$} \\
     &                & [\kms]      & [\kms] & & [\kms] & [$'$]       & [pc] \\
\hline
242 & B0 V-IV        & 216.0\,$\pm$\,\o3.6 &\o$-$55.6\,$\pm$\,12.7 & 4.6 &\o$\le$40$\phantom{\le}$\p & 5.96 & \o86.6 \\
298 & B1-2 V-IIIe$+$ & 332.1\,$\pm$\,\o4.0 &\pp\o60.5\,$\pm$\,12.8 & 5.0 &  431\,$\pm$\,15           & 3.92 & \o57.1 \\
358 & B0.5: V        & 164.7\,$\pm$\,10.7  & $-$106.9\,$\pm$\,16.2 & 8.8 &  345\,$\pm$\,22           & 3.02 & \o44.0 \\
368 & B1-3 V         & 229.6\,$\pm$\,12.4  &\o$-$42.0\,$\pm$\,17.4 & 3.4 &  380:                     & 1.20 & \o17.5 \\
467 & B1-2 Ve$+$     & 322.3\,$\pm$\,30.2  &\pp\o50.7\,$\pm$\,32.6 & 4.2 &  355\,$\pm$\,29           & 9.54 &  138.7 \\
616 & B0.5: V        & 227.4\,$\pm$\,10.7  &\o$-$44.2\,$\pm$\,16.2 & 3.6 &\o$\le$40$\phantom{\le}$\p & 1.26 & \o18.3 \\
625 & B1.5 V         & 310.7\,$\pm$\,\o9.2 &\pp\o39.1\,$\pm$\,15.3 & 3.2 &\o64\,$\pm$\,12            & 2.57 & \o37.4 \\
831 & B5 Ia          & 210.4\,$\pm$\,\o6.2 &\o$-$61.2\,$\pm$\,13.7 & 5.0 &\o41\p                     & 7.91 &  115.0 \\
851 & B2 III         & 229.4\,$\pm$\,\o0.3 &\o$-$42.2\,$\pm$\,12.2 & 3.5 &\o$\le$40$\phantom{\le}$\p & 7.18 &  104.4 \\
\hline
\end{tabular}
\tablefoot{\footnotesize Estimates of \vsini\ are from \citet{duf13} except for
  VFTS\,831, which is taken from \citet{cmm}. 
Projected distances from the core of R136 are from Paper~I.}
\end{center}
\end{table*}

\begin{figure}
\begin{center}
\includegraphics[scale=0.52]{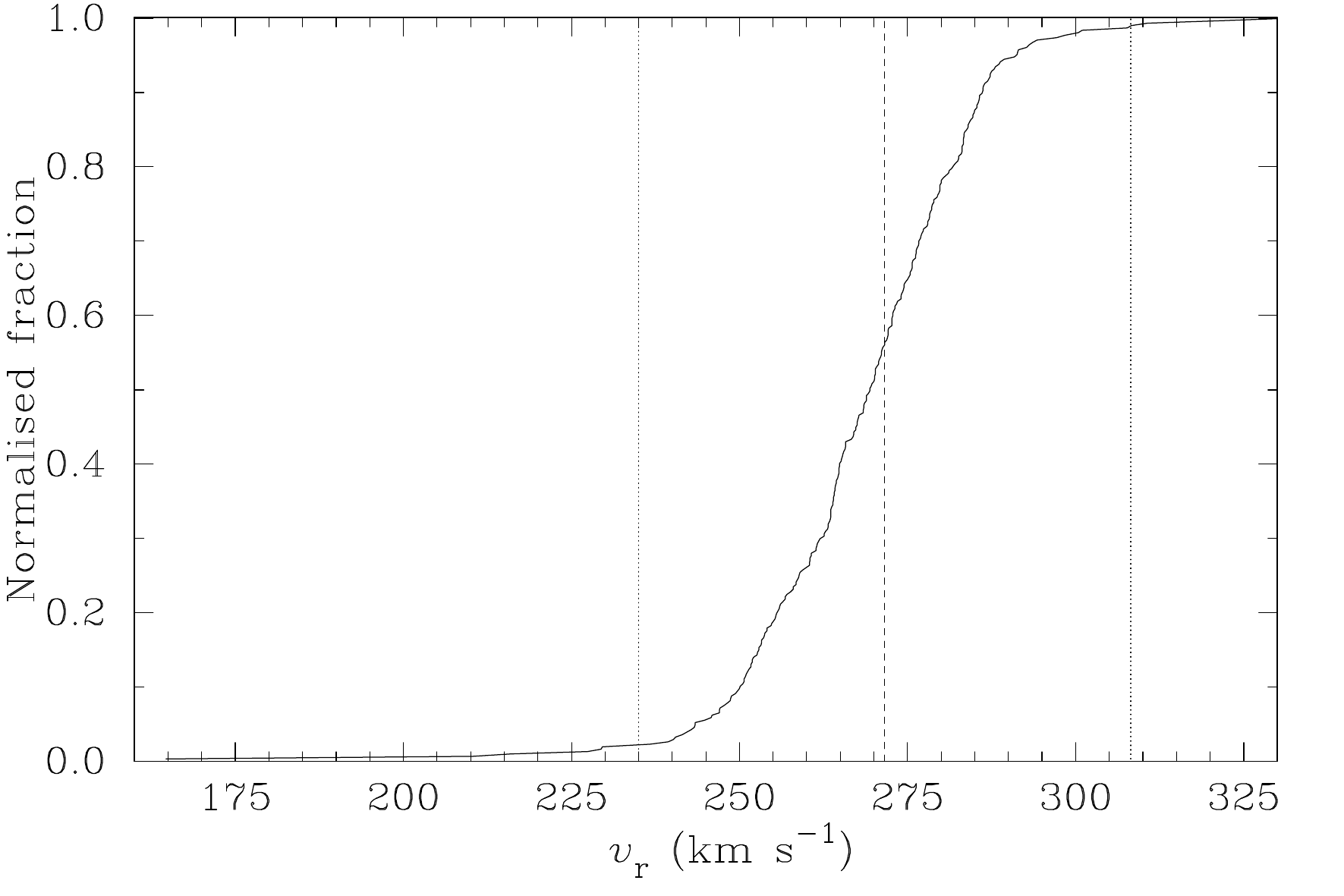}
\caption{\footnotesize Cumulative distribution of estimated radial velocities for
  the 307 single B-type stars. The dashed vertical line indicates the
  weighted mean velocity of the sample, $v_{\rm r}$\,$=$\,271.6\,\kms\
  (calculated excluding stars in the two older clusters and the
  candidate runaways). The vertical dotted lines are the 3$\sigma$
  thresholds adopted for identification of potential runways
  ($\pm$36.6\,\kms).}\label{vr_all}
\end{center}
\end{figure}

To identify candidate runaways in the B-type sample we compared the
radial-velocity estimates for each star with the initial weighted mean
and standard deviation calculated for the stars in NGC\,2060,
NGC\,2070, and the surrounding field population (i.e. excluding
Hodge\,301 and SL\,639). Employing a threshold of $|\delta v_{\rm r}|$
\,$>$\,3$\sigma$ we identified eight candidate runaways and then
recalculated the mean velocities (and dispersions) excluding these
objects. Reapplying a 3-$\sigma$ threshold provided one additional
candidate, while a second iteration did not reduce the velocity
dispersion further. The mean velocities and dispersions for the
different spatial samples (excluding potential runaways) are presented
in Table~\ref{results_spatial}.  The cumulative distribution of the
velocity estimates for the 307 single stars is shown in
Figure~\ref{vr_all}, with the $\pm$3$\sigma$ threshold (i.e.
$\pm$36.6\,\kms) for identification of runaway stars indicated by the
vertical dotted lines.

Our adopted velocity threshold is comparable in magnitude to the limit
used by \citet{b61}, which was that candidate runaways have a {\em
  space velocity} which differs by more than 40\,\kms\ from the
systemic velocity of the region in which they formed. (In this
context, our radial-velocity estimates should serve as lower limits to
the 3-D space velocities.)  In contrast, \citet{spz00} adopted a lower
threshold of 25\,\kms\ in his theoretical study of the origins and
characteristics of runaways.  Given the limitations of the B-type data
and the relatively small number of lines available for some objects,
we did not consider a comparably low threshold.

Details of the nine candidate runaways are summarised in
Table~\ref{runaways}, and their locations are shown in
Figure~\ref{runaways_spatial}.  Four of the candidates have
differential velocities in the range of 3.2-3.6\,$\sigma$, thus their
status as candidates will depend strongly on the adopted velocity
criteria.  Moreover, our adopted velocity threshold (the dispersion of
the weighted mean velocities in the sample) was employed without
consideration of the uncertainty on the radial-velocity estimate for a
given star.  Thus, VFTS\,467, although formally with a radial velocity
of 4.2$\sigma$ from the systemic value, has a large uncertainty on its
estimated velocity. Indeed, only three He~\1 lines were available for
this star and the \lam4144 estimate (277.5\,\kms) is comparable to the
systemic value.

The choice to use weighted means to investigate the
spatially-different samples in Section~\ref{rvs_spatial} was motivated
by the fact that approximately 10\,\% of our velocity estimates have
uncertainties of $\ge$\,15\,\kms\ (each of which is a broad-lined
star, i.e. used the Set~2 lines from Table~\ref{vr_lines}).  To
investigate the dependence of the number and identity of the candidate
runaways on these adopted means, we repeated the above steps for
both the (unweighted) mean and median velocities of the sample.  The
mean velocity for the 281 stars (not including those in Hodge\,301 and
SL\,639) was 269.8\,$=$\,16.4\,\kms, with a median value of
270.7\,\kms.  Adopting either of these systemic values and a
3-$\sigma$ threshold (taking the dispersion as 16.4\,\kms) led to
identification of the five most significant runaways from
Table~\ref{runaways} (i.e.  VFTS\,242, 298, 358, 467, and 831), with a
second iteration providing one additional candidate (VFTS\,616).  The
final mean velocity from these calculations is 270.1\,$\pm$\,13.9, in
good agreement with the equivalent subsample in the final row of
Table~\ref{results_spatial}. From these checks we conclude that the four
most significant candidates (VFTS\,242, 298, 358, and 831) are secure
as potential runaways, with five more tenative candidates as
discussed above.

For the members of Hodge\,301, VFTS\,309 is a potential outlier (at
3.6$\sigma$) compared to the mean velocity for the cluster in
Table~\ref{results_spatial}. Indeed, excluding its $v_{\rm r}$
estimate from the calculated mean and dispersion for Hodge\,301 gives
$\overline{v_{\rm r}}$\,$=$\,261.5\,$\pm$\,5.0\,\kms, such that
VFTS\,309 is then a 4.1$\sigma$ outlier.  However, its velocity is
also in reasonable agreement with those for stars in NGC\,2060 and
NGC\,2070, so we are unable to confirm it as a potential runaway from
Hodge\,301.  Finally, we note that the velocity estimate for VFTS\,310
is not significant as a potential runaway from Hodge\,301.

\medskip

With our adopted velocity criteria, the nine candidate runaways
correspond to 2.9\,\% of the sample of 307 single stars with
radial-velocity estimates.  This lower fractional share of runaways
compared to the O-type stars (Sana et al., in prep.) 
is in qualitative agreement with recent N-body predictions for the
mass-spectrum of runaways \citep{bko12,bk12}, although quantitative
comparison with the predictions is complicated by the different
criteria employed (e.g. \citeauthor{bko12} used a radial distance
threshold for a star to be considered a runaway).  We note that
  the distribution in Figure~\ref{vr_all} is slightly asymmetric, with
  two thirds of the candidates having negative differential
  velocities.  This could point to veiling of more distant runaways
  by the nebula, although this is not seen in the results for the O-type
  stars (in which a greater fraction are seen to have positive
  differential velocities, Sana et al., in prep.), suggesting that
  small-number statistics likely explain the asymmetry.

\subsection{Rotational velocities of candidate runaways}\label{sec_rw_vsini}

\begin{figure}
\begin{center}
\includegraphics[scale=0.52]{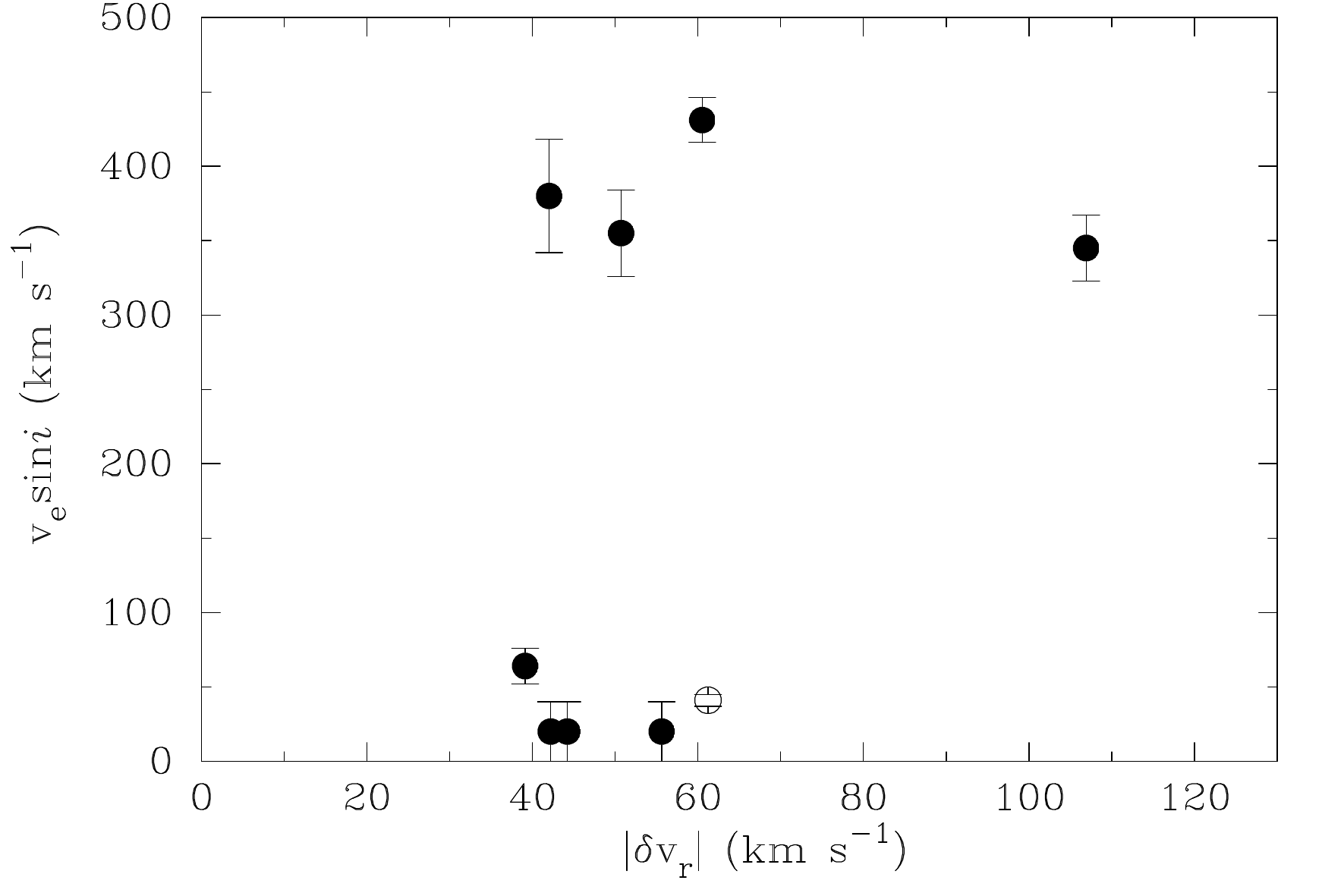}
\caption{\footnotesize Comparison of the projected rotational velocities (\vsini)
  with the differential radial velocities ($\delta v_{\rm r}$) compared
  to the mean systemic velocity of the B-type sample. Uncertainties on
  \vsini\ of $\pm$10\,\% are adopted for VFTS\,368 and 831 (with the
  latter plotted as an open symbol given its supergiant
  classification).}\label{rw_vsini}
\end{center}
\end{figure}

Estimated projected rotational velocities \citep[from][] {duf13,cmm}
for our candidate B-type runaways are included in
Table~\ref{runaways}.  There appears to be a dichotomy in their
rotation rates, as highlighted by Fig.~\ref{rw_vsini}.  Five stars
have small rotational velocities, with the estimates for four stars
effectively limited by the velocity resolution of the data (i.e.
$\sim$40\,\kms) and the fifth with \vsini\,$=$\,64\,$\pm$\,12\,\kms.
Meanwhile, the remaining four stars each have estimates of
\vsini\,$\ge$\,345\,\kms.  To investigate the significance of these
results, we ran Monte Carlo simulations using the \vsini\
distributions from \citet{duf13}. We note that the results from
\citeauthor{duf13} were for the unevolved B-type sample (i.e. dwarf
and giant objects), so we exclude VFTS\,831 from the following
discussion.

We generated random rotational velocities for eight stars drawn from
the distribution of \citeauthor{duf13}, and repeated this 10$^6$ times
to estimate the probability of recovering rotational velocities
consistent with our eight (non-supergiant) runaway candidates (i.e.
four stars with \vsini\,$\le$\,65, and four with
\vsini\,$\ge$\,345\,\kms). We then repeated these runs 10$^4$
times to estimate the dispersion on the probabilities. In a second
test we limited the low velocity subsample to the three stars with
\vsini\,$\le$\,40\,\kms\ (i.e. excluding VFTS\,625, which has the
least significant runaway velocity), together with the four rapid
rotators.  We found low probabilities for recovering the observed
\vsini\ values from both tests (as summarised in
Table~\ref{vsini_cf_rw}), with significance levels of ~3.0 and
3.2$\sigma$, respectively.

We also investigated the probabilities compared to the cumulative
distribution function of unprojected rotational velocities ($v_e$)
from \citeauthor{duf13} (their Table~6).  In this case we are limited
by the unknown inclination angles of our targets but, assuming some
relatively conservative lower-limits, we repeated tests similar to
those above (in which we adopt the nearest available value from Dufton
et al. for our upper limit, i.e. 340\,\kms).  These tests are formally
less significant (Table~\ref{vsini_cf_rw}), but provide further
support that the rotational velocities of our candidate runaways do
not appear to be randomly drawn from the distribution for the larger
B-type sample.  We note that this agrees with similar conclusions
regarding the rotational velocities of the O-type runaways
\citep[][Sana et al., in prep.]{w14}.

\begin{table}
\begin{center}
  \caption{\footnotesize Probability tests to investigate the rotational velocities
    of the candidate runaway stars compared to those for the full
    (unevolved) B-type sample from \citet{duf13}.}\label{vsini_cf_rw}
\begin{tabular}{lcc}
\hline\hline
Test (from 10$^4$ repeats of 10$^6$ runs) & $n$ & $P$ \\
\hline
4\,$\times$\,\vsini\,$\le$\,65~~\&~~4\,$\times$\,\vsini\,$\ge$\,345\,\kms & \o10\,$\pm$\,3\o & 0.0010 \\
3\,$\times$\,\vsini\,$\le$\,40~~\&~~4\,$\times$\,\vsini\,$\ge$\,345\,\kms & \o\o6\,$\pm$\,2\o & 0.0006 \\
& & \\
4\,$\times$\,$v_e$\,$\le$\,100~~\&~~4\,$\times$\,$v_e$\,$\ge$\,340\,\kms & 395\,$\pm$\,20 & 0.0395 \\
3\,$\times$\,$v_e$\,$\le$\,\v\,60~~\&~~4\,$\times$\,$v_e$\,$\ge$\,340\,\kms & 181\,$\pm$\,13 & 0.0181 \\
\hline
\end{tabular}
\tablefoot{\footnotesize `$n$': number of Monte Carlo runs (from 10$^6$ sets)
  for which the simulated rotational velocities are consistent with
  those of the candidate runaways (and translated into a fractional
  probability, $P$, in the final column); the quoted uncertainties are
  the standard deviations from 10$^4$ repeat runs.}
\end{center}
\end{table}

\subsection{Atmospheric analysis of two candidate runaways}

Atmospheric analysis of the B-type spectra is now underway within the
VFTS project.  In the context of the current study, we were
particularly motivated to investigate the properties of the two most
noteworthy runaway candidates, VFTS\,358 and 831; we now comment
briefly on our findings for these two objects.

\subsubsection{VFTS\,358}

This star has the largest differential velocity from the estimated systemic
value ($\delta v_{\rm r}$\,$=$\,$-$106.9\,$\pm$\,16.2\,\kms) and also
has a relatively large projected rotational velocity
\citep[\vsini\,$=$\,345\,$\pm$\,22\,\kms;][]{duf13}. This prompted us
to investigate its physical properties, in part to see if there was
any evidence of past binary interaction \citep[see e.g.][]{sdm14}.

Detailed analysis was complicated by the rapid rotation, which made
estimation of the atmospheric parameters and element abundances
difficult. However, using similar methods as those used by
\citet{ihchem}, we undertook a preliminary analysis using a grid of
model atmospheres calculated using the {\sc tlusty} and {\sc synspec}
codes \citep{hub88,hub95,hub98,lan07}; further details of the grids
are given by \citet{ryans03} and \citet{duf05}\footnote{See also
  http://star.pst.qub.ac.uk}.

We estimated the effective temperature ($T_{\rm eff}$) and gravity
(log\,$g$) of VFTS\,358 by fitting rotationally-broadened profiles to
the H$\delta$, H$\gamma$ and He~\2 \lam4686 lines, assuming solar
helium abundances. This gave estimates of $T_{\rm
  eff}$\,$\simeq$\,29\,kK and log\,$g$\,$\simeq$\,3.5\,dex, although
we were unable to estimate the microturbulence because of the paucity
of observable metal lines (which also precluded use of the silicon
ionisation balance to estimate the temperature).

\begin{figure}
\begin{center}
\includegraphics[]{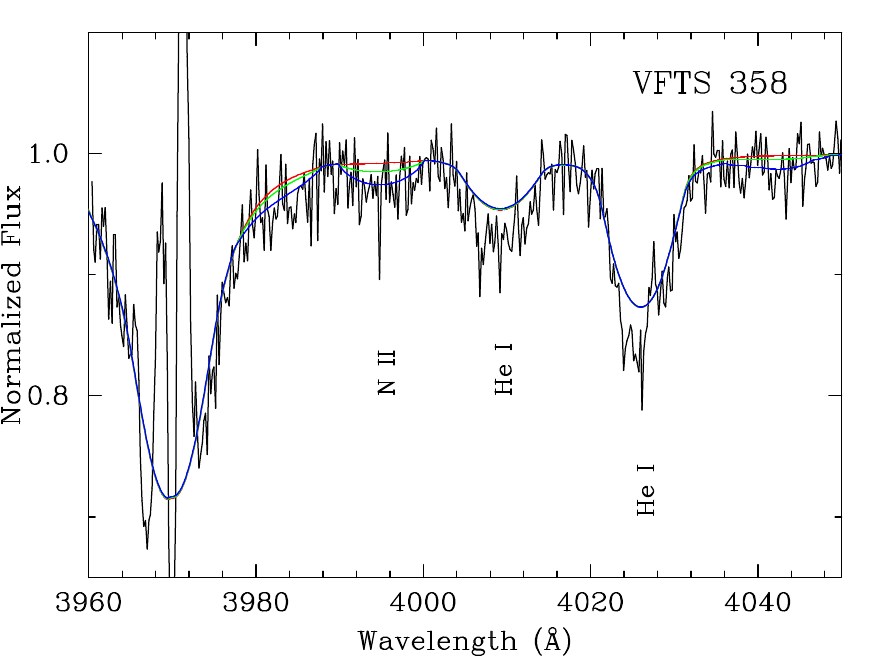}
\caption{\footnotesize Section of the VFTS\,358 spectrum showing, from
  left-to-right, the H$\epsilon$, N~{\scriptsize II} \lam3995, and
  He~{\scriptsize I} \lam\lam4009, 4026 lines; we note that H$\epsilon$
  is complicated by absorption from the interstellar Ca~$H$ line (as
  well as nebular emission).  Model spectra for our adopted physical
  parameters and 12\,$+$\,log\,(N/H)\,$=$\,6.9, 7.7, and 8.5, are
  overplotted in red, green, and blue, respectively.}\label{spec_358}
\end{center}
\end{figure}

Absorption from N~\2 \lam3995 appears to be present in the spectrum
(with an equivalent width of $\sim$120\,m\AA), although its
identification is marginal, as shown in Figure~\ref{spec_358}.
Rotationally-broadened model spectra for our adopted physical
parameters (and an assumed microturbulence of 5\,\kms) are shown for
three nitrogen abundances: 12\,$+$\,log\,(N/H)\,$=$\,6.9 \citep[the approximate
baseline N-abundance for the LMC, see discussion by][]{ihchem}, 7.7, and 8.5\,dex
plotted in red, green, and blue, respectively. This comparison
suggests significant nitrogen enrichment, with an abundance of
$\gtrsim$\,8.5\,dex (the maximum abundance considered in our grid).
Tests for theoretical profiles with a microturbulence of 20\,\kms\
\citep[which is larger than normally found for stars with this
gravity, e.g.][]{ihchem} implied a similarly large nitrogen
abundance.  Fitting a rotationally-broadened profile to the N~\2
\lam3995 line leads to \vsini\,$=$\,342\,\kms\ (cf.  345\,\kms\ from
the helium lines) and a central wavelength of 3995.06\,\AA\ (cf. the
laboratory wavelength of 3995.00\,\AA). This agreement provides some
support that the feature is indeed from N~\2 absorption.

The neutral helium lines in the observed spectrum appear somewhat stronger
than those in the best-fitting model (e.g. the He~\1 \lam\lam4009,
4026 lines in Figure~\ref{spec_358}) suggesting that the helium
abundance might also be enhanced. In turn, this would tend to decrease
the $T_{\rm eff}$ estimated from the He~\2 line (and the surface
gravity). Tests using cooler models indicated that the change in
nitrogen abundance would be modest. For example, a change in 
effective temperature of $-$2\,kK leads to a change in 
12\,$+$\,log\,(N/H) of approximately $-$0.2\,dex.  Hence, the
nitrogen abundance would still be significantly greater than the
baseline value for the LMC.

At present we are limited by the quality of the VFTS spectrum but from
this preliminary analysis there is evidence of significant nitrogen
enrichment (and maybe also helium) in the photosphere of VFTS\,358.
Indeed, the estimated nitrogen abundance is the largest obtained so
far from the VFTS survey (including results from the B-type
supergiants from \citet{cmm}.  The large velocity offset,
high rotation rate, enhanced nitrogen (and possibly helium) all fit,
at least qualitatively, with a scenario in which VFTS\,358 is the
former secondary of an interacting binary that became unbound at the
explosion of the primary star.

\subsubsection{VFTS\,831}

VFTS\,831 is remarkable in two regards. First as a luminous B-type
supergiant, likely to be the evolved descendent of a formerly more massive
O-type star, and secondly because, from the definition used in
Table~\ref{results_spatial}, it is located within SL\,639
($\approx$12\farcs5 from the visual centre). With $v_{\rm
  r}$\,$=$\,210.4\,$\pm$\,6.2\,\kms\ it also appears to be runaway
compared to the other targets in SL639. A model-atmosphere analysis of
the spectrum of VFTS\,831 has been presented by \citet{cmm}, 
who classified it as an apparently single star.  It is one of the cooler
objects in the supergiant sample, with $T_{\rm eff}$\,$=$\,14,000\,K
and log\,$g$\,$=$\,2.1.  

Following the discussion in Section~\ref{sec_rw_vsini}, VFTS\,831 does
not appear particularly remarkable in terms of stellar rotation.
Indeed, it has a relatively low projected rotational velocity and a
substantial contribution from macroturbulent broadening (\vsini,
$\Theta_{\rm RT}$) $\sim$(40, 65)\,\kms, as found in most B-type
supergiants \citep[see e.g.][]{mar14,sh14}. From the intrinsically
strong N~\2 \lam3995 singlet transition, \citeauthor{cmm} estimated a
nitrogen abundance of 12\,$+$\,log\,(N/H)\,$=$\,7.94.  Although large,
this is consistent with the other single supergiants in the sample,
which have abundances spanning approximately 6.8 to 8.1 dex.

\subsection{Single-lined binaries}

For completeness we also investigated the velocities of single-lined
binaries for potential runaway systems.  In the final column of
Table~\ref{rv_binaries} we include the observed {\em range} of
velocities, $\delta v_{\rm r\,max}$, from Dunstall et al. (in prep.).
Following the arguments laid out in Section~\ref{com_binaries} of the Appendix, we
adopt an estimated error on the true systemic velocity compared to the
single-epoch estimate of 0.15\,$\times$\,$\delta v_{\rm r\,max}$.  For
stars with $v_{\rm r}$ above (or below) the systemic velocity, we then
subtract (or add) this estimated error, to give a corrected velocity
estimate, $v_{\rm corr}$.  Using the same 3$\sigma$ criteria
discussed above we then looked for potential runaways.

Only one single-lined system, VFTS\,730, was found to be a potential
runaway, with $v_{\rm corr}$\,$=$\,324.0\,\kms\ (a 4.3\,$\sigma$
outlier cf. the adopted systemic velocity). Its range of observed
velocities is relatively small 
($\delta v_{\rm  r\,max}$\,$=$\,16.6\,\kms, thus only just above the detection
threshold employed by Dunstall et al.)
but it could be, for example, a long-period system with an eccentric
orbit. Indeed, inspection of its velocity estimates as a function of
epoch show them to be monotonically decreasing from 332 to 316\,\kms,
suggesting it as a long-period system.  Further investigation of this
(and other) binary systems will require comprehensive spectroscopic
monitoring to determine their centre-of-mass velocities and other
orbital parameters.

\subsection{Origins of candidate runaways}

The production of runaways is thought to be from one of two channels,
with evidence in support of both mechanisms \citep{hbz00}. In the
dynamical-ejection scenario the runaway is ejected via gravitational
interaction between single-binary or binary-binary systems in a dense
star cluster \citep{pra67}. In the binary-ejection scenario, the
primary star explodes as a supernova, leaving behind a neutron star or
black hole.  In most cases the compact object leaves the system
depending on the magnitude and direction of the birth kick, and the
companion is ejected with a velocity approximately equal to its
orbital velocity \citep{b61,eld11}.

The complex star-formation history of 30~Dor makes it hard to cleanly
attribute our candidate runaways to either of the two scenarios.  R136 is
sufficiently young \citep[1--2\,Myr;][]{dk98,mh98} that none of its
members is thought to have undergone a supernova explosion. Thus, any
runaways from R136 would have to be ejected via dynamical
interactions.  However, as noted above, the VFTS region
contains several older subclusters and associations, including
NGC\,2060, which contains a supernova remnant \citep[e.g.][]{cksl92}, 
and Hodge\,301 which was argued by \citet{gc00} as a likely site of 
tens of past supernovae from consideration of its mass function.

One of the important diagnostics might prove to be the connection with the
observed rotational velocities (Fig.~\ref{rw_vsini}), in which the
apparently rapid rotators could have a binary origin. For example, the
theoretical discussions by \citet{sdm13}, and the suggestion that
the high-\vsini\ tail of the distribution of the rotational velocities
for the VFTS O-type stars could be accounted for by past binary
interactions \citep{ora13}.  Detailed dynamical analysis of the O-type
stars in the survey has been underway in parallel to the current work,
and the broader question of the origins of runaway stars from 30~Dor
is discussed in more depth by Sana et al. (in prep.). 

\section{Summary}

We have presented comprehensive spectral classifications and
radial-velocity estimates for the B-type stars observed in the VFTS.
We now briefly summarise our main findings:
\begin{itemize}
\item{The members of the two older clusters, Hodge\,301 and SL\,639,
    appear kinematically distinct from those in the younger clusters and
    general field population.}
\item{The preliminary H-R diagram for stars in Hodge\,301 and SL\,639
    (Figure~\ref{clusters_fig}) suggests an age of $\sim$10-15\,Myr
    for the population of SL\,639, slightly younger than that of
    Hodge\,301 (15\,$\pm$\,5\,Myr).}
\item{The systemic velocity for 273 single stars, excluding those in
    the two older clusters and candidate runaways, is
    $\overline{v_{\rm r}}$\,$=$\,271.6\,$\pm$\,12.2~(s.d.)~\kms.}
\item{Employing a 3-$\sigma$ velocity threshold, we identify nine
    single stars as candidate runaways (2.9\,\% of the single stars
    with radial-velocity estimates).  These are mostly unevolved,
    early B-type objects, but include a B5-type supergiant
    (VFTS\,831).}
\item{There appears to a bimodal distribution of rotation rates in the
    candidate runaways.  Five stars have \vsini\,$\le$\,65\,\kms,
    while the others have \vsini\,$\ge$\,345\,\kms.}
\item{Excluding VFTS\,831 (because of its evolved nature), there is a
    remarkably small probability ($\sim$0.001\%) that the projected
    rotational velocities for our other eight candidate runaways could
    be randomly drawn from those for the full (unevolved) B-type
    sample.}
\item{VFTS\,358 has the largest velocity offset from the systemic
    velocity of 30~Dor ($\delta v_{\rm
      r}$\,$=$\,$-$106.9\,$\pm$\,16.2\,\kms) and appears to be
    rotating rapidly (\vsini\,$=$\,345\,\kms). Preliminary spectral
    analysis suggests evidence of significant nitrogen enrichment,
    perhaps indicative of past binary interaction.}
\end{itemize}

The classifications presented here will underpin the quantitative
analyses in progress on this substantial part of the VFTS data. For
example, they will enable investigation of the mass/temperature
dependence (if any) of the estimated \vsini\ results from
\citet{duf13}, particularly when also combined with the results for
the O-type stars from \citet{ora13}. In the longer-term, the ongoing
proper-motion studies using {\em HST} imaging (programme GO12499, P.I.
Lennon) will add another dimension to our studies of the complex
dynamics in this region, by providing tangential velocities
\citep[see][]{sabbi13}.

Our radial-velocity estimates should also lend themselves to detailed
investigation of the structural properties of the VFTS sample
\citep[e.g. using the techniques discussed by][]{p14,wright14},
particularly when combined with the results for the O-type stars from
\citet{s13}; this should provide further insights into the formation
and evolutionary history of the massive-star population in the region.

Finally, there is an ongoing monitoring programme of FLAMES
spectroscopy to characterise a subset of the O-type binaries (PI:
Sana); a similar monitoring programme for the $\sim$100 binaries
discussed here and by Dunstall et al. (in prep.) 
would be valuable to determine the
full orbital parameters of the lower-mass population of binaries,
while also providing robust estimates of centre-of-mass velocities to
better explore their dynamics in the context of ejected runaways.

\begin{acknowledgements}
  Based on observations at the European Southern Observatory Very
  Large Telescope in programme 182.D-0222. We thank Mark Gieles for
  helpful discussions regarding the candidate runaways, and the
  referee for their constructive comments. STScI is operated by AURA
  Inc., under NASA contract NAS 5-26555.  SdM acknowledges support for
  this work by NASA through an Einstein Fellowship grant, PF3-140105.
  JMA acknowledges support by grants AYA2010-17631 and AYA2010-15081
  of the Spanish Ministry of Economy and Competitiveness. SS-D
  acknowledges funding from the Spanish Government Ministerio de
  Econom\'{i}a y Competitividad (MINECO) through grants
  AYA2010-21697-C05-04, AYA2012-39364-C02-01 and Severo Ochoa
  SEV-2011-0187, and the Canary Islands Government under grant
  PID2010119.
\end{acknowledgements}

\bibliographystyle{aa}
\bibliography{24414}

\onecolumn
\small
\begin{landscape}
\begin{center}
\begin{longtable}{llcccccccll}
  \caption{\leftline{Spectral classifications and radial velocities for B-type stars in the VFTS.\label{classifications}}}\\
  \hline
  VFTS & Spectral Type & Mult.? & H$\alpha$ & $V$ & $B-V$ & $v_{\rm r}$ & $\sigma$ & Set & Alt.\,ID & Comments \\
  & & & & & & \multicolumn{2}{c}{[\kms]} & & & \\
  \hline
\endfirsthead
  \caption{\leftline{\it{continued}}}\\
\hline
  VFTS & Spectral Type & Mult.? & H$\alpha$ & $V$ & $B-V$ & $v_{\rm r}$ & $\sigma$ & Set & Alt.\,ID & Comments \\
  & & & & & & \multicolumn{2}{c}{[\kms]} & & & \\
\hline
\endhead
\hline
\multicolumn{11}{r}{\it{continued on next page}}\\
\endfoot
\hline
\endlastfoot
001 & B1.5: V             & \ldots       &  a  & 16.93 & \pp0.05 &  282.1 & \o4.2 & 2    & \ldots & \ldots \\
003 & B1 Ia$^+$           & \ldots       &\q q?& $-$   & \pp$-$  &  265.2 & \o2.1 & 1    & Sk$-$69$^\circ$~224 & \ldots \\
004 & B2 V                & \ldots       &  a  & 16.87 & \pp0.16 &  283.1 &  10.8 & 2    & \ldots & \ldots \\
005 & B2 V(n)             & \ldots       &  a  & 16.25 & \pp0.02 &  282.6 & \o2.6 & 2    & \ldots & \ldots \\
007 & B1-2 V              & \ldots       &  a  & 16.74 & \pp0.03 &  281.1 & \o1.6 & 2    & \ldots & \ldots \\
008 & B0.5: V(n)          & \ldots       &  a  & 16.92 & \pp0.07 &  271.3 & \o6.9 & 2    & \ldots & \ldots \\
009 & B1-1.5 V            & $\checkmark$ &  a  & 16.18 & \pp0.11 &  293.2 &  21.6 & B    & \ldots & \ldots \\ 
010 & B2 V                & \ldots       &  a  & 16.79 & \pp0.04 &  287.1 & \o3.9 & 1    & \ldots & \ldots \\
\hline
\end{longtable}
\tablefoot{Column entries are: (1) VFTS identifier; (2) Spectral
  classification; (3) Binary status (from Dunstall et al., in prep.)
(4) H$\alpha$
  morphology: `a'\,$=$\,(photospheric) absorption, `e'\,$=$\,(stellar)
  emission; `f'\,$=$\,flat, with no indication of the nature of the
  line profiles outside the region of nebular emission (i.e. no
  strong, broad emission); `q'\,$=$\,P-Cygni profile (used classically
  to distinguish from a `p' suffix, which simply indicates
  peculiar); (5)\,\&\,(6) $V$ and $B-V$ photometry from Paper~I;
  (7)\,\&\,(8) estimated radial velocity ($v_{\rm r}$) and standard
  deviation ($\sigma$); (9) method/lines used to estimate $v_{\rm r}$:
  `1' and `2' refer to the lines listed in Table~\ref{vr_lines}, `3'
  indicates the three stars for which other metal lines were used, `B'
  indicates that the estimate is the mean value from the multi-epoch
  data available (see Section~\ref{methods_binaries}, with the
  limitations of these as estimates of the centre-of-mass velocities
  discussed in Section~\ref{com_binaries}); (10) Alternative
  identifiers used in the literature; (11) Previous
  classifications and other relevant comments.  Aliases/past
  classifications are from: R/F60 \citep{f60}; AL \citep{al64}; Sk
  \citep{sk70}; BE74 \citep{be74}; BI \citep{bi75}; W84 \citep{w84};
  Mk \citep{m85}; W86 \citep{w86}; T88 \citep{tld88}; ST92
  \citep{st92}; P93 \citep{p93}; WB97 \citep{wb97}; B99 \citep{b99}.
  The crosschecks with the catalogues from \citet{al64} and
  \citet{be74} employed the updated astrometry from \citet{h12,h13}.}
\end{center}
\end{landscape}

\twocolumn
\normalsize

\appendix

\section{Ancillary material}\label{rv_checks}

\subsection{Spectral montages}
\begin{figure*}
\begin{center}
\includegraphics{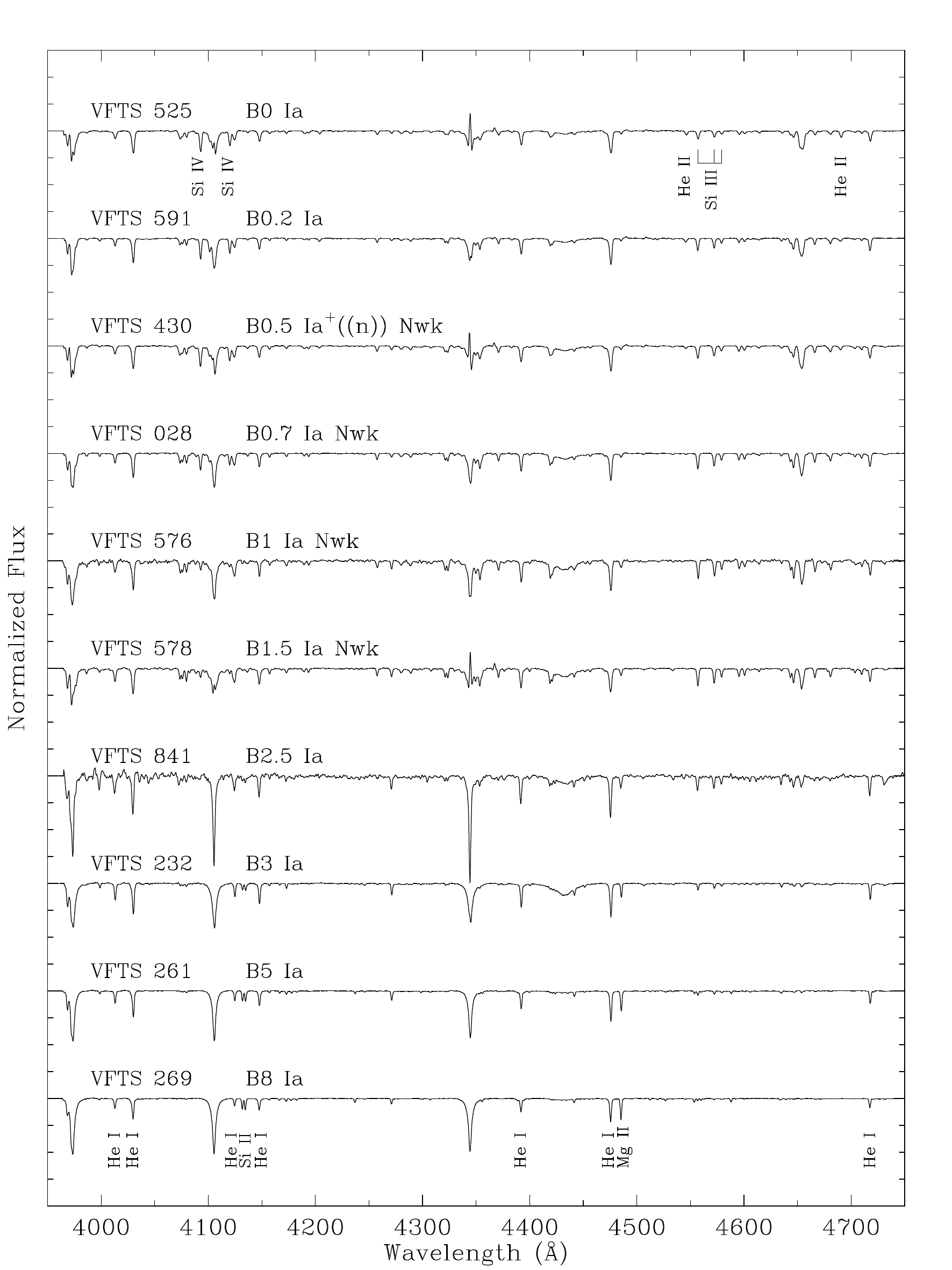}
\caption{\footnotesize Example B-type supergiants (Class Ia).  The primary
  diagnostic lines for spectral classification are identified.  In the
  spectrum of VFTS\,525 these are: He~{\scriptsize II} \lam\lam4542,
  4686; Si~{\scriptsize III} \lam\lam4553-68-74; Si~{\scriptsize IV}
  \lam\lam4089, 4116; those identified in VFTS\,269 are:
  He~{\scriptsize I} \lam\lam4009, 4026, 4121, 4144, 4388, 4471, 4713;
  Mg~{\scriptsize II} \lam4481; Si~{\scriptsize II} \lam\lam4128-32.
  The broad \lam4428 diffuse interstellar band is also evident in some
  sightlines.}\label{bsgs_fig}
\end{center}
\end{figure*}

\begin{figure*}
\begin{center}
\includegraphics{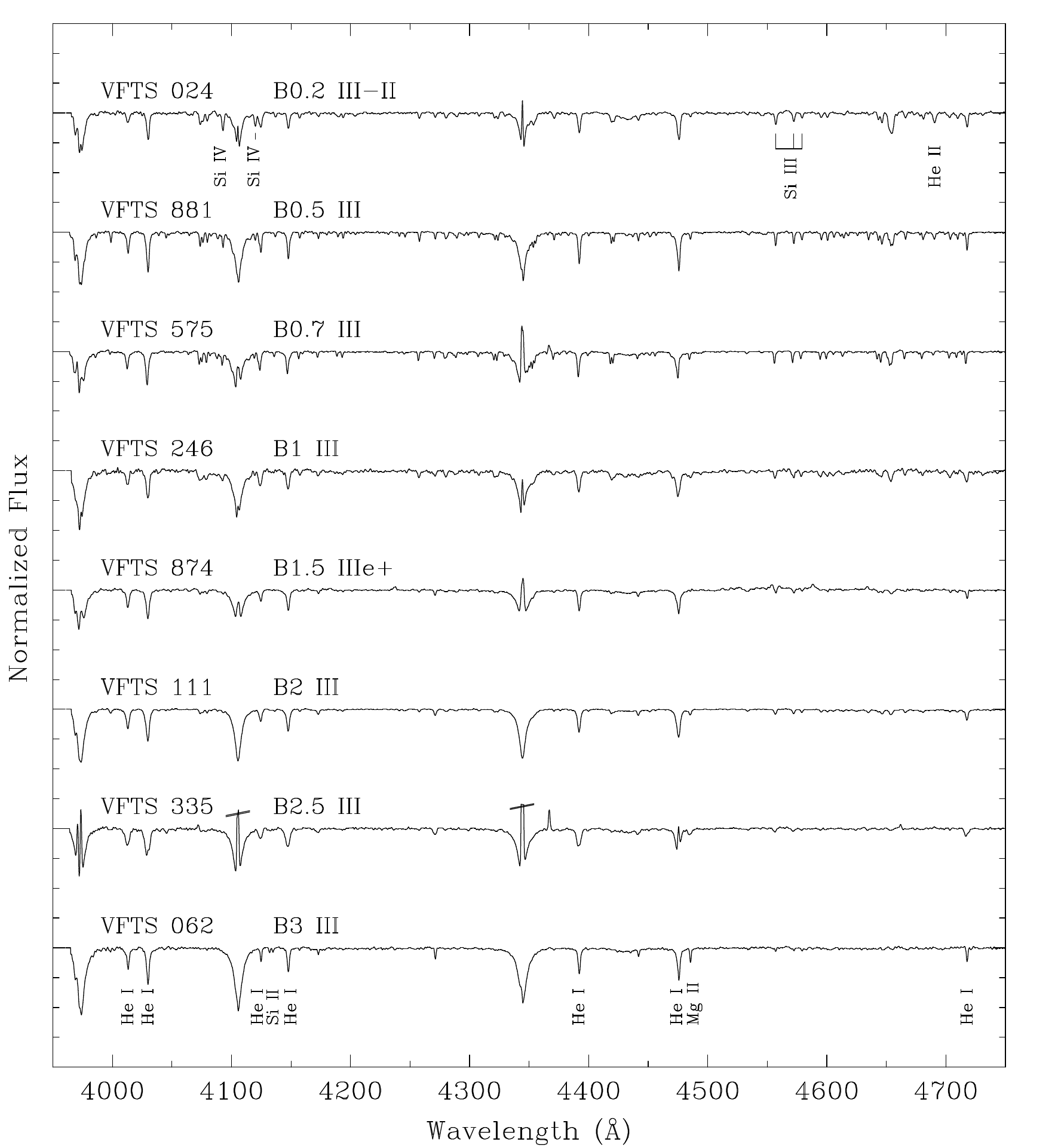}
\caption{\footnotesize Example B-type giants (Class III).  The identified lines are
  the same as those in Figure~\ref{bsgs_fig}; nebular lines have been
  truncated in VFTS\,335 as indicated.}\label{giants_fig}
\end{center}
\end{figure*}

\begin{figure*}
\begin{center}
\includegraphics{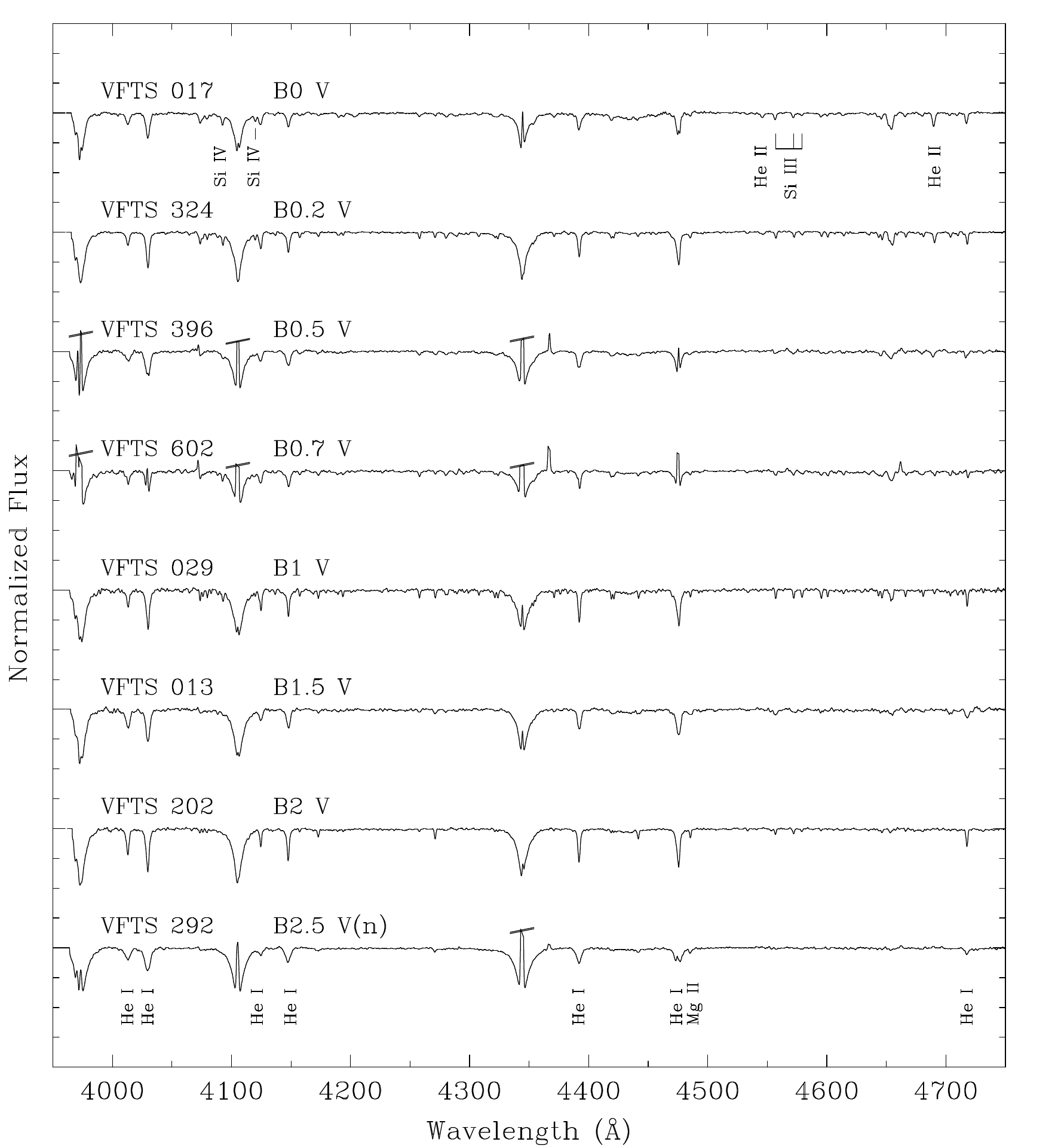}
\caption{\footnotesize Example B-type dwarfs (Class V).  The identified lines are
  the same as those in Figure~\ref{bsgs_fig}; nebular lines have been
  truncated in three spectra as indicated.}\label{dwarfs_fig}
\end{center}
\end{figure*}

\begin{figure*}
\begin{center}
\includegraphics{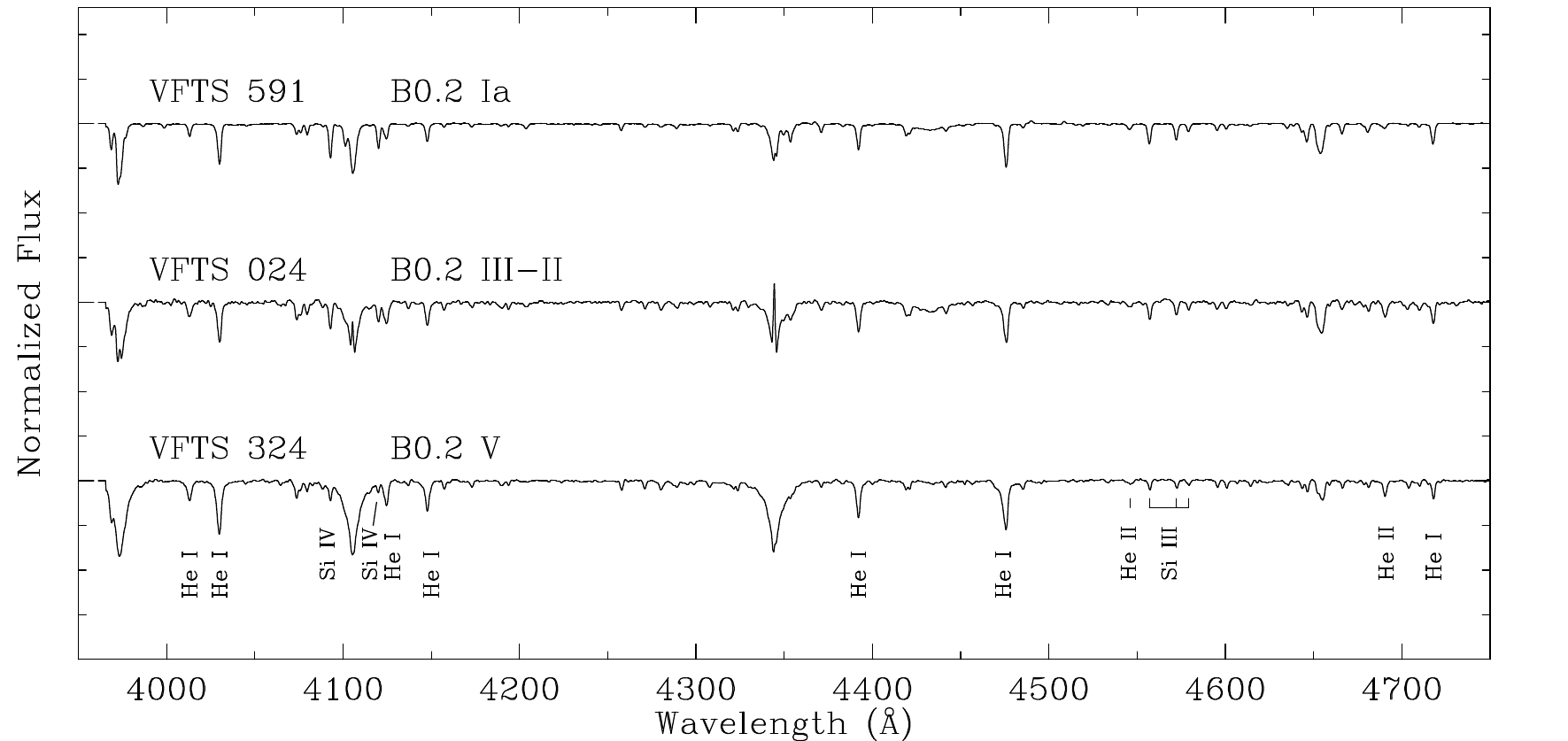}
\caption{\footnotesize Example luminosity sequence at B0.2. The identified lines are
  the same as those in Figure~\ref{bsgs_fig}; we note the greater
  intensity of the metallic lines with increasing luminosity
  (including the C\,\siii\,$+$\,O\,\sii\ features at
  $\sim$\lam4650).}\label{seq_b02} \vspace*{1cm}
\end{center}
\end{figure*}

\begin{figure*}
\begin{center}
\includegraphics{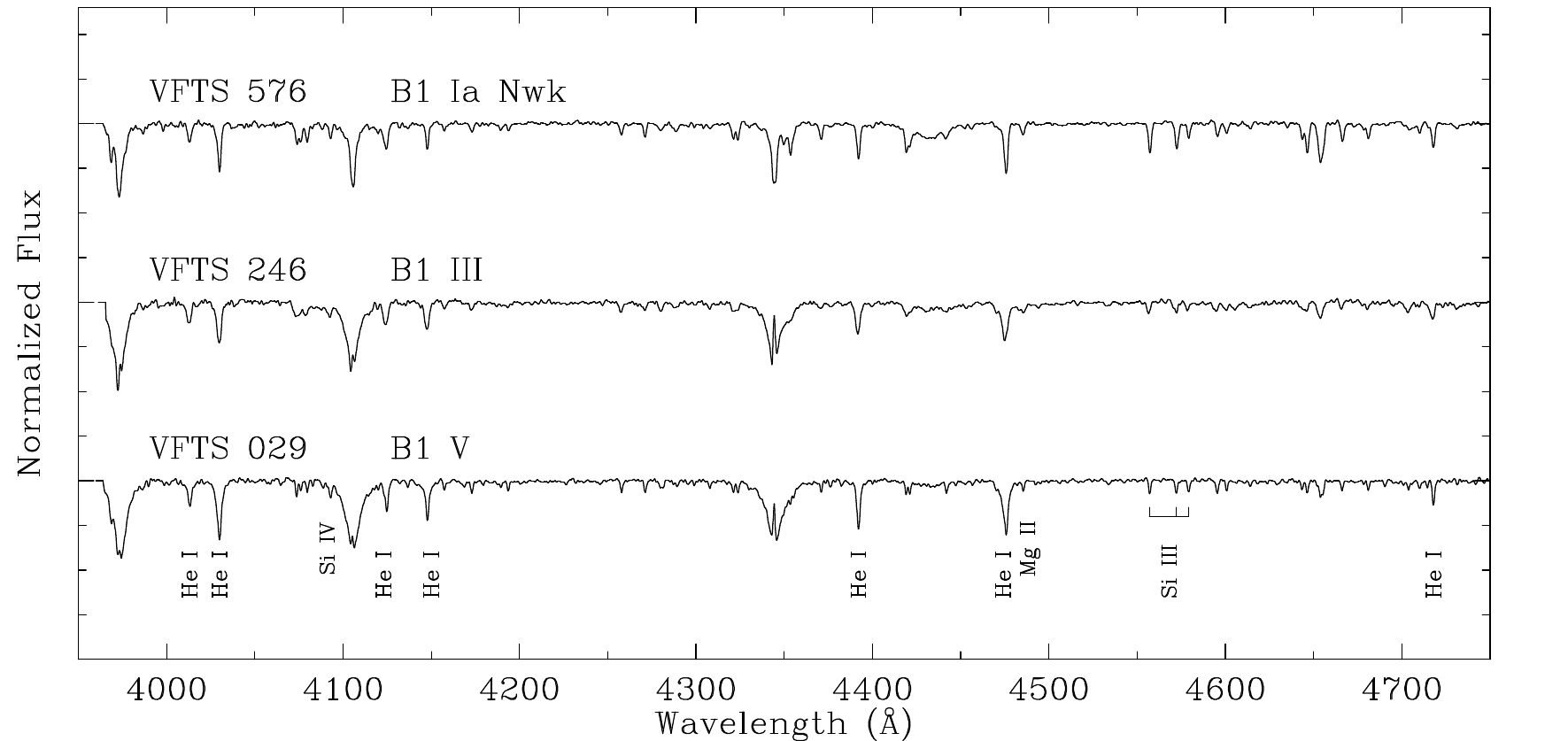}
\caption{\footnotesize Example luminosity sequence at B1. The identified He and Si
lines are the same as those in Figure~\ref{bsgs_fig}.}\label{seq_b1}
\end{center}
\end{figure*}

\begin{figure*}
\begin{center}
\includegraphics{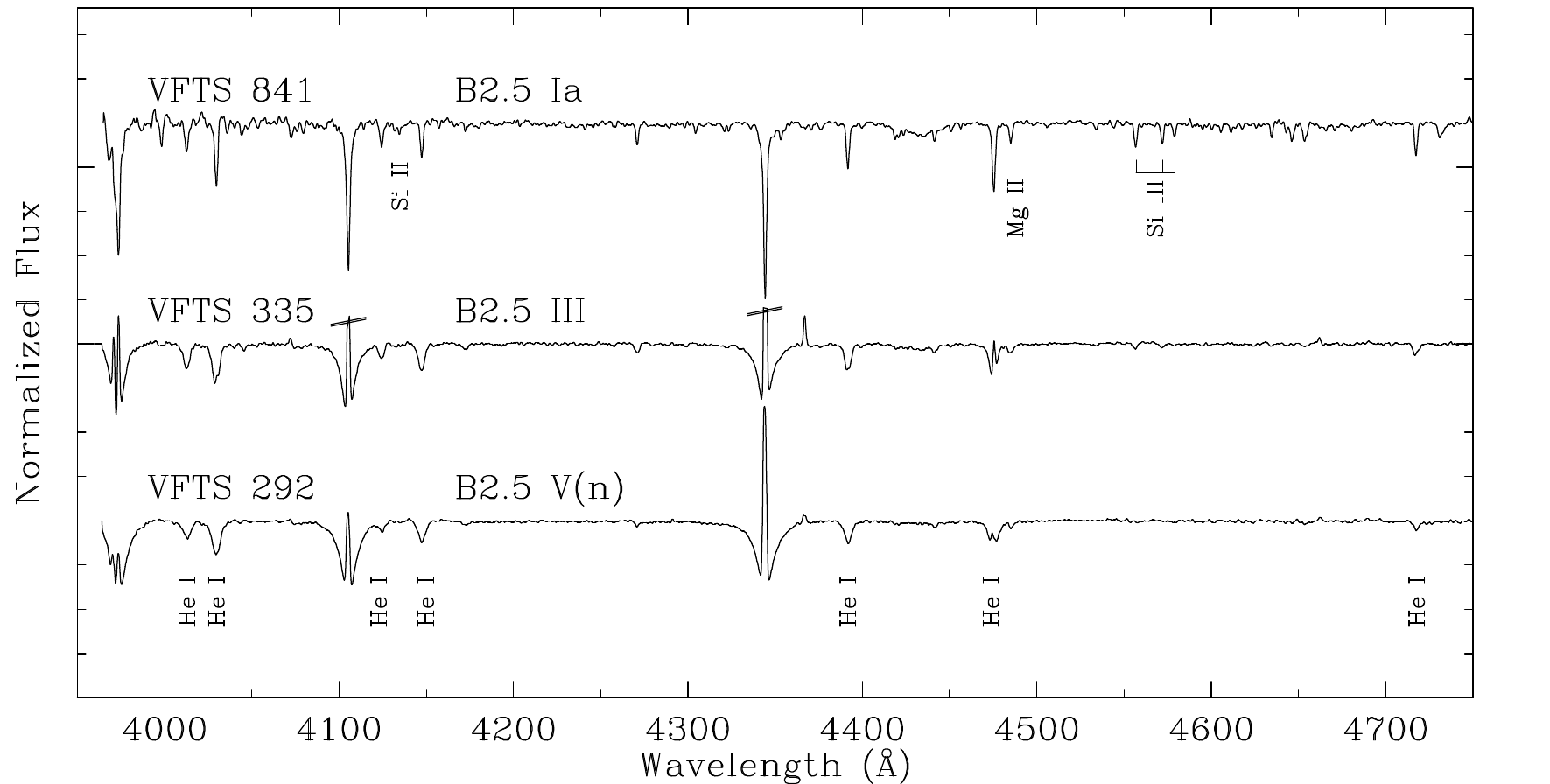}
\caption{\footnotesize Example luminosity sequence at B2.5. The identified He and Si
  lines are the same as those in Figure~\ref{bsgs_fig} and the nebular
  lines have been truncated in VFTS\,335 as indicated.}\label{seq_b25}
\end{center}
\end{figure*}

To illustrate the spectral sequences as a function of luminosity
class, examples of VFTS B-type spectra are shown in
Figures~\ref{bsgs_fig}, \ref{giants_fig}, and \ref{dwarfs_fig}.  The
principal lines used in classification, as summarised in
Tables~\ref{criteria_sg} and \ref{criteria_dwarfs} are indicated in
each figure.  Luminosity sequences at spectral types B0.2, B1 and B2.5
are shown in Figures~\ref{seq_b02}, \ref{seq_b1} and \ref{seq_b25},
respectively.  For clarity, the spectra in all six figures have been
smoothed and rebinned to an effective resolving power of
$R$\,$=$\,4\,000.

\subsection{Detailed velocity estimates}
For completeness, we list the individual velocity estimates from each
line, calculated using the rest wavelengths given in
Table~\ref{vr_lines}.  Tables~\ref{rv_set1} and \ref{rv_set2} present
the results for stars using the Set~1 and Set~2 diagnostic lines,
respectively; results for the binary stars (following the discussion
outlined in Section~\ref{methods_binaries}) are given in
Table~\ref{rv_binaries}.

\subsubsection{Comparison with published velocity estimates}\label{xcheck_sana}

The methods and line sets adopted to estimate radial velocities
($v_{\rm r}$) for the B-type stars differ to those employed by
\citet{s13} for the O-type objects from the survey.  As a consistency
check between the two studies, we analysed 45 of the
apparently-single, late O-type stars (with O9.5 and O9.7 types) and
compared our velocity estimates with those from \citeauthor{s13} We
find a mean and standard deviation between the two methods of $\Delta
v$\,$=$\,0.5\,$\pm$\,6.2\,\kms\ (where $\Delta v$\,$=$\,$v_{\rm
  r}$\,$-$\,$v_{\rm Sana}$).

\begin{figure}
\begin{center}
\includegraphics[scale=0.52]{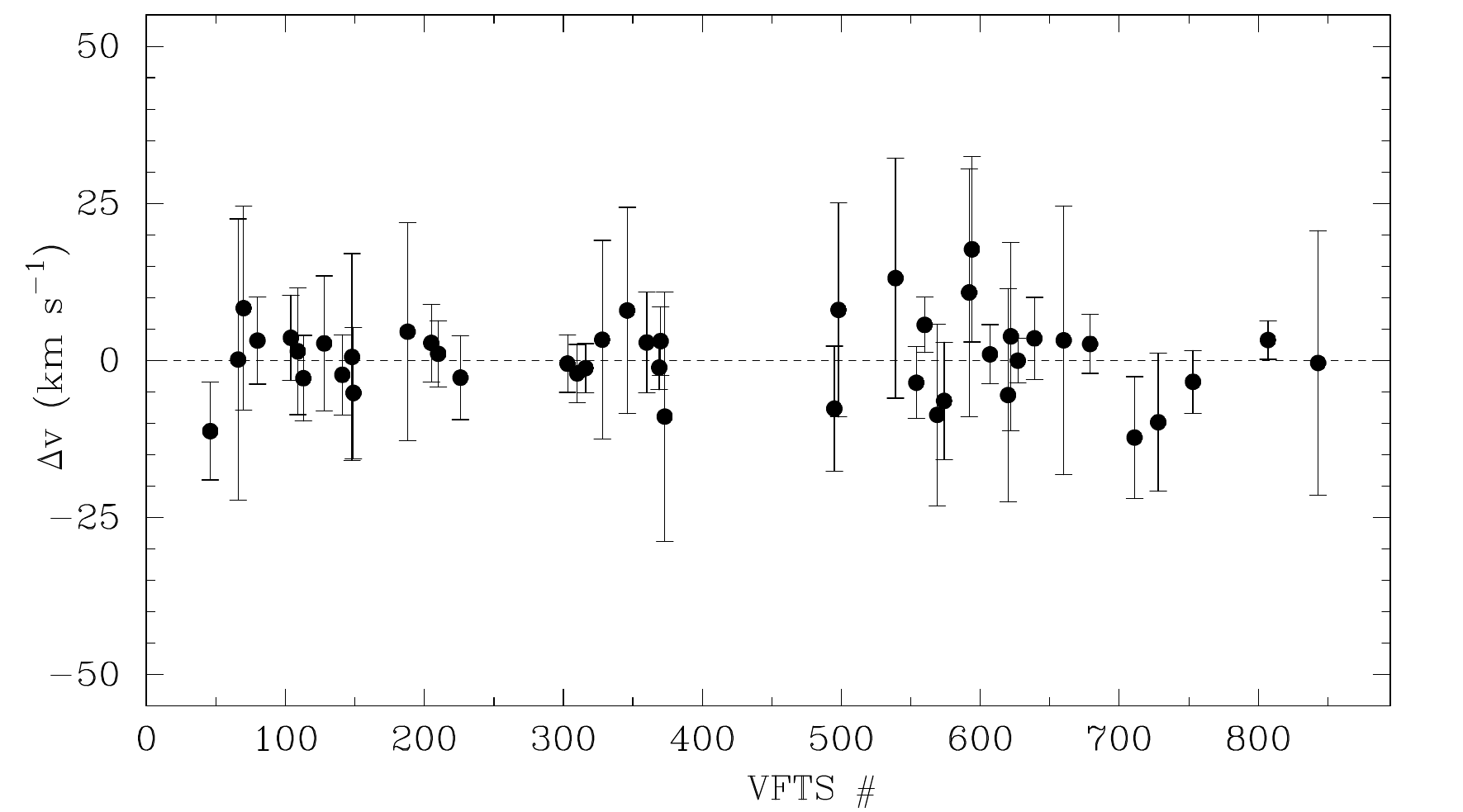}\vspace{0.2cm}
\includegraphics[scale=0.52]{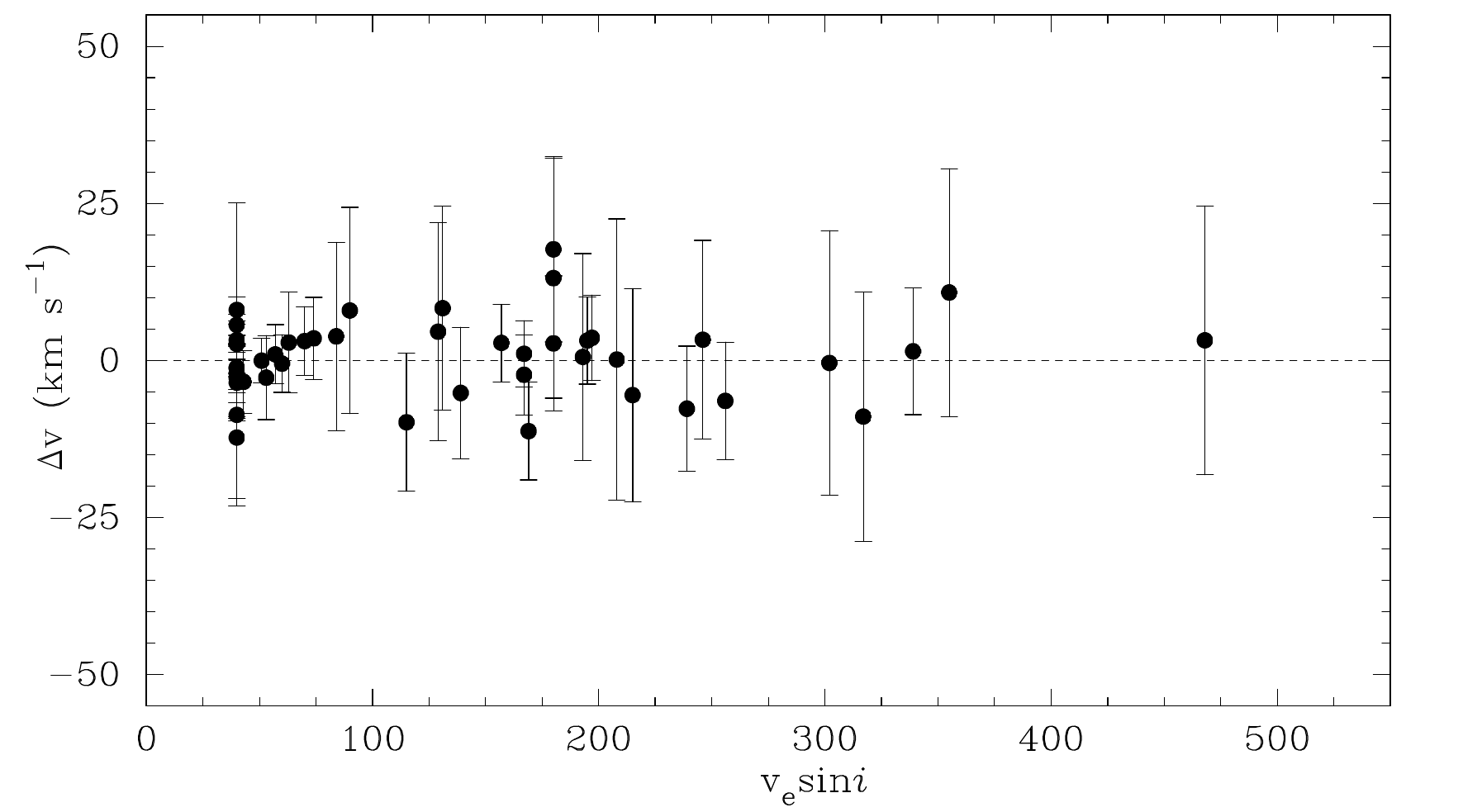}
\caption{\footnotesize Comparison of radial velocities ($v_{\rm r}$) estimated using
  the methods in this paper with results from \citet{s13}, in which
  $\Delta v$\,$=$\,$v_{\rm r}$\,$-$\,$v_{\rm Sana}$, and the plotted
  uncertainties are the standard deviations of both estimates added in
  quadrature. The upper panel shows the $\Delta v$ results in the
  sequence of VFTS~identifiers (i.e. increasing right ascension),
  while the lower panel shows $\Delta v$ as a function of
  \vsini.}\label{rv_overlap}
\end{center}
\end{figure}

The differential results ($\Delta v$) for the 45 stars are shown in
Figure~\ref{rv_overlap} (with the standard deviations added in
quadrature). For the purposes of global analysis of the sample and
identification of potential runaways, these results are in excellent
agreement.  As a further check of any systematic trends arising from
the different methods, the lower panel of Figure~\ref{rv_overlap}
shows the same differential velocities, but now as a function of
estimated \vsini\ \citep[from][]{duf13}; no obvious systematic difference
is present.

\subsection{Centre-of-mass velocities}\label{com_binaries}

We have investigated the expected differences between our mean values
and centre-of-mass velocities using a Monte Carlo technique.  We
assumed a sinusoidal velocity curve (i.e.  circular orbit) and,
adopting the observational sampling of Field~A of the survey
(Table~A.1 of Paper~I), we calculated ranges of observed velocities
(i.e. $\delta v_{\rm r\,max}$) and their resulting mean velocity.  We
assumed a random initial phase, and considered periods ranging from
one day up to several years.

\begin{figure}
\begin{center}
\includegraphics[scale=0.52]{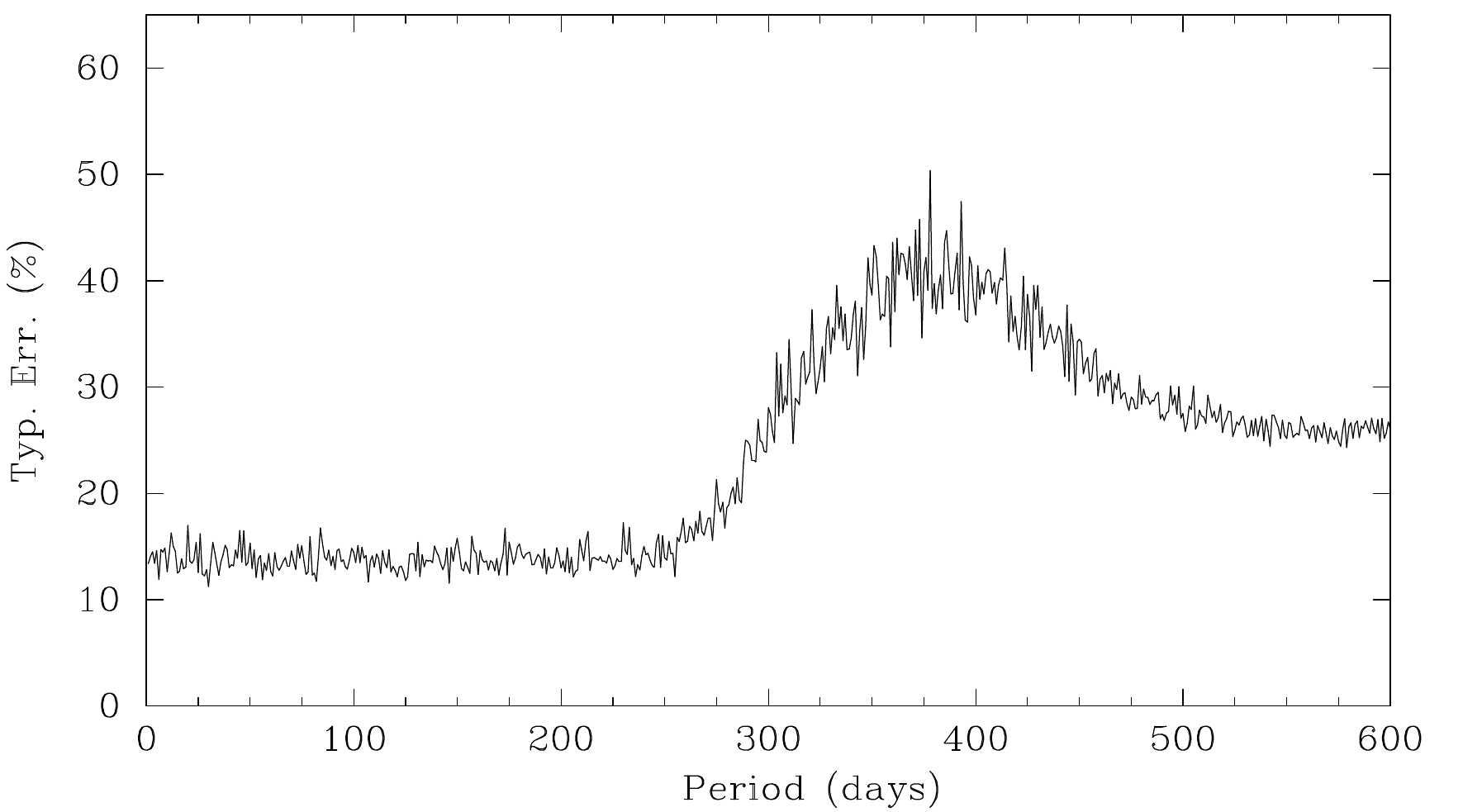}
\caption{\footnotesize Typical errors in the average radial velocities (calculated
  for the observational sampling of Field~A of the VFTS) for simulated
  binaries compared to their true centre-of-mass velocities (as a
  percentage of the expected range in radial velocities, $\delta
  v_{\rm r max}$, see text for details).}\label{com_fig}
\end{center}
\end{figure}

These simulated mean velocities provide us with an estimate of the
expected error compared to the genuine centre-of-mass velocity.  For a
given period, we can then average these differences over the range of
assumed phases to find a typical error.  The ratio of this typical
error to $\delta v_{\rm r\,max}$ is shown in Figure~\ref{com_fig}.
For binaries with periods of $\lesssim$\,250\,d, the expected error is
approximately 12-15\,\% of $\delta v_{\rm r\,max}$, with this increasing
for longer periods, and then tending to a value of $\sim$25\,\% for
periods $\gtrsim$\,350\,d (greater than the maximum time-sampling of
the real data).

These simplistic simulations (e.g. not taking into account potential
eccentric orbits) have probably underestimated the
uncertainties but, given the observational cadence of the survey, they
illustrate that the mean multi-epoch $v_{\rm r}$ values should provide
reasonable estimates of the centre-of-mass velocities for short-period
systems ($\lesssim$\,250\,d), which display relatively small
peak-to-peak variations.

\onecolumn
\scriptsize

\begin{landscape}
\begin{center}
\begin{longtable}{lccccccccccc}
  \caption{\leftline{Line-by-line radial-velocity estimates for single stars 
using the Set~1 absorption lines.\label{rv_set1}}}\\
  \hline
  VFTS & He\,I$_{\lambda4121}$ & C\,II$_{\lambda4267}$
  & He\,I$_{\lambda4438}$ & Si\,III$_{\lambda4553}$ & Si\,III$_{\lambda4553}$ 
  & Si\,III$_{\lambda4568}$ & Si\,III$_{\lambda4575}$ & He\,I$_{\lambda4713}$ & $v_{\rm r}$ & $\sigma$ & n \\
  & & & & (LR02) & (LR03) & & & & & & \\
  \hline
\endfirsthead
  \caption{\leftline{\it{continued}}}\\
\hline
  VFTS & He\,I$_{\lambda4121}$ & C\,II$_{\lambda4267}$
  & He\,I$_{\lambda4438}$ & Si\,III$_{\lambda4553}$ & Si\,III$_{\lambda4553}$ 
  & Si\,III$_{\lambda4568}$ & Si\,III$_{\lambda4575}$ & He\,I$_{\lambda4713}$ & $v_{\rm r}$ & $\sigma$ & n \\
  & & & & (LR02) & (LR03) & & & & & & \\
\hline
\endhead
\hline
\multicolumn{11}{r}{\it{continued on next page}}\\
\endfoot
\hline
\endlastfoot
003 & 268.3 & 271.3 & 269.2 & 265.8 & 263.9 & 264.0 & 264.8 & 262.8 & 265.2 & \o2.1 & 8 \\
010 & 289.9 & \ldots& \ldots& \ldots& 281.2 & 291.5 & \ldots& 283.7 & 287.1 & \o3.9 & 4 \\
024 & \ldots& 282.8 & 284.2 & 279.7 & 291.2 & 281.5 & 285.9 & 289.7 & 285.7 & \o4.4 & 7 \\
028 & 263.3 & 280.0 & 268.0 & 277.7 & 273.9 & 274.2 & 271.2 & 275.4 & 273.5 & \o4.4 & 8 \\
029 & 288.0 & 284.5 & 289.5 & 282.7 & 284.6 & 280.6 & 283.0 & 288.6 & 285.6 & \o3.0 & 8 \\
031 & 275.0 & 292.6 & 292.9 & 271.4 & \ldots& \ldots& \ldots& 275.5 & 280.0 & \o8.4 & 5 \\
034 & \ldots& 296.6 & \ldots& 308.7 & 329.1 & 297.7 & \ldots& \ldots& 308.3 &  13.8 & 4 \\
044 & 292.9 & 285.3 & 287.6 & 284.6 & 271.9 & \ldots& \ldots& 295.5 & 288.1 & \o7.2 & 6 \\
046 & 242.3 & 248.5 & 230.1 & 238.7 & 236.0 & 261.0 & \ldots& 239.8 & 242.2 & \o7.6 & 7 \\
050 & 282.3 & 283.8 & 282.7 & 279.4 & 281.4 & 284.5 & 282.2 & 285.7 & 283.1 & \o2.0 & 8 \\
052 & 276.4 & 278.7 & 280.7 & \ldots& 280.5 & 279.0 & 279.0 & 282.4 & 279.8 & \o2.0 & 7 \\
053 & 275.1 & 277.0 & 279.2 & 274.3 & 278.6 & 274.4 & 274.8 & 278.9 & 276.4 & \o2.0 & 8 \\
060 & 254.9 & 286.1 & 299.4 & 287.5 & 298.6 & 295.2 & 287.9 & 284.4 & 284.2 &  14.3 & 8 \\
062 & 289.8 & 286.1 & 288.2 & 282.5 & 281.6 & 289.1 & 285.5 & 289.5 & 287.5 & \o2.7 & 8 \\
069 & 280.2 & 285.7 & 280.1 & \ldots& 280.1 & 279.7 & 278.5 & 282.6 & 280.7 & \o1.7 & 7 \\
075 & 303.9 & 292.7 & 284.1 & \ldots& \ldots& 278.5 & \ldots& 307.0 & 297.2 &  10.5 & 5 \\
082 & 280.1 & 288.9 & 282.7 & 283.8 & 283.6 & 281.2 & 281.5 & 286.2 & 283.3 & \o2.4 & 8 \\
095 & 293.1 & 296.5 & 296.1 & \ldots& 294.7 & 284.5 & 297.7 & 296.7 & 294.3 & \o3.8 & 7 \\
111 & 268.1 & 276.7 & 273.4 & 269.0 & 265.4 & 273.3 & 270.0 & 277.4 & 272.0 & \o4.2 & 8 \\
113 & 277.7 & 276.1 & 277.0 & \ldots& 277.5 & 281.9 & \ldots& 285.4 & 280.2 & \o3.7 & 6 \\
119 & \ldots& 277.8 & 281.9 & \ldots& 272.8 & 270.3 & 272.0 & 278.0 & 275.8 & \o3.9 & 6 \\
121 & 282.5 & 282.3 & 285.8 & 281.3 & 279.8 & 282.2 & 283.5 & 288.4 & 283.4 & \o2.8 & 8 \\
124 & 289.0 & 290.1 & 289.9 & 285.3 & 287.1 & 287.4 & 287.4 & 292.6 & 289.4 & \o2.2 & 8 \\
126 & 252.8 & 252.0 & 255.3 & 251.0 & 251.2 & 254.0 & 256.8 & 258.1 & 254.1 & \o2.6 & 8 \\
152 & 276.7 & 278.2 & 274.5 & 271.5 & 271.7 & 268.0 & 268.7 & 275.1 & 274.3 & \o2.9 & 8 \\
159 & 265.4 & 274.9 & 265.3 & 264.0 & 265.0 & 253.6 & 283.3 & 267.7 & 267.5 & \o6.1 & 8 \\
161 & \ldots& 268.5 & \ldots& 269.1 & 252.6 & \ldots& \ldots& \ldots& 262.6 & \o7.9 & 3 \\
164 & \ldots& 296.7 & \ldots& 316.2 & 292.7 & \ldots& \ldots& \ldots& 300.4 &  10.0 & 3 \\
170 & 281.6 & 275.6 & 280.5 & 277.3 & 276.6 & 275.1 & 275.8 & 281.8 & 278.7 & \o2.7 & 8 \\
178 & 287.9 & 293.5 & 285.0 & 286.8 & 281.9 & 281.0 & 279.4 & 284.6 & 284.5 & \o3.2 & 8 \\
183 & 257.9 & \ldots& 253.9 & 255.4 & 257.0 & 257.8 & 251.9 & 255.6 & 256.1 & \o1.7 & 7 \\
186 & 256.8 & 280.0 & 274.5 & 265.4 & 254.2 & 267.7 & 271.1 & 268.5 & 264.3 &  10.6 & 8 \\
200 & 248.6 & 247.7 & \ldots& \ldots& \ldots& \ldots& \ldots& 265.6 & 253.6 & \o8.0 & 3 \\
202 & 276.5 & 274.3 & 274.9 & 273.1 & 275.8 & 277.5 & 275.7 & 280.3 & 276.6 & \o2.3 & 8 \\
209 & 275.7 & 273.6 & 275.2 & \ldots& 268.1 & 271.3 & 269.3 & 274.4 & 273.1 & \o2.6 & 7 \\
214 & 276.6 & 272.2 & 275.8 & 274.9 & 274.2 & 273.8 & 273.7 & 277.6 & 275.4 & \o1.7 & 8 \\
220 & \ldots& 276.1 & \ldots& 286.5 & 282.9 & 281.2 & \ldots& \ldots& 282.6 & \o3.3 & 4 \\
226 & \ldots& 197.1 & 182.9 & 196.6 & 185.2 & 186.0 & 178.6 & 196.4 & 191.2 & \o6.5 & 7 \\
232 & 290.3 & 291.0 & 283.8 & 287.5 & 287.2 & 287.1 & 282.2 & 290.1 & 288.6 & \o2.5 & 8 \\
234 & 244.5 & \ldots& 259.4 & 246.6 & \ldots& 285.1 & 295.5 & 255.3 & 260.9 &  17.3 & 7 \\
235 & 266.0 & 269.5 & 269.0 & 262.5 & 264.3 & 263.0 & 262.0 & 267.4 & 265.6 & \o2.4 & 8 \\
237 & 306.5 & 279.3 & 278.6 & 291.1 & 279.9 & 287.8 & 281.4 & 282.6 & 287.9 &  10.5 & 8 \\
241 & 257.0 & 275.3 & 268.4 & 283.7 & 268.2 & 260.8 & 259.0 & 273.3 & 267.6 & \o8.7 & 8 \\
242 & 215.8 & \ldots& 218.4 & 220.6 & 214.5 & 217.0 & 206.0 & 217.2 & 216.0 & \o3.6 & 7 \\
253 & 272.4 & \ldots& 269.1 & 265.3 & 270.7 & \ldots& \ldots& 271.8 & 270.7 & \o2.1 & 5 \\
258 & 285.5 & 288.2 & 299.0 & 294.7 & 277.3 & 281.9 & \ldots& 289.2 & 287.3 & \o5.9 & 7 \\
261 & 284.1 & 286.4 & 280.6 & \ldots& 282.6 & 284.2 & 286.3 & 287.3 & 285.0 & \o2.2 & 7 \\
268 & \ldots& \ldots& \ldots& 282.0 & 280.9 & 249.5 & \ldots& \ldots& 272.1 &  14.5 & 3 \\
269 & 267.4 & 269.2 & 263.8 & 263.4 & 262.5 & 265.5 & 268.6 & 268.1 & 266.9 & \o2.1 & 8 \\
270 & 266.3 & 265.1 & 263.8 & 260.9 & 261.8 & 262.3 & 262.9 & 267.6 & 264.9 & \o2.3 & 8 \\
273 & 256.3 & 255.9 & 253.4 & 254.5 & 250.9 & 250.7 & 238.1 & 258.5 & 253.8 & \o4.9 & 8 \\
276 & 252.7 & 256.2 & 252.1 & 273.2 & 258.3 & 250.7 & 253.1 & 261.6 & 256.8 & \o5.9 & 8 \\
284 & 278.4 & 276.0 & 278.0 & 273.2 & 268.5 & 267.4 & 269.5 & 272.0 & 272.7 & \o3.9 & 8 \\
293 & 259.0 & 260.6 & 261.7 & 257.2 & 261.7 & \ldots& 262.8 & 261.3 & 260.4 & \o1.9 & 7 \\
296 & \ldots& 246.6 & 243.2 & 241.8 & 245.2 & \ldots& \ldots& 260.5 & 250.0 & \o7.8 & 5 \\
297 & 252.6 & 252.8 & 255.7 & 249.9 & 250.9 & 248.0 & 246.9 & 252.2 & 251.7 & \o2.2 & 8 \\
302 & 265.1 & 264.1 & 266.0 & 264.0 & 264.4 & 262.6 & 265.2 & 267.9 & 264.8 & \o1.6 & 8 \\
303 & 269.7 & \ldots& 272.2 & 284.8 & 266.6 & 266.8 & \ldots& 269.7 & 270.5 & \o4.2 & 6 \\
307 & 274.2 & 278.2 & 273.6 & 273.9 & 269.8 & 270.9 & 270.9 & 273.9 & 272.7 & \o2.0 & 8 \\
308 & 264.1 & 266.0 & 264.2 & 261.8 & 258.7 & 261.8 & \ldots& 263.7 & 263.5 & \o1.9 & 7 \\
310 & 272.8 & 264.2 & 268.7 & 269.9 & 265.4 & 268.1 & 269.2 & 275.4 & 270.2 & \o3.7 & 8 \\
313 & 265.5 & 269.8 & 270.5 & 269.4 & 271.5 & 268.7 & 275.2 & 274.4 & 270.7 & \o3.3 & 8 \\
315 & 259.5 & 263.7 & 261.8 & 262.1 & 260.2 & 261.1 & 260.6 & 264.1 & 261.6 & \o1.5 & 8 \\
316 & 269.1 & 265.3 & 275.6 & 267.1 & 267.0 & 263.0 & 270.0 & 271.9 & 269.1 & \o3.4 & 8 \\
331 & 286.5 & 274.1 & 278.6 & 270.0 & 277.9 & 274.7 & 267.5 & 289.5 & 279.4 & \o7.3 & 8 \\
346 & 296.8 & 268.9 & 260.4 & 263.4 & 265.1 & 268.3 & 245.7 & 281.0 & 274.7 &  15.4 & 8 \\
347 & 267.6 & 263.3 & 268.4 & 265.2 & 270.7 & 267.4 & 273.2 & 265.8 & 267.5 & \o2.5 & 8 \\
353 & 296.8 & 290.1 & 289.8 & 289.0 & 290.4 & 290.8 & \ldots& 296.7 & 292.7 & \o3.3 & 7 \\
360 & 260.0 & \ldots& \ldots& \ldots& 269.1 & 266.2 & 258.1 & \ldots& 264.1 & \o4.4 & 4 \\
363 & 258.5 & 262.8 & 260.3 & 264.7 & 264.3 & 265.9 & 263.8 & 267.8 & 264.1 & \o2.9 & 8 \\
369 & 263.0 & 260.0 & 263.5 & 267.1 & 258.8 & 259.7 & 259.0 & 265.5 & 262.7 & \o2.9 & 8 \\
370 & 237.2 & 241.5 & 227.5 & 231.8 & 239.5 & 229.6 & 223.6 & 235.6 & 234.6 & \o4.6 & 8 \\
376 & \ldots& \ldots& \ldots& 283.0 & 266.6 & 267.4 & \ldots& \ldots& 271.6 & \o7.2 & 3 \\
381 & 246.9 & 249.6 & 241.8 & 246.2 & 224.0 & 248.1 & \ldots& 235.4 & 242.1 & \o8.0 & 7 \\
384 & 275.3 & 264.8 & 256.2 & 264.4 & 264.4 & 257.4 & 259.2 & 269.1 & 264.8 & \o5.5 & 8 \\
413 & 277.2 & 286.1 & 283.7 & \ldots& \ldots& \ldots& \ldots& 287.7 & 283.5 & \o4.4 & 4 \\
417 & 271.1 & 258.0 & 254.5 & 252.3 & 251.1 & 255.0 & 249.1 & 285.5 & 260.7 &  12.5 & 8 \\
420 & \ldots& 280.1 & 274.5 & 275.7 & 271.5 & 271.6 & 265.8 & 274.6 & 273.1 & \o3.4 & 7 \\
423 & \ldots& 275.9 & 271.8 & 274.1 & 282.0 & 281.0 & 279.8 & 285.5 & 278.3 & \o4.3 & 7 \\
424 & 265.1 & 264.8 & 259.3 & 256.9 & 260.0 & 265.4 & \ldots& 264.0 & 263.2 & \o2.6 & 7 \\
426 & \ldots& \ldots& \ldots& 261.8 & 279.2 & 274.3 & 263.4 & \ldots& 270.7 & \o7.6 & 4 \\
431 & 265.2 & 271.3 & 266.1 & 269.3 & 272.2 & 271.7 & 271.3 & 271.0 & 270.2 & \o2.4 & 8 \\
452 & 268.2 & 263.8 & 268.9 & 263.8 & 271.3 & \ldots& \ldots& 259.3 & 265.4 & \o4.0 & 6 \\
458 & 276.3 & 277.3 & 274.0 & 273.1 & 274.8 & 275.6 & 274.6 & 279.4 & 276.3 & \o2.1 & 8 \\
469 & 277.8 & 279.1 & 277.3 & 273.7 & 274.5 & 271.4 & 270.6 & 277.0 & 275.6 & \o2.6 & 8 \\
471 & 254.2 & 271.3 & 261.9 & 263.0 & 268.1 & 266.0 & 294.4 & 274.4 & 267.0 &  11.2 & 8 \\
478 & 273.4 & 268.8 & 267.6 & 270.8 & 269.1 & 270.6 & 275.5 & 278.2 & 272.2 & \o3.3 & 8 \\
480 & \ldots& 284.7 & \ldots& 279.3 & 279.3 & 281.8 & \ldots& \ldots& 278.3 & \o4.6 & 4 \\
498 & 278.7 & \ldots& \ldots& \ldots& 264.0 & 271.5 & 279.7 & \ldots& 273.9 & \o6.3 & 4 \\
504 & 240.6 & \ldots& \ldots& 273.9 & \ldots& 246.8 & 276.3 & \ldots& 251.4 &  14.2 & 4 \\
516 & 262.6 & 241.1 & 264.6 & 273.7 & \ldots& \ldots& \ldots& 247.7 & 256.8 &  10.9 & 5 \\
533 & 257.2 & 260.7 & 256.0 & 259.8 & 252.1 & 253.4 & 252.8 & 246.0 & 254.7 & \o4.4 & 8 \\
535 & \ldots& \ldots& \ldots& \ldots& 281.4 & 266.0 & 265.1 & \ldots& 271.1 & \o7.5 & 3 \\
540 & 246.3 & 252.0 & 240.6 & 246.9 & 245.6 & 251.0 & \ldots& 246.3 & 247.0 & \o2.7 & 7 \\
541 & \ldots& 274.7 & 271.6 & 273.3 & 265.6 & 266.8 & 268.5 & 271.0 & 269.6 & \o3.1 & 7 \\
553 & 278.8 & 274.9 & 277.6 & 272.1 & 279.2 & 284.5 & 269.7 & 270.8 & 276.3 & \o4.6 & 8 \\
560 & 263.5 & 264.3 & 266.6 & 263.7 & 269.5 & 267.9 & \ldots& 264.8 & 265.3 & \o1.9 & 7 \\
569 & 229.8 & \ldots& 228.3 & \ldots& 254.6 & 259.1 & \ldots& \ldots& 247.1 &  13.1 & 4 \\
572 & 255.7 & 246.8 & 249.3 & 251.0 & 246.6 & 245.1 & 241.8 & 250.3 & 249.4 & \o4.1 & 8 \\
578 & 264.5 & 272.5 & 265.9 & 273.2 & 277.7 & 276.4 & 277.1 & 280.6 & 274.5 & \o4.9 & 8 \\
590 & \ldots& 266.5 & 262.0 & 263.3 & 248.8 & 249.0 & 249.0 & 253.7 & 254.1 & \o6.2 & 7 \\
593 & 292.6 & 291.3 & 292.9 & 281.3 & 289.6 & 286.5 & \ldots& 294.9 & 291.5 & \o3.7 & 7 \\
594 & \ldots& \ldots& \ldots& 269.9 & 269.7 & 264.0 & \ldots& \ldots& 268.0 & \o2.7 & 3 \\
598 & \ldots& \ldots& \ldots& 263.8 & 256.9 & 258.3 & \ldots& \ldots& 259.4 & \o3.0 & 3 \\
607 & 256.4 & 251.8 & 267.7 & 259.1 & 254.9 & 261.0 & 252.6 & 262.1 & 258.8 & \o4.5 & 8 \\
616 & 231.3 & 237.6 & 237.4 & 243.4 & 213.1 & 215.2 & 220.5 & 225.6 & 227.4 &  10.7 & 8 \\
622 & 298.6 & 277.7 & 282.2 & 267.6 & 264.6 & 258.6 & 255.9 & 273.9 & 275.8 &  14.5 & 8 \\
623 & 266.7 & 268.3 & \ldots& 267.8 & 266.4 & 267.8 & 270.1 & 271.5 & 268.5 & \o1.9 & 7 \\
625 & 306.7 & 316.9 & \ldots& 302.8 & 299.5 & \ldots& \ldots& 324.5 & 310.7 & \o9.2 & 5 \\
627 & 270.0 & 265.3 & 270.5 & \ldots& 267.3 & 269.5 & 258.8 & 270.8 & 268.8 & \o2.9 & 7 \\
630 & 292.9 & 288.1 & 289.3 & 273.4 & 273.5 & 278.0 & 279.8 & 301.8 & 285.6 &  10.7 & 8 \\
639 & 277.4 & 270.2 & 269.6 & 267.1 & 264.0 & 274.4 & 253.7 & 268.4 & 269.5 & \o5.8 & 8 \\
650 & 288.0 & 287.7 & 285.7 & 283.3 & 282.8 & 282.7 & 283.0 & 290.2 & 286.2 & \o3.0 & 8 \\
666 & 255.5 & 256.8 & 258.0 & 255.4 & 252.4 & 253.4 & 252.5 & 257.8 & 255.5 & \o2.1 & 8 \\
668 & 271.6 & 266.2 & 267.2 & 269.0 & 266.9 & 271.8 & 266.3 & 271.3 & 269.2 & \o2.3 & 8 \\
672 & \ldots& 278.0 & 279.0 & 274.1 & 280.2 & 280.5 & 279.2 & 284.4 & 279.5 & \o3.2 & 7 \\
673 & 278.8 & 275.7 & 275.2 & 277.1 & 274.6 & 277.5 & 277.9 & 278.7 & 277.1 & \o1.5 & 8 \\
681 & 243.8 & 280.1 & 275.4 & 280.1 & 266.7 & \ldots& \ldots& 268.2 & 264.0 &  14.5 & 6 \\
690 & 252.5 & 283.0 & 270.3 & 269.4 & 283.9 & 300.8 & 268.2 & 280.3 & 272.7 &  14.7 & 8 \\
692 & 269.1 & 267.5 & 274.6 & 272.3 & 271.1 & 271.4 & 277.3 & 269.6 & 271.2 & \o2.5 & 8 \\
696 & 261.0 & 265.9 & 263.7 & 265.8 & 265.3 & 266.5 & 267.2 & 269.1 & 265.8 & \o2.3 & 8 \\
707 & 277.3 & 276.9 & 276.1 & 276.0 & 271.5 & 275.5 & 272.1 & 274.8 & 275.0 & \o2.0 & 8 \\
711 & 235.8 & 253.7 & 258.0 & 254.5 & 250.0 & 234.0 & 235.2 & 245.8 & 244.4 & \o8.7 & 8 \\
712 & 249.2 & 240.2 & \ldots& 274.0 & 276.9 & 246.8 & 259.3 & 252.5 & 255.1 &  10.9 & 7 \\
714 & 279.4 & 285.1 & 283.0 & 284.9 & 283.3 & 284.2 & 281.6 & 285.6 & 283.4 & \o1.9 & 8 \\
725 & 286.1 & 285.9 & 286.3 & 285.9 & 283.3 & 282.3 & 284.5 & 287.8 & 285.3 & \o1.8 & 8 \\
727 & 274.7 & 275.7 & 266.5 & 274.4 & \ldots& \ldots& \ldots& 271.4 & 272.9 & \o3.0 & 5 \\
728 & 256.4 & 267.7 & 252.6 & 266.3 & 260.0 & 284.5 & 260.6 & 275.6 & 265.8 & \o9.6 & 8 \\
729 & 254.4 & 261.1 & 245.9 & 268.2 & 281.4 & 272.4 & 279.8 & 277.3 & 268.6 &  10.9 & 8 \\
732 & 264.1 & 268.0 & 267.5 & 269.0 & 267.1 & 267.9 & 267.7 & 266.7 & 267.3 & \o1.4 & 8 \\
735 & 267.0 & 275.7 & 271.2 & 261.9 & \ldots& \ldots& \ldots& \ldots& 268.9 & \o4.9 & 4 \\
740 & 265.6 & 271.3 & 270.0 & 266.2 & 265.8 & 266.3 & 266.7 & 269.6 & 267.3 & \o1.9 & 8 \\
745 & 279.4 & 274.4 & 271.6 & 283.2 & 271.5 & 271.6 & 270.1 & 277.5 & 275.7 & \o3.7 & 8 \\
748 & 258.3 & 259.6 & 261.0 & 264.4 & 258.2 & 265.5 & 267.5 & 256.4 & 260.4 & \o3.6 & 8 \\
753 & 267.9 & 278.1 & 264.5 & 269.2 & 265.8 & 271.3 & 261.5 & 273.7 & 269.6 & \o4.2 & 8 \\
756 & 261.6 & 273.1 & 294.0 & 289.8 & 288.8 & 299.6 & 258.4 & 286.2 & 281.6 &  13.5 & 8 \\
762 & 276.1 & 253.0 & 260.3 & 259.1 & \ldots& \ldots& \ldots& 242.8 & 259.0 &  12.6 & 5 \\
772 & 244.5 & 254.1 & \ldots& \ldots& \ldots& \ldots& \ldots& 259.5 & 253.7 & \o6.1 & 3 \\
795 & 266.5 & \ldots& 270.7 & 257.2 & 284.4 & 262.2 & \ldots& 276.9 & 270.1 & \o9.0 & 6 \\
801 & 283.8 & 284.9 & 292.6 & 272.1 & \ldots& \ldots& \ldots& 265.6 & 278.0 & \o9.7 & 5 \\
807 & 292.4 & \ldots& 287.3 & 288.9 & 290.2 & 290.9 & \ldots& 292.2 & 291.0 & \o1.7 & 6 \\
811 & 278.2 & 276.2 & 276.3 & 273.0 & 255.8 & 301.8 & \ldots& 280.3 & 277.4 &  10.7 & 7 \\
814 & 262.2 & 278.6 & 274.5 & 289.2 & \ldots& \ldots& \ldots& 267.2 & 271.6 & \o9.1 & 5 \\
817 & 234.7 & 269.9 & 245.0 & 261.1 & 249.6 & 255.1 & 264.6 & \ldots& 251.9 &  11.8 & 7 \\
823 & 235.0 & 276.0 & \ldots& \ldots& \ldots& \ldots& \ldots& 248.5 & 248.0 &  13.1 & 4 \\
829 & \ldots& 267.4 & 261.4 & 259.9 & 256.5 & 266.5 & 249.8 & 263.1 & 261.4 & \o4.9 & 7 \\
831 & 223.6 & 211.0 & 203.4 & 206.8 & 202.6 & 206.7 & 215.6 & 210.5 & 210.4 & \o6.2 & 8 \\
835 & 257.0 & 252.5 & 254.3 & 250.2 & 249.7 & 255.2 & 251.7 & 246.8 & 251.9 & \o3.4 & 8 \\
841 & 260.2 & 258.6 & 253.7 & 255.9 & 253.9 & 257.7 & 263.8 & 260.3 & 258.0 & \o3.0 & 8 \\
845 & 259.5 & 258.9 & 259.6 & 259.6 & 257.9 & 259.9 & 260.9 & 262.6 & 259.9 & \o1.4 & 8 \\
851 & 229.5 & 228.7 & 229.4 & \ldots& \ldots& 229.8 & \ldots& 229.5 & 229.4 & \o0.3 & 5 \\
855 & 246.5 & 251.4 & 249.6 & 246.4 & 243.1 & 246.3 & 248.8 & 252.4 & 248.8 & \o3.0 & 8 \\
860 & 255.6 & 252.0 & 249.4 & 256.6 & 256.6 & 250.9 & 243.8 & 250.7 & 252.6 & \o3.0 & 8 \\
864 & 278.2 & 275.3 & 271.0 & 271.8 & 279.0 & 281.0 & \ldots& 285.0 & 278.2 & \o4.6 & 7 \\
866 & 258.6 & 268.5 & 260.9 & 255.5 & 255.5 & 267.3 & 277.5 & 261.6 & 261.8 & \o5.7 & 8 \\
867 & 271.4 & 275.6 & 274.4 & 275.3 & 275.1 & 275.0 & 278.8 & 280.5 & 275.7 & \o2.7 & 8 \\
868 & 264.6 & 258.2 & 261.6 & 260.1 & 259.8 & 264.3 & 257.7 & 262.5 & 261.5 & \o2.3 & 8 \\
872 & 276.3 & 280.1 & 289.9 & 287.6 & 281.9 & 291.4 & \ldots& 293.7 & 285.5 & \o6.9 & 7 \\
879 & 281.2 & 277.2 & 277.9 & \ldots& 264.8 & \ldots& \ldots& 273.6 & 275.7 & \o4.9 & 5 \\
881 & 283.3 & 282.9 & 282.6 & 281.7 & 281.8 & 282.5 & 281.7 & 285.9 & 283.0 & \o1.5 & 8 \\
885 & 270.1 & 266.3 & 250.3 & 265.3 & \ldots& 262.9 & 260.5 & 263.5 & 264.0 & \o5.4 & 7 \\
886 & 253.9 & 264.5 & 255.5 & 260.0 & 263.9 & 248.3 & 253.0 & 263.5 & 258.4 & \o5.5 & 8 \\
\hline
\end{longtable}
\tablefoot{Column entries are: (1) VFTS identifier; (2-10) estimated
  radial velocities from each line; (11) weighted mean
  velocity for each star ($v_{\rm r}$, from eqn.~\ref{mean_vr}); (12)
  standard deviation ($\sigma$, from eqn.~\ref{sigma_vr}); (13) number
  of lines ($n$) used in calculation of $v_{\rm r}$ and $\sigma$.}
\end{center}
\end{landscape}

\begin{landscape}
\begin{center}
\begin{longtable}{lcccccccc}
  \caption{\leftline{Line-by-line radial-velocity estimates for single stars 
using the Set~2 absorption lines.\label{rv_set2}}}\\
  \hline
  VFTS & He\,I$_{\lambda4009}$ & He\,I$_{\lambda4144}$ & He\,I$_{\lambda4388}$ 
  & He\,I$_{\lambda4713}$ & He\,I$_{\lambda4922}$ &  $v_{\rm r}$ & $\sigma$ & n \\
  \hline
\endfirsthead
  \caption{\leftline{\it{continued}}}\\
\hline
  VFTS & He\,I$_{\lambda4009}$ & He\,I$_{\lambda4144}$ & He\,I$_{\lambda4388}$ 
  & He\,I$_{\lambda4713}$ & He\,I$_{\lambda4922}$ &  $v_{\rm r}$ & $\sigma$ & n \\
\hline
\endhead
\hline
\multicolumn{8}{r}{\it{continued on next page}}\\
\endfoot
\hline
\endlastfoot
001 & 287.9 & 279.4 & 283.3 & \ldots& 276.5 & 282.1 & \o4.2 & 4 \\
004 & 299.2 & 293.7 & 280.1 & 281.2 & 269.0 & 283.1 &  10.8 & 5 \\
005 & 281.7 & 284.0 & 285.6 & 282.0 & 278.4 & 282.6 & \o2.6 & 5 \\
007 & 282.1 & 282.6 & 279.2 & \ldots& \ldots& 281.1 & \o1.6 & 3 \\
008 & 282.6 & 266.1 & 276.2 & \ldots& 264.5 & 271.3 & \o6.9 & 4 \\
012 & \ldots& 306.2 & 310.8 & 295.5 & 307.2 & 306.1 & \o4.9 & 4 \\
013 & 290.5 & 295.5 & 284.5 & 297.1 & 291.3 & 290.9 & \o4.3 & 5 \\
020 & 254.5 & 271.9 & 272.9 & \ldots& 235.3 & 258.8 &  16.3 & 4 \\
036 & 244.0 & 273.4 & 289.4 & \ldots& \ldots& 270.8 &  18.4 & 3 \\
038 & 274.3 & 278.4 & 272.5 & 279.6 & 283.0 & 278.0 & \o4.1 & 5 \\
040 & 303.5 & 273.7 & 297.8 & \ldots& \ldots& 291.4 &  12.6 & 3 \\
043 & 271.4 & 260.1 & 259.7 & \ldots& 264.9 & 263.6 & \o4.4 & 4  \\
048 & 274.3 & 278.9 & 273.8 & \ldots& 269.6 & 274.1 & \o3.3 & 4 \\
054 & 246.8 & 283.3 & 289.0 & 266.5 & 242.6 & 265.7 &  19.9 & 5 \\
066 & 289.9 & 274.1 & 282.7 & 242.3 & 254.1 & 272.9 &  13.6 & 5 \\
068 & 297.9 & 273.7 & 243.2 & \ldots& 232.5 & 254.8 &  24.0 & 4 \\
070 & 302.7 & 286.2 & 280.6 & 291.9 & 312.1 & 294.7 &  12.5 & 5 \\
071 & 265.6 & 274.1 & 270.9 & \ldots& 263.5 & 268.4 & \o4.3 & 4 \\
078 & 278.7 & 285.7 & 279.0 & 289.0 & 285.2 & 283.1 & \o3.7 & 5 \\
080 & 291.7 & 289.7 & 279.2 & 285.4 & 279.7 & 284.1 & \o5.1 & 5 \\
083 & 290.2 & 290.5 & 282.8 & \ldots& \ldots& 287.3 & \o3.7 & 3 \\
084 & 284.0 & 276.3 & 273.5 & 277.4 & 273.2 & 276.2 & \o3.8 & 5 \\
085 & 284.5 & 289.9 & 281.0 & 287.3 & \ldots& 285.3 & \o3.6 & 4 \\
088 & 284.2 & 283.1 & 293.4 & \ldots& 272.7 & 284.0 & \o7.4 & 4 \\
100 & 255.2 & 258.1 & 258.5 & \ldots& \ldots& 257.6 & \o1.4 & 3 \\
101 & 279.5 & 280.1 & 269.5 & \ldots& 281.2 & 277.3 & \o4.8 & 4 \\
104 & 267.5 & 277.1 & 268.1 & 280.3 & 268.5 & 271.2 & \o4.7 & 5 \\
109 & 283.9 & 293.3 & 283.5 & \ldots& 278.2 & 284.1 & \o5.6 & 4 \\
122 & \ldots& 259.8 & 281.2 & \ldots& 238.2 & 260.7 &  17.6 & 3 \\
127 & 235.5 & 249.4 & 242.7 & \ldots& \ldots& 243.3 & \o5.5 & 3 \\
128 & 294.5 & 279.5 & 272.9 & \ldots& 268.7 & 277.4 & \o8.7 & 4 \\
134 & 253.7 & 256.2 & 274.1 & \ldots& \ldots& 262.8 & \o9.4 & 3 \\
137 & 288.9 & 279.7 & 284.3 & \ldots& 284.2 & 284.0 & \o3.1 & 4 \\
141 & 271.3 & 284.4 & 285.2 & 287.3 & 278.2 & 281.4 & \o5.2 & 5 \\
148 & 258.0 & 268.9 & 257.1 & 289.0 & 263.8 & 264.5 & \o8.8 & 5 \\
149 & 259.7 & 287.0 & 282.1 & 265.1 & 270.3 & 274.6 & \o9.6 & 5 \\
156 & 272.0 & 269.0 & 308.9 & \ldots& \ldots& 283.8 &  18.3 & 3 \\
158 & 264.7 & 276.4 & 310.1 & \ldots& \ldots& 285.7 &  19.3 & 3 \\
166 & 275.3 & 267.0 & 280.4 & \ldots& \ldots& 274.4 & \o5.6 & 3 \\
167 & 275.4 & 290.6 & 285.1 & 290.2 & 289.8 & 286.9 & \o5.0 & 5 \\
181 & 267.4 & 271.9 & 273.8 & \ldots& 297.4 & 278.9 &  11.9 & 4 \\
187 & 284.1 & 294.1 & 270.7 & \ldots& 279.1 & 280.8 & \o8.5 & 4 \\
188 & 312.8 & 284.3 & 280.9 & 269.1 & 268.6 & 281.7 &  14.8 & 5 \\
194 & 270.3 & 275.7 & 294.8 & 288.8 & 259.2 & 277.9 &  13.0 & 5 \\
196 & 264.5 & 267.1 & 278.4 & 283.0 & \ldots& 271.9 & \o7.1 & 4 \\
201 & 290.2 & 293.4 & 276.8 & \ldots& \ldots& 284.9 & \o7.8 & 3 \\
203 & 278.5 & 272.8 & 275.7 & 251.7 & 272.7 & 272.8 & \o6.8 & 5 \\
205 & 286.0 & 278.1 & 279.6 & 280.4 & 273.9 & 278.7 & \o3.7 & 5 \\
207 & 285.8 & 287.4 & 283.2 & 265.9 & 265.0 & 277.9 & \o9.7 & 5 \\
210 & 286.3 & 296.1 & 289.4 & 282.6 & 292.6 & 290.5 & \o4.3 & 5 \\
212 & 284.6 & 288.0 & 284.1 & 292.4 & 285.5 & 286.2 & \o2.5 & 5 \\
219 & 278.4 & 298.5 & 270.2 & \ldots& \ldots& 282.5 &  12.6 & 3 \\
221 & 249.6 & 266.9 & 263.3 & 276.7 & 254.6 & 261.4 & \o8.2 & 5 \\
224 & 235.1 & 258.4 & 299.2 & \ldots& \ldots& 263.9 &  25.8 & 3 \\
228 & 273.9 & 285.7 & 279.2 & 297.2 & 277.9 & 281.2 & \o6.5 & 5 \\
229 & 256.5 & 283.6 & 276.4 & \ldots& \ldots& 272.8 &  11.0 & 3 \\
230 & 282.1 & 286.4 & 271.3 & 267.4 & 285.7 & 279.7 & \o7.5 & 5 \\
233 & 257.3 & 251.1 & 256.3 & \ldots& 227.9 & 248.2 &  11.9 & 4 \\
234 & 244.4 & 256.9 & 247.5 & 262.3 & 229.9 & 247.0 &  10.3 & 5 \\
238 & 254.8 & 292.5 & 269.4 & 261.7 & 258.9 & 268.5 &  13.5 & 5 \\
239 & 267.2 & 279.2 & 282.1 & \ldots& 268.4 & 274.7 & \o6.5 & 4 \\
247 & 293.9 & 289.5 & 291.6 & 304.5 & 296.5 & 293.6 & \o4.1 & 5 \\
254 & 251.6 & 256.3 & 252.2 & \ldots& 240.8 & 250.7 & \o5.6 & 4 \\
263 & 250.0 & 267.8 & 291.8 & \ldots& \ldots& 270.1 &  16.8 & 3 \\
274 & 278.8 & 256.3 & 251.3 & \ldots& 235.6 & 252.9 &  14.1 & 4 \\
279 & 242.1 & 266.4 & 278.5 & 256.7 & \ldots& 262.6 &  14.1 & 4 \\
282 & 241.5 & 253.5 & 242.3 & 265.8 & 289.3 & 260.5 &  19.7 & 5 \\
283 & 260.0 & 248.2 & 257.3 & 241.7 & 248.6 & 252.1 & \o6.0 & 5 \\
286 & 277.0 & 279.6 & 272.9 & \ldots& 284.5 & 278.4 & \o4.4 & 4 \\
287 & 282.5 & 254.0 & 262.4 & \ldots& \ldots& 264.7 &  11.4 & 3 \\
288 & 264.2 & 246.4 & 261.0 & 269.5 & 257.6 & 258.0 & \o7.2 & 5 \\
292 & 256.9 & 261.9 & 264.5 & 275.3 & 264.7 & 263.6 & \o4.9 & 5 \\
295 & \ldots& 295.8 & 300.2 & \ldots& 322.7 & 307.7 &  12.0 & 3 \\
298 & 334.0 & 327.3 & 336.2 & \ldots& \ldots& 332.1 & \o4.0 & 3 \\
300 & 280.6 & 259.7 & 285.5 & 294.4 & 268.0 & 276.7 &  12.1 & 5 \\
301 & 267.7 & 258.9 & 280.9 & 262.6 & \ldots& 268.4 & \o9.0 & 4 \\
309 & 288.2 & 281.0 & 282.4 & 294.1 & 269.2 & 281.8 & \o7.6 & 5 \\
320 & 290.0 & 265.7 & 318.1 & \ldots& \ldots& 291.3 &  22.1 & 3 \\
321 & 263.4 & 251.2 & 262.3 & \ldots& \ldots& 258.6 & \o5.6 & 3 \\
322 & 284.0 & 262.5 & 277.9 & 297.7 & \ldots& 276.9 &  11.1 & 4 \\
326 & 236.0 & 244.0 & 256.1 & \ldots& 258.9 & 250.1 & \o8.8 & 4 \\
328 & 297.8 & 320.2 & 307.5 & 328.2 & 316.5 & 313.7 & \o8.8 & 5 \\
330 & 261.1 & 269.3 & 272.9 & 272.3 & 264.7 & 267.7 & \o4.4 & 5 \\
335 & 250.2 & 247.2 & 251.8 & 245.2 & 247.1 & 248.7 & \o2.2 & 5 \\
340 & 254.2 & 265.0 & 258.9 & \ldots& 221.4 & 251.7 &  16.2 & 4 \\
343 & 238.1 & 254.0 & 241.0 & \ldots& 227.7 & 240.5 & \o9.3 & 4 \\
348 & 261.8 & 267.5 & 272.0 & \ldots& \ldots& 267.8 & \o4.0 & 3 \\
349 & 258.2 & 280.0 & 272.3 & 259.2 & 268.4 & 269.4 & \o7.7 & 5 \\
354 & 250.1 & 265.1 & 259.5 & \ldots& 234.8 & 250.7 &  12.3 & 4 \\
358 & 148.3 & 165.1 & 175.4 & \ldots& \ldots& 164.7 &  10.7 & 3 \\
365 & 273.1 & 266.0 & 262.9 & 301.9 & 265.4 & 270.0 &  10.9 & 5 \\
367 & 275.8 & 271.7 & 272.7 & 268.6 & 259.1 & 269.4 & \o6.0 & 5 \\
368 & 249.2 & 215.7 & 227.1 & \ldots& \ldots& 229.6 &  12.4 & 3 \\
373 & 262.5 & 274.2 & 275.1 & \ldots& \ldots& 271.4 & \o5.4 & 3 \\
387 & 246.6 & 294.3 & 308.0 & \ldots& \ldots& 286.3 &  24.2 & 3 \\
394 & 247.1 & 245.9 & 228.1 & \ldots& \ldots& 239.4 & \o8.9 & 3 \\
397 & 279.4 & 259.1 & 290.2 & \ldots& \ldots& 276.7 &  13.2 & 3 \\
401 & 262.1 & 246.4 & 301.2 & \ldots& \ldots& 269.7 &  23.3 & 3 \\
403 & 250.9 & 247.6 & 255.7 & \ldots& \ldots& 251.2 & \o3.4 & 3 \\
414 & 285.2 & 261.9 & 254.3 & \ldots& \ldots& 264.2 &  11.6 & 3 \\
421 & 273.0 & 264.5 & 261.1 & \ldots& \ldots& 265.8 & \o4.8 & 3 \\
425 & 296.2 & 260.5 & 279.4 & \ldots& \ldots& 276.2 &  13.6 & 3 \\
428 & 277.0 & 284.6 & 297.9 & \ldots& \ldots& 286.5 & \o8.5 & 3 \\
447 & 264.6 & 272.5 & 266.6 & 260.3 & 256.5 & 264.6 & \o5.8 & 5 \\
448 & 229.0 & 281.0 & 269.0 & \ldots& 256.9 & 267.0 &\ o9.2 & 4 \\
449 & 276.0 & 269.6 & 275.1 & \ldots& 246.7 & 265.8 &  12.3 & 4 \\
452 & 262.9 & 261.5 & 258.0 & 258.9 & 252.7 & 258.6 & \o3.6 & 5 \\
453 & 263.2 & 266.0 & 263.6 & \ldots& \ldots& 264.4 & \o1.3 & 3 \\
461 & 252.8 & 283.9 & 279.7 & \ldots& 276.7 & 275.2 &  10.9 & 4 \\
463 & 276.8 & 286.0 & 275.9 & \ldots& \ldots& 279.8 & \o4.7 & 3 \\
467 & 331.7 & 277.5 & 355.6 & \ldots& \ldots& 322.3 &  30.2 & 3 \\
473 & 225.4 & 264.1 & 276.4 & \ldots& 254.5 & 257.1 &  19.3 & 4 \\
474 & 244.9 & 248.8 & 274.3 & \ldots& 276.2 & 264.1 &  14.0 & 4 \\
485 & 282.5 & 269.7 & 275.7 & \ldots& \ldots& 275.4 & \o4.8 & 3 \\
486 & 276.4 & 264.0 & 300.7 & \ldots& \ldots& 280.0 &  15.6 & 3 \\
489 & 282.6 & 247.2 & 277.2 & \ldots& \ldots& 268.9 &  15.1 & 3 \\
495 & 277.7 & 273.4 & 280.7 & 254.9 & 265.6 & 272.6 & \o7.8 & 5 \\
499 & 308.2 & 288.8 & 284.5 & \ldots& \ldots& 293.1 &  10.1 & 3 \\
523 & 281.8 & 269.5 & 264.7 & \ldots& \ldots& 271.0 & \o6.9 & 3 \\
530 & 313.5 & 276.0 & 289.2 & \ldots& \ldots& 290.3 &  14.2 & 3 \\
539 & 294.8 & 258.7 & 269.2 & \ldots& \ldots& 271.5 &  11.4 & 3 \\
543 & 249.6 & 257.5 & 251.2 & \ldots& \ldots& 253.1 & \o3.3 & 3 \\
547 & 265.9 & 259.6 & 240.9 & \ldots& \ldots& 255.9 &  10.3 & 3 \\
551 & 294.9 & 282.5 & 277.6 & \ldots& \ldots& 284.7 & \o7.1 & 3 \\
554 & 266.9 & 274.9 & 272.5 & \ldots& \ldots& 272.1 & \o2.9 & 3 \\
556 & 251.9 & 268.7 & 268.4 & \ldots& 245.0 & 260.5 &  10.2 & 4 \\
563 & \ldots& 247.3 & 278.6 & \ldots& 271.2 & 267.5 &  12.4 & 3 \\
567 & 283.5 & 277.7 & 279.4 & \ldots& 250.2 & 272.1 &  13.3 & 4 \\
568 & \ldots& 280.9 & 287.8 & \ldots& 270.3 & 280.3 & \o6.9 & 3 \\
570 & 253.9 & 261.7 & 322.8 & \ldots& \ldots& 279.4 &  30.4 & 3 \\
574 & 268.2 & 267.9 & 269.5 & 240.6 & 265.3 & 264.7 & \o8.4 & 5 \\
580 & 279.1 & 285.7 & 284.0 & \ldots& \ldots& 283.3 & \o2.6 & 3 \\
584 & 250.8 & 248.3 & 261.3 & \ldots& \ldots& 253.3 & \o5.7 & 3 \\
592 & 234.4 & 268.0 & 261.2 & \ldots& \ldots& 257.4 &  13.2 & 3 \\
600 & 246.4 & 277.0 & 290.7 & \ldots& 297.9 & 279.8 &  18.4 & 4 \\
602 & 300.0 & 299.4 & 302.9 & \ldots& \ldots& 301.0 & \o1.6 & 3 \\
605 & 316.7 & 293.6 & 287.9 & \ldots& 243.2 & 278.6 &  26.3 & 4 \\
610 & 245.3 & 249.6 & 254.4 & \ldots& 232.8 & 245.8 & \o8.4 & 4 \\
615 & 246.9 & 275.6 & 254.6 & \ldots& 266.9 & 262.7 &  10.0 & 4 \\
620 & 286.9 & 276.1 & 292.2 & 279.0 & 261.0 & 278.0 &  11.8 & 5 \\
624 & 260.9 & 266.9 & 259.9 & \ldots& 271.9 & 265.3 & \o4.9 & 4 \\
629 & 280.4 & 237.7 & 278.9 & \ldots& \ldots& 265.1 &  19.9 & 3 \\
632 & 285.4 & 281.4 & 282.9 & 312.9 & 274.7 & 284.7 &  11.1 & 5 \\
633 & 274.8 & 277.2 & 279.4 & 281.3 & 261.0 & 274.0 & \o7.2 & 5 \\
636 & \ldots& 237.8 & 248.7 & \ldots& 237.7 & 241.5 & \o5.2 & 3 \\
640 & 264.9 & 252.4 & 253.3 & 262.1 & 253.4 & 256.0 & \o5.0 & 5 \\
643 & 247.3 & 252.5 & 242.6 & \ldots& 266.3 & 250.9 & \o8.2 & 4 \\
644 & 216.8 & 266.9 & 250.1 & \ldots& \ldots& 248.6 &  20.2 & 3 \\
646 & 255.4 & 261.5 & 270.1 & \ldots& 264.6 & 263.9 & \o4.8 & 4 \\
659 & 265.5 & 264.3 & 265.7 & 252.2 & 269.0 & 264.9 & \o4.6 & 5 \\
660 & \ldots& 278.3 & 260.2 & \ldots& 242.0 & 257.1 &  14.7 & 3 \\
670 & 285.5 & 269.7 & 278.7 & 304.1 & 305.2 & 286.1 &  13.8 & 5 \\
671 & 251.3 & 274.6 & 254.4 & \ldots& 237.3 & 255.3 &  13.3 & 4 \\
676 & 295.0 & 249.7 & 277.5 & \ldots& \ldots& 272.1 &  18.7 & 3 \\
678 & 265.7 & 266.0 & 287.6 & \ldots& 258.9 & 270.2 &  11.2 & 4 \\
679 & 262.9 & 268.2 & 263.8 & 265.8 & 265.9 & 265.5 & \o1.8 & 5 \\
683 & 261.5 & 261.9 & 263.6 & \ldots& 299.4 & 271.1 &  15.7 & 4 \\
684 & \ldots& 270.2 & 269.4 & \ldots& 298.0 & 278.6 &  13.1 & 3 \\
685 & 229.7 & 225.1 & 278.5 & 253.4 & \ldots& 243.4 &  22.5 & 4 \\
699 & 280.4 & 249.5 & 267.5 & \ldots& \ldots& 264.4 &  12.0 & 3 \\
701 & 281.6 & 263.6 & 256.2 & \ldots& \ldots& 266.6 &  10.1 & 3 \\
709 & 258.1 & 271.6 & 259.3 & \ldots& 272.2 & 265.5 & \o6.5 & 4 \\
720 & 294.0 & 301.0 & 292.4 & 306.1 & 302.0 & 298.4 & \o4.8 & 5 \\
726 & 267.5 & 240.7 & 261.4 & \ldots& 239.2 & 249.7 &  12.1 & 4 \\
734 & 247.7 & 269.1 & 265.7 & \ldots& 270.9 & 263.7 & \o8.9 & 4 \\
738 & 272.6 & 273.8 & 302.4 & \ldots& \ldots& 284.4 &  14.1 & 3 \\
741 & 269.1 & 270.5 & 255.5 & 284.1 & 248.6 & 263.2 &  10.9 & 5 \\
747 & 274.6 & 269.9 & 271.7 & 266.7 & 267.9 & 270.2 & \o2.5 & 5 \\
749 & 257.2 & 274.6 & 258.0 & 276.7 & 264.0 & 264.8 & \o7.2 & 5 \\
754 & 234.5 & 244.0 & 242.9 & 244.5 & 237.2 & 240.2 & \o4.0 & 5 \\
757 & 264.9 & 275.3 & 271.4 & 285.8 & 279.5 & 273.4 & \o6.3 & 5 \\
780 & 265.4 & 245.0 & 242.3 & 238.7 & 230.7 & 243.4 &  11.4 & 5 \\
781 & 268.0 & 260.9 & 287.2 & \ldots& \ldots& 272.7 &  11.5 & 3 \\
786 & 283.0 & 290.2 & 264.2 & \ldots& 262.0 & 274.0 &  12.0 & 4 \\
789 & 245.2 & 283.6 & 261.0 & \ldots& \ldots& 263.6 &  15.5 & 3 \\
794 & 267.8 & 271.6 & 294.8 & \ldots& \ldots& 278.9 &  12.1 & 3 \\
796 & 274.7 & 288.0 & 290.7 & \ldots& \ldots& 284.9 & \o6.9 & 3 \\
798 & 279.1 & 254.3 & 250.4 & 245.5 & 219.1 & 247.1 &  19.5 & 5 \\
800 & 277.7 & 247.2 & 286.3 & \ldots& \ldots& 268.9 &  17.2 & 3 \\
804 & \ldots& 266.1 & 260.4 & \ldots& 245.3 & 256.5 & \o9.0 & 3 \\
813 & 271.1 & 271.8 & 268.5 & \ldots& 286.3 & 274.5 & \o7.1 & 4 \\
815 & 252.1 & 259.9 & 248.9 & 266.3 & 257.4 & 255.7 & \o5.4 & 5 \\
823 & 259.8 & 253.1 & \ldots& 255.9 & 272.1 & 262.0 & \o8.2 & 4 \\
824 & 223.0 & 258.8 & 248.9 & 249.2 & 250.1 & 245.9 &  12.4 & 5 \\
825 & \ldots& 244.1 & 241.8 & \ldots& 277.1 & 252.9 &  15.6 & 3 \\
826 & 241.6 & 243.1 & 256.5 & 239.5 & 230.8 & 242.8 & \o9.0 & 5 \\
832 & 259.6 & 262.4 & 258.5 & 289.5 & 262.9 & 263.6 & \o8.5 & 5 \\
833 & 286.6 & 277.0 & 276.4 & \ldots& 268.5 & 277.0 & \o5.9 & 4 \\
836 & 252.7 & 251.8 & 255.1 & \ldots& \ldots& 253.3 & \o1.4 & 3 \\
838 & 270.3 & 260.2 & 258.2 & \ldots& 265.0 & 263.4 & \o4.5 & 4 \\
840 & 246.9 & 225.1 & 244.7 & \ldots& 234.1 & 236.7 & \o8.9 & 4 \\
842 & 192.8 & 260.6 & 277.9 & 210.8 & \ldots& 244.8 &  34.7 & 4 \\
843 & 295.3 & 281.0 & 275.2 & \ldots& 269.0 & 277.9 & \o8.8 & 4 \\
846 & \ldots& 260.4 & 262.7 & \ldots& 252.7 & 258.4 & \o4.4 & 3 \\
848 & 242.5 & 246.4 & 249.9 & 257.2 & \ldots& 247.7 & \o4.6 & 4 \\
853 & 237.0 & 264.5 & 272.8 & \ldots& \ldots& 258.9 &  15.1 & 3 \\
857 & 253.6 & 256.7 & 260.5 & 257.3 & 250.3 & 255.4 & \o3.8 & 5 \\
869 & 298.0 & 283.1 & 283.1 & \ldots& 294.0 & 288.8 & \o6.4 & 4 \\
875 & 245.8 & 242.8 & 259.1 & \ldots& 261.7 & 252.7 & \o8.3 & 4 \\
876 & 254.1 & 261.1 & 235.9 & \ldots& 254.4 & 251.0 & \o9.5 & 4 \\
880 & \ldots& 285.1 & 280.6 & \ldots& 242.4 & 270.4 &  18.7 & 3 \\
882 & 280.1 & 296.1 & 287.0 & \ldots& 271.2 & 283.3 & \o9.4 & 4 \\
889 & 284.5 & 255.1 & 265.9 & \ldots& 251.8 & 263.1 &  12.1 & 4 \\
\hline
\end{longtable}
\tablefoot{Column entries are: (1) VFTS identifier; (2-6) estimated
  radial velocity from each line; (7) weighted mean velocity for each
  star ($v_{\rm r}$, from eqn.~\ref{mean_vr}); (8) standard deviation
  ($\sigma$, from eqn.~\ref{sigma_vr}); (9) number of lines ($n$) used
  in calculation of $v_{\rm r}$ and $\sigma$.}
\end{center}
\end{landscape}

\begin{landscape}
\begin{center}
\begin{longtable}{lccccccccccccc}
  \caption{\leftline{Line-by-line radial-velocity estimates for single-lined binaries
using absorption lines in the LR02 observations.\label{rv_binaries}}}\\
  \hline
  VFTS & MJD & He\,I$_{\lambda4009}$ & He\,I$_{\lambda4144}$ & He\,I$_{\lambda4169}$ 
  & C\,II$_{\lambda4267}$ & He\,I$_{\lambda4388}$ & He\,I$_{\lambda4438}$ 
  & Si\,III$_{\lambda4553}$ &  $v_{\rm single}$ & $\sigma$ & n & $v_{\rm r}$ & $\delta v_{\rm r\,max}$ \\
  \hline
\endfirsthead
  \caption{\leftline{\it{continued}}}\\
\hline
  VFTS & MJD & He\,I$_{\lambda4009}$ & He\,I$_{\lambda4144}$ & He\,I$_{\lambda4169}$ 
  & C\,II$_{\lambda4267}$ & He\,I$_{\lambda4388}$ & He\,I$_{\lambda4438}$ 
  & Si\,III$_{\lambda4553}$ & $v_{\rm single}$ & $\sigma$ & n & $v_{\rm r}$ & $\delta v_{\rm r\,max}$ \\
\hline
\endhead
\hline
\multicolumn{14}{r}{\it{continued on next page}}\\
\endfoot
\hline
\endlastfoot
009 & 54817 & 302.0 & 293.5 & \ldots& 293.8 & 288.7 & \ldots& 291.5 & 293.6 & \o4.8 & 5 & 293.2\,$\pm$\,21.6 & \o52.6 \\
015 & 54817 & 289.3 & 310.7 & \ldots& \ldots& 309.9 & \ldots& \ldots& 305.3 & \o9.0 & 3 & 278.0\,$\pm$\,30.5 & \o65.7 \\
017 & 54828 & 250.3 & 261.8 & \ldots& \ldots& 247.8 & \ldots& \ldots& 253.3 & \o6.3 & 3 & 280.9\,$\pm$\,31.0 & \o72.1 \\
018 & 54860 & 292.8 & 295.5 & \ldots& 288.8 & 295.6 & 287.7 & 288.5 & 292.9 & \o3.2 & 6 & 279.2\,$\pm$\,11.9 & \o21.4 \\
025 & 54828 & 300.0 & 308.3 & \ldots& \ldots& 299.0 & \ldots& \ldots& 302.2 & \o4.1 & 3 & 290.3\,$\pm$\,11.3 & \o28.4 \\
027 & 54748 & 327.1 & 335.3 & 340.5 & 347.0 & 329.8 & \ldots& 374.6 & 343.3 &  18.0 & 6 & 273.1\,$\pm$\,42.5 &  111.4 \\
033 & 54748 & 255.2 & 252.5 & \ldots& \ldots& 263.7 & \ldots& \ldots& 257.5 & \o4.9 & 3 & 270.2\,$\pm$\,22.8 & \o51.2 \\
037 & 54828 & 297.7 & 288.2 & \ldots& \ldots& 303.3 & \ldots& \ldots& 297.0 & \o6.1 & 3 & 299.4\,$\pm$\,18.4 & \o56.3 \\
041 & 54748 & 295.9 & 297.2 & 303.5 & 294.2 & 285.9 & 299.9 & 318.0 & 297.7 & \o9.3 & 7 & 287.0\,$\pm$\,19.0 & \o40.1 \\
097 & 54748 & \ldots& 291.6 & \ldots& \ldots& 303.6 & \ldots& 318.1 & 308.7 &  10.6 & 3 & 296.7\,$\pm$\,11.5 & \o23.8 \\
106 & 54822 & 290.8 & 289.2 & \ldots& \ldots& 282.9 & \ldots& \ldots& 287.0 & \o3.5 & 3 & 281.8\,$\pm$\,13.3 & \o32.4 \\
107 & 54822 & 289.0 & 282.0 & \ldots& \ldots& 281.2 & \ldots& \ldots& 283.2 & \o3.0 & 3 & 294.5\,$\pm$\,23.7 & \o56.5 \\
118 & 54824 & 272.8 & 281.7 & 294.7 & 281.6 & 273.4 & 286.2 & 286.0 & 279.2 & \o6.4 & 7 & 302.2\,$\pm$\,15.9 & \o39.7 \\
133 & 54824 & 294.8 & 302.8 & \ldots& 292.5 & 299.2 & \ldots& 290.8 & 298.2 & \o4.0 & 5 & 293.5\,$\pm$\,15.2 & \o38.7 \\
135 & 54817 & 257.9 & 263.1 & \ldots& \ldots& 280.6 & \ldots& \ldots& 269.0 & \o9.9 & 3 & 270.9\,$\pm$\,52.5 &  136.6 \\
144 & 54815 & 277.8 & 274.6 & \ldots& 270.6 & 273.2 & \ldots& \ldots& 274.3 & \o2.2 & 4 & 281.8\,$\pm$\,14.5 & \o34.2 \\
146 & 54748 & 238.1 & 270.0 & \ldots& \ldots& 256.5 & \ldots& \ldots& 255.4 &  12.8 & 3 & 259.6\,$\pm$\,12.6 & \o34.0 \\
157 & 54828 & 256.4 & 243.8 & \ldots& \ldots& 290.5 & \ldots& \ldots& 266.5 &  20.7 & 3 & 284.0\,$\pm$\,24.4 & \o59.7 \\
162 & 54815 & 302.1 & 285.0 & \ldots& 297.5 & 292.9 & 290.9 & 300.8 & 293.7 & \o6.1 & 6 & 283.7\,$\pm$\,14.8 & \o32.5 \\
179 & 54824 & 324.7 & 302.2 & 304.1 & 296.8 & 299.1 & 294.3 & 291.7 & 303.1 &  10.1 & 7 & 294.5\,$\pm$\,\o6.4& \o16.9 \\
189 & 54824 & 194.6 & 180.5 & \ldots& \ldots& 172.4 & \ldots& \ldots& 181.5 & \o9.0 & 3 & 274.0\,$\pm$\,69.9 &  185.9 \\
195 & 54824 & 277.1 & 281.5 & \ldots& 275.5 & 279.6 & 283.7 & 270.7 & 279.0 & \o3.7 & 6 & 268.1\,$\pm$\,12.9 & \o28.7 \\
199 & 54828 & 261.5 & 280.7 & \ldots& \ldots& 282.9 & \ldots& \ldots& 275.9 & \o9.4 & 3 & \ldots & \o\ldots \\
204 & 54815 & 252.5 & 254.7 & 258.2 & 241.0 & 247.3 & 249.7 & 241.8 & 249.7 & \o5.0 & 7 & 254.9\,$\pm$\,12.7 & \o31.0 \\
206 & 54815 & 259.2 & 283.1 & \ldots& \ldots& 266.4 & 291.7 & \ldots& 272.1 &  11.5 & 4 & 257.3\,$\pm$\,36.8 & \o84.5 \\
211 & 54817 & 301.0 & 280.3 & \ldots& \ldots& 279.2 & \ldots& \ldots& 284.7 & \o9.1 & 3 & 272.0\,$\pm$\,13.2 & \o33.6 \\
213 & 54804 & 270.7 & 273.4 & 264.4 & 279.9 & 273.2 & \ldots& 268.5 & 272.4 & \o3.3 & 6 & 276.4\,$\pm$\,12.8 & \o30.0 \\
215 & 54822 & 342.7 & 358.3 & \ldots& 364.9 & 353.5 & 338.3 & \ldots& 352.6 & \o7.6 & 5 & 287.5\,$\pm$\,63.0 &  121.7 \\
218 & 54815 & 265.4 & 264.9 & 233.0 & 255.6 & 253.6 & 253.1 & 247.3 & 257.6 & \o8.5 & 7 & 249.2\,$\pm$\,\o7.9& \o20.0 \\
225 & 54815 & 272.8 & 265.9 & 272.6 & 264.3 & 259.8 & 263.8 & 264.6 & 265.4 & \o4.5 & 7 & 289.2\,$\pm$\,15.8 & \o32.0 \\
227 & 54804 & 284.9 & 274.8 & \ldots& \ldots& 280.4 & 270.8 & 267.9 & 277.4 & \o5.2 & 5 & 277.4\,$\pm$\,\o9.1& \o23.1 \\
240 & 54809 & 112.2 & 130.1 & \ldots& \ldots& 125.3 & \ldots& \ldots& 123.2 & \o7.0 & 3 & 224.2\,$\pm$\,52.6 &  152.1 \\
246 & 54809 & 269.6 & 254.3 & \ldots& \ldots& 239.3 & \ldots& 279.2 & 255.1 &  14.5 & 4 & 275.8\,$\pm$\,28.8 & \o63.2 \\
248 & 54824 & 249.6 & 313.5 & \ldots& \ldots& 308.1 & \ldots& \ldots& 296.7 &  25.7 & 3 & 297.6\,$\pm$\,35.8 & \o89.5 \\
255 & 54804 & 276.3 & 255.7 & \ldots& \ldots& 238.6 & \ldots& \ldots& 257.7 &  14.8 & 3 & 268.5\,$\pm$\,65.4 &  163.8 \\
257 & 54817 & 285.8 & 252.7 & 272.2 & \ldots& 254.9 & \ldots& 256.4 & 263.4 &  13.3 & 5 & 269.2\,$\pm$\,11.9 & \o27.8 \\
278 & 54828 & 291.6 & 269.0 & 284.4 & 270.6 & 269.0 & \ldots& \ldots& 275.5 & \o9.5 & 5 & 270.8\,$\pm$\,12.8 & \o33.0 \\
291 & 54748 & 205.3 & 206.1 & 201.1 & 200.5 & 200.7 & 187.5 & \ldots& 201.9 & \o5.3 & 6 & 234.7\,$\pm$\,46.1 &  103.5 \\
299 & 54822 & 346.1 & 357.3 & 374.6 & 362.5 & 357.5 & 367.1 & 355.6 & 357.8 & \o7.9 & 7 & 288.9\,$\pm$\,45.8 &  107.8 \\
305 & 54817 & 304.1 & 297.7 & 309.2 & 276.2 & 292.1 & 305.8 & \ldots& 297.2 & \o9.0 & 6 & 321.7\,$\pm$\,50.4 &  123.3 \\
324 & 54815 & 296.0 & 297.7 & 317.0 & 301.5 & 295.7 & 307.1 & 305.6 & 299.7 & \o5.9 & 7 & 274.6\,$\pm$\,25.1 & \o64.3 \\
325 & 54748 & 261.6 & 278.9 & \ldots& \ldots& 279.6 & \ldots& 277.5 & 275.0 & \o7.3 & 4 & 287.4\,$\pm$\,46.0 &  118.0 \\
334 & 54817 & 244.2 & 273.3 & \ldots& 233.2 & 281.6 & \ldots& 309.6 & 268.7 &  21.5 & 5 & 266.2\,$\pm$\,17.3 & \o40.8 \\
336 & 54824 & 280.4 & 261.5 & \ldots& \ldots& 267.7 & \ldots& 252.4 & 267.4 & \o8.3 & 4 & 266.7\,$\pm$\,14.7 & \o36.7 \\
337 & 54809 & 287.4 & 297.1 & \ldots& \ldots& 260.4 & \ldots& \ldots& 282.2 &  15.6 & 3 & 278.6\,$\pm$\,11.1 & \o31.8 \\
342 & 54794 & 299.8 & 313.4 & 318.7 & 302.5 & 304.7 & 309.8 & 300.9 & 306.7 & \o5.9 & 7 & 292.2\,$\pm$\,43.4 &  103.6 \\
351 & 54794 & 231.6 & 238.3 & 237.8 & 227.6 & 225.4 & 231.4 & 226.3 & 230.8 & \o5.3 & 7 & 246.8\,$\pm$\,24.9 & \o57.8 \\
359 & 54824 & 295.8 & 295.4 & 301.1 & \ldots& 286.6 & 276.5 & 275.3 & 288.6 & \o8.6 & 6 & 295.0\,$\pm$\,10.5 & \o25.6 \\
364 & 54822 & 250.8 & 224.4 & 247.7 & 232.6 & 227.9 & 202.2 & \ldots& 232.1 &  13.5 & 6 & 266.6\,$\pm$\,29.9 & \o67.5 \\
374 & 54809 & 293.2 & 248.9 & \ldots& \ldots& 236.6 & \ldots& \ldots& 254.8 &  22.2 & 3 & 291.5\,$\pm$\,21.7 & \o65.5 \\
375 & 54809 & 280.4 & 320.3 & \ldots& \ldots& 324.0 & \ldots& \ldots& 311.6 &  18.2 & 3 & 296.4\,$\pm$\,22.1 & \o58.5 \\
383 & 54809 & 315.9 & 325.0 & \ldots& \ldots& 319.9 & \ldots& \ldots& 320.5 & \o3.7 & 3 & 224.5\,$\pm$\,64.2 &  168.0 \\
388 & 54824 & 270.7 & 260.9 & \ldots& \ldots& 255.9 & 274.5 & 254.4 & 261.0 & \o6.8 & 5 & 276.4\,$\pm$\,\o9.5& \o25.4 \\
396 & 54809 & 224.2 & 281.6 & \ldots& \ldots& 257.4 & \ldots& \ldots& 258.0 &  21.5 & 3 & 270.9\,$\pm$\,17.7 & \o38.9 \\
430 & 54824 & 193.2 & 207.7 & \ldots& \ldots& 211.7 & \ldots& 216.0 & 208.1 & \o8.0 & 4 & 218.3\,$\pm$\,48.7 &  120.8 \\
459 & 54828 & 265.7 & 249.9 & \ldots& \ldots& 247.8 & \ldots& \ldots& 252.9 & \o7.3 & 3 & 260.5\,$\pm$\,10.9 & \o29.1 \\
496 & 54794 & 264.4 & 274.8 & \ldots& \ldots& 293.9 & \ldots& \ldots& 278.6 &  12.2 & 3 & 279.9\,$\pm$\,14.2 & \o35.5 \\
501 & 54815 & \ldots& 303.9 & 327.9 & \ldots& 328.5 & 313.5 & 297.8 & 317.0 &  12.7 & 5 & 287.2\,$\pm$\,68.8 &  141.4 \\
520 & 54822 & 280.6 & 274.4 & \ldots& 254.2 & 251.5 & 246.6 & 263.8 & 262.7 &  12.3 & 6 & 278.2\,$\pm$\,73.5 &  180.0 \\
525 & 54748 & 299.7 & 293.9 & 304.2 & 296.9 & 289.0 & 291.0 & 288.9 & 293.0 & \o4.8 & 7 & 289.6\,$\pm$\,\o9.5& \o25.4 \\
534 & 54828 & 275.1 & 292.4 & 270.1 & \ldots& 281.3 & 275.4 & \ldots& 277.9 & \o6.3 & 5 & 278.2\,$\pm$\,23.1 & \o58.7 \\
548 & 54822 & 262.4 & 292.5 & \ldots& \ldots& 299.4 & \ldots& 295.7 & 288.4 &  14.4 & 4 & 279.9\,$\pm$\,\o8.5& \o20.7 \\
575 & 54828 & 229.8 & 228.6 & 232.3 & 225.9 & 224.9 & 219.1 & \ldots& 226.9 & \o3.4 & 6 & 263.4\,$\pm$\,24.9 & \o55.5 \\
576 & 54815 & 285.4 & 287.6 & 288.5 & 282.1 & 284.9 & 274.9 & 289.8 & 285.7 & \o3.8 & 7 & 262.0\,$\pm$\,34.9 & \o74.6 \\
589 & 54794 & \ldots& \ldots& 223.8 & 204.8 & 201.3 & 208.8 & 206.7 & 205.8 & \o6.3 & 5 & 314.5\,$\pm$\,94.0 &  178.7 \\
591 & 54804 & 285.6 & 288.1 & 285.8 & 294.3 & 286.9 & 278.7 & \ldots& 286.8 & \o2.8 & 6 & 277.7\,$\pm$\,\o8.8& \o21.3 \\
606 & 54828 & 255.1 & 264.2 & \ldots& 264.0 & 270.4 & 266.7 & 286.5 & 266.7 & \o8.0 & 6 & 265.6\,$\pm$\,10.3 & \o28.8 \\
628 & 54824 & 236.6 & 265.7 & \ldots& \ldots& 263.7 & \ldots& \ldots& 257.9 &  11.9 & 3 & 250.8\,$\pm$\,62.7 &  152.7 \\
637 & 54809 & 274.1 & 235.3 & \ldots& \ldots& 237.1 & \ldots& \ldots& 246.0 &  16.5 & 3 & 299.3\,$\pm$\,32.3 &  103.1 \\
662 & 54804 & 235.9 & 287.1 & \ldots& \ldots& 253.2 & \ldots& \ldots& 256.3 &  20.3 & 3 & 251.3\,$\pm$\,39.1 &  100.8 \\
665 & 54804 & 264.5 & 275.7 & 267.4 & \ldots& 265.4 & 264.1 & 264.5 & 267.7 & \o4.5 & 6 & 274.2\,$\pm$\,16.1 & \o37.2 \\
675 & 54748 & 260.0 & 267.4 & 273.9 & 261.2 & 262.6 & 265.0 & 270.7 & 266.0 & \o4.5 & 7 & 276.7\,$\pm$\,23.8 & \o59.5 \\
686 & 54804 & 297.8 & 292.8 & 311.1 & 298.1 & 293.9 & 302.7 & 303.8 & 298.2 & \o5.4 & 7 & 276.8\,$\pm$\,38.9 & \o90.0 \\
687 & 54748 & 316.7 & 312.1 & 285.6 & 319.0 & 315.2 & 293.2 & 321.6 & 314.0 & \o8.8 & 7 & 295.8\,$\pm$\,23.0 & \o56.8 \\
697 & 54794 & 264.8 & 280.6 & 274.2 & 265.5 & 267.8 & \ldots& \ldots& 271.1 & \o6.7 & 5 & 263.7\,$\pm$\,\o8.3& \o20.8 \\
705 & 54809 & 168.6 & 183.5 & \ldots& \ldots& 172.1 & \ldots& 163.7 & 173.8 & \o7.0 & 4 & 268.3\,$\pm$\,80.8 &  174.9 \\
713 & 54794 & 239.4 & 233.7 & \ldots& \ldots& 234.9 & \ldots& \ldots& 235.6 & \o2.3 & 3 & 250.6\,$\pm$\,38.7 &  100.9 \\
715 & 54809 & 262.6 & 283.0 & \ldots& 292.5 & 268.0 & 299.3 & 271.7 & 275.5 &  11.6 & 6 & 270.0\,$\pm$\,22.1 & \o53.1 \\
718 & 54828 & 323.0 & 320.8 & \ldots& \ldots& 327.8 & \ldots& \ldots& 323.9 & \o2.9 & 3 & 306.8\,$\pm$\,16.2 & \o39.4 \\
719 & 54804 & 230.7 & 253.5 & \ldots& 239.9 & 226.9 & 232.5 & 240.1 & 237.0 &  10.4 & 6 & 265.0\,$\pm$\,18.9 & \o52.4 \\
723 & 54794 & 279.6 & 283.7 & 294.0 & 279.3 & 280.5 & 271.6 & 279.8 & 281.3 & \o4.8 & 7 & 239.2\,$\pm$\,26.3 & \o62.1 \\
730 & 54822 & 326.8 & 333.0 & \ldots& 377.0 & 326.4 & \ldots& \ldots& 332.1 &  12.6 & 4 & 326.5\,$\pm$\,\o6.8& \o16.6 \\
752 & 54817 & 261.7 & 260.5 & \ldots& \ldots& 267.6 & 247.8 & 241.8 & 260.7 & \o7.7 & 5 & 258.8\,$\pm$\,40.3 & \o87.4 \\
779 & 54809 & 248.0 & 248.7 & 244.2 & 246.7 & 245.2 & 249.4 & 250.8 & 247.7 & \o2.1 & 7 & 258.9\,$\pm$\,27.2 & \o61.0 \\
784 & 54748 & 339.3 & 259.2 & \ldots& \ldots& 274.6 & \ldots& \ldots& 286.2 &  31.9 & 3 & 312.4\,$\pm$\,27.6 & \o70.7 \\
788 & 54824 & 295.6 & 314.4 & \ldots& \ldots& 292.9 & \ldots& 308.1 & 302.0 & \o9.4 & 4 & 280.7\,$\pm$\,27.8 & \o70.1 \\
792 & 54815 & 265.0 & 281.9 & 285.5 & 265.3 & 267.0 & 276.5 & \ldots& 271.9 & \o7.7 & 6 & 278.3\,$\pm$\,\o8.0& \o18.0 \\
799 & 54809 & 281.3 & 276.6 & 281.6 & 270.2 & 263.3 & 271.2 & 265.3 & 271.6 & \o7.0 & 7 & 270.7\,$\pm$\,\o7.3& \o19.9 \\
827 & 54828 & 244.8 & 245.6 & 260.4 & 243.1 & 238.2 & 236.9 & 241.2 & 243.1 & \o5.5 & 7 & 254.3\,$\pm$\,14.4 & \o34.7 \\
834 & 54809 & 211.0 & 207.1 & \ldots& 219.4 & 219.3 & \ldots& 214.9 & 213.8 & \o5.0 & 5 & 240.3\,$\pm$\,22.1 & \o43.6 \\
837 & 54828 & 306.0 & 301.8 & \ldots& \ldots& 315.2 & \ldots& \ldots& 308.4 & \o5.8 & 3 & 282.5\,$\pm$\,27.0 & \o66.0 \\
847 & 54822 & 317.6 & 299.9 & \ldots& \ldots& 300.7 & \ldots& 288.6 & 302.0 & \o9.6 & 4 & 298.7\,$\pm$\,\o9.6& \o26.2 \\
850 & 54828 & 255.4 & 249.0 & \ldots& 242.5 & 234.6 & 232.3 & 232.2 & 241.8 & \o9.4 & 6 & 256.9\,$\pm$\,11.6 & \o27.1 \\
874 & 54828 & 277.5 & 277.4 & 247.9 & 265.2 & 280.4 & \ldots& 283.4 & 274.8 & \o9.5 & 6 & 274.9\,$\pm$\,\o8.3& \o20.0 \\
877 & 54815 & 234.4 & 240.2 & \ldots& \ldots& 207.4 & \ldots& \ldots& 228.2 &  14.2 & 3 & 236.7\,$\pm$\,11.2 & \o26.6 \\
883 & 54828 & 254.6 & 301.7 & \ldots& \ldots& 295.5 & \ldots& \ldots& 289.9 &  17.3 & 3 & \ldots & \o\ldots \\
888 & 54815 & 317.6 & 326.7 & 334.2 & 316.7 & 314.2 & \ldots& 330.5 & 321.4 & \o7.1 & 6 & 238.3\,$\pm$\,50.8 &  121.0 \\
890 & 54828 & \ldots& 304.9 & \ldots& \ldots& 302.2 & \ldots& 316.2 & 305.8 & \o5.3 & 3 & \ldots & \o\ldots \\
891 & 54815 & 289.1 & 278.7 & 278.3 & 283.3 & 271.3 & 279.9 & 275.6 & 279.0 & \o6.1 & 7 & 273.6\,$\pm$\,\o8.3& \o20.5 \\
\hline
\end{longtable}
\tablefoot{Column entries are: (1) VFTS identifier; (2) Modified
  Julian Date (MJD) of the co-added observations used for these measurements
  (see Appendix of Paper~I for details of exposures); (3-9)
  estimated radial velocity from each line; (10) weighted mean
  (single-epoch) velocity for each star ($v_{\rm single}$, from
  eqn.~\ref{mean_vr}); (11) standard deviation ($\sigma$, from
  eqn.~\ref{sigma_vr}); (12) number of lines ($n$) used in calculation
  of $v_{\rm single}$ and $\sigma$; (13) estimated mean $v_{\rm r}$
  (and standard deviation) calculated from the multi-epoch
  measurements from Dunstall et al. (in prep.); (14) maximum difference ($\delta
  v_{\rm r\,max}$) in the radial velocity estimates between the
  available epochs.}
\end{center}
\end{landscape}

\end{document}